\pdfoutput=1
\documentclass[11pt, a4paper]{article}
\usepackage{relsize}
\usepackage{array}
\usepackage{a4wide}
\usepackage[hidelinks]{hyperref}
\usepackage{latexsym}
\usepackage{cite}
\usepackage{mathrsfs}
\usepackage{amssymb}
\usepackage{amsmath}
\usepackage{amsfonts}
\usepackage{amsthm}
\usepackage{graphicx}
\usepackage{enumerate}
\usepackage[usenames,dvipsnames]{xcolor}
\usepackage{todonotes}
\usepackage{bm}
\usepackage{comment}
\usepackage{cancel}
\usepackage{float}
\usepackage{verbatim}
\usepackage{mathtools}
\usepackage{subcaption}
\usepackage{xcolor}

\usepackage{tikz}\usepackage{tikz-feynman}
\graphicspath{ {Images/} }

\renewcommand{\vec}{\bs}  
\newcommand{\bs}[1]{\boldsymbol{#1}}

\usepackage{etoolbox}
\apptocmd{\sloppy}{\hbadness 10000\relax}{}{}

\newcommand{\cA}{\mathcal{A}}

\newcommand{\be}{\begin{equation}\label}
\newcommand{\ee}{\end{equation}}
\newcommand{\bea}{\begin{eqnarray}\label}
\newcommand{\eea}{\end{eqnarray}}

\newcommand{\la}{\langle}
\newcommand{\ra}{\rangle}

\newcommand{\cD}{\mathcal{D}}

\makeatletter
\newcommand*{\textoverline}[1]{$\overline{\hbox{#1}}\m@th$}
\newcommand*\bigcdot{\mathpalette\bigcdot@{.65}}
\newcommand*\bigcdot@[2]{\mathbin{\vcenter{\hbox{\scalebox{#2}{$\m@th#1\bullet$}}}}}
\makeatother

\newcommand{\eq}[1]{\begin{equation}#1\end{equation}}
\newcommand{\eqs}[1]{\begin{equation}\begin{split}#1\end{split}\end{equation}}

\newcommand{\eqn}{&=&}
\newcommand{\non}{\notag \\}
\newcommand{\eqnmy}{&=&}
\newcommand{\nonmy}{\notag \\}
\newcommand{\gr}{\text{GR}}

\date{}

\begin{document}

\title{From on-shell amplitude in AdS to cosmological correlators: gluons and gravitons}

\author{Jiajie Mei${}^{a,b}$ and Yuyu Mo${}^c$ \vspace{7pt}\\ \normalsize \textit{
${}^a$
Department of Mathematical Sciences}\\\normalsize\textit{Durham University, Durham, DH1 3LE, United Kingdom}\\
\textit{
${}^b$ Institute of Physics}\\\normalsize\textit{University of Amsterdam, Amsterdam, 1098 XH, The Netherlands}\\ 
\textit{
${}^c$
Higgs Centre for Theoretical Physics}\\\normalsize\textit{School of Physics and Astronomy},\\\normalsize\textit{
The University of Edinburgh, Edinburgh EH9 3FD, Scotland, UK}}

{\let\newpage\relax\maketitle}
\begin{abstract}

We recently introduced a recursive bootstrap method for constructing $n$-point gluon and graviton Mellin-momentum amplitudes in (A)dS spacetime. The power of this approach was illustrated by our successful computation of the first five-point graviton amplitude in (A)dS. In this work, we provide further details of these calculations and start with a more in-depth review of the formalism by offering detailed insights into the motivation for defining on-shell amplitudes in AdS, along with various explicit examples. Furthermore, we demonstrate how cosmological correlators can be effortlessly obtained once the amplitude is determined.
\end{abstract}

\pagebreak
\tableofcontents

\newpage

\section{Introduction}
In recent decades, mounting evidence suggests that directly understanding observables can be more insightful than focusing on the underlying theory's Lagrangian. The simplicity of a theory may be obscured by its Lagrangian formulation, whereas a direct approach to observables can reveal fundamental insights \cite{Arkani-Hamed:2008owk}. Significant progress has been made in bootstrapping the S-matrix using fundamental physical principles such as Lorentz invariance, locality, and unitarity \cite{Benincasa:2007xk, Elvang:2015rqa, Cheung:2017pzi}.

Gravity stands out as one of the most compelling examples of this shift in perspective. Einstein gravity is notoriously difficult to compute, even at the perturbative level. In contrast, the modern scattering amplitudes approach in flat space has achieved remarkable success by using the on-shell methods to compute gravity amplitudes \cite{Britto:2005fq,Arkani-Hamed:2008bsc,Benincasa:2007qj,Cachazo:2005ca}.

Despite the success of these methods in flat space, many mysteries remain. One might wonder whether these successes are merely coincidences of flat space, especially considering that our universe is not flat. A natural question arises: can these approaches be implemented or understood in curved backgrounds? (Anti) de Sitter space provides a simple yet meaningful playground for exploring these questions. In particular, de Sitter (dS) space is relevant to cosmology, as it can approximate the spacetime during inflation. This area of study, known as the cosmological bootstrap, has been an active research area \cite{Baumann:2022jpr,Arkani-Hamed:2018kmz,Baumann:2020dch,Pajer:2020wxk,Goodhew:2020hob,Melville:2021lst,Jazayeri:2021fvk}.

Our next question is: What are the observables? In cosmology, the known observables are the cosmological correlators (in-in correlators). However, these correlators are not invariant under field redefinition or gauge transformation\footnote{The wavefunction coefficients also suffer the same problem.}, which has limited our understanding of higher point amplitude, especially regarding spinning particles \cite{Albayrak:2018tam,Chu:2023kpe,Armstrong:2020woi,Albayrak:2019yve,Albayrak:2020fyp,Chu:2023pea,Albayrak:2019asr,Chen:2023xlt,Albayrak:2023jzl,Bissi:2022wuh,Caron-Huot:2021kjy,Albayrak:2023kfk}. To achieve similar success as seen with the S-matrix, we employ a new representation \cite{Mei:2023jkb}, where the cosmological correlators can be directly obtained from computations involving AdS Witten diagrams as we will explain in the paper.

The first goal of this paper is to demonstrate that the Mellin-momentum amplitude \cite{Mei:2023jkb} exhibits similar analytic properties to the S-matrix in flat space, remaining invariant under field redefinition and gauge transformation. This simple structure makes use of the differential representation\cite{Eberhardt:2020ewh,Roehrig:2020kck,Cheung:2022pdk,Herderschee:2022ntr,Li:2023azu,Lee:2022fgr,Lee:2023qqx,Li:2022tby} which allows us to mimic the calculations of scattering amplitudes in flat space.
Along the way, we introduce a novel and efficient algorithm for bootstrapping $n$-point amplitudes, incorporating the modern on-shell amplitude approach. The key concept involves recycling lower-point on-shell amplitudes to recursively construct higher-point amplitudes. By taking the residue of OPE poles, we fix the amplitude up to contact terms. Finally, by comparing with the soft limit and flat space limit, we determine all the contact terms. It is crucial to emphasize that our algorithm is entirely automated and requires no guesswork, making it possible to explore unknown higher-point functions. Our method proves valuable not only for understanding the structure of higher-point Mellin-momentum amplitudes but also for easily mapping the results from amplitudes to cosmological correlators.

As the most important and compelling example, we consider the gravity correlator. In \cite{Bonifacio:2022vwa}, the four-point gravity correlator in $dS_4$ was first determined in full generality. Using this new method, as shown in \cite{Mei:2024abu}, we present the first five-graviton amplitude in (A)dS. This paper provides further illustrations of the bootstrap technique and offers more examples to manifest the properties of Mellin-momentum amplitudes.

The paper is organized as follows.   
In section \ref{section2}, we review Mellin-momentum formalism in detail, covering its motivation, definition, and some simple scalar examples. Section \ref{section3} introduces a general setup for the spinning particles and provides flat-space-like Feynman rules for gluons. We then use the Feynman rules to compute all the Witten diagrams up to five-point and discuss the general pole structure for gluon amplitudes. In section \ref{section4} we give detailed explanation of the amplitude bootstrap procedure and explicitly demonstrate the bootstrap process for four and five-point Yang-Mills (YM) and General Relativity (GR) Mellin-momentum amplitude. In section \ref{backtomomentum}, we present an efficient and recursive method for computing the remaining scalar integrals to obtain wavefunction coefficients for YM and GR. Finally, we offer graphical rules for deriving cosmological correlators from the wavefunction coefficients.
In section \ref{section6} we present our conclusions with potential future directions. There are appendices giving technical details on field redefinition, conformal ward identities, the derivation of YM and GR propagator, and specific bootstrapping process, such as determining double OPE poles for five-point YM and single OPE poles for five-point GR amplitudes.

\textbf{Notation and conventions}: Our boundary momentum is $k_{\mu}=\vec{k}$, and $\mu$ runs over the boundary coordinates, while $m$ runs over both the boundary and radial coordinates $z,\mu$. All the dot products are defined by boundary coordinates: $\eta_{\mu \nu} k^{\mu} \varepsilon^{\nu}= k \cdot \varepsilon$.\\
We will often use the following short-hand notation to write the dot product of polarizations.
\eqs{
\varepsilon_{ij,kl}:= \varepsilon_i \cdot \varepsilon_j \varepsilon_k \cdot \varepsilon_l \quad \mathcal{K}_{ij,kl}:= k_i \cdot \varepsilon_j k_k \cdot \varepsilon_l
}

\textbf{Bootstrap codes:} The codes documenting the bootstrap process—\texttt{4ym.nb} for the four-point Yang-Mills (YM), \texttt{4gr.nb} for the four-point General Relativity (GR), \texttt{5ym.nb} for the five-point YM, and \texttt{5gr.nb} for the five-point GR—are available in the
\href{https://github.com/YuyuMo-UoE/On-shell-bootstrap-for-Mellin-momentum-amplitude}{GitHub repository}.

\section{Mellin-Momentum space}
\label{section2}
\subsection{Motivation for on-shell amplitude}
Classically, a conserved spin-1 current obeys, $\partial^{\mu}J_{\mu}=0$, but when inserted inside the correlation function it holds only on the separated points which leads to the well-known Ward-Takahashi identity:
\begin{equation}\label{WTIDENTITY1}
\partial^{\mu}\left\langle J_{\mu}\left(\vec{x}_1\right) O\left(\vec{x}_2\right) \dots O\left(\vec{x}_n\right)\right\rangle=-\sum_{a=2}^n \delta\left(\vec{x}_1-\vec{x}_a\right)\left\langle O\left(\vec{x}_2\right) \dots \delta O\left(\vec{x}_a\right) \dots O\left(\vec{x}_n\right)\right\rangle .
\end{equation}
The right-hand side, characterized by a delta function support, will be referred to as local terms, as it represents a local effect that is non-zero only when two operators are close to each other \footnote{These are also referred to as contact terms, but to distinguish them from the bulk contact interaction terms, we will be referred them as local terms.}. 
\\
Taking $U(1)$ symmetry as a simple example, after Fourier transforms the Ward identity becomes:
\eqs{
k_1^{\mu} \left\langle J_{\mu}\left(\vec{k}_1\right) O\left(\vec{k}_2\right) \dots O\left(\vec{k}_n\right)\right\rangle=-\sum_{a=2}^n e_a \left\langle O\left(\vec{x}_2\right) \dots O\left(\vec{k}_1+\vec{k}_a\right) \dots O\left(\vec{x}_n\right)\right\rangle .
}
So unlike scattering amplitude or S-matrix in flat space, the correlator (cosmological correlator or wavefunction coefficient) is not invariant under gauge ward identity but only up to local terms.

 Moreover, the correlator is not a unique object. Under a field redefinition, the correlator remains invariant only up to local terms. For example, consider the wavefunction coefficient or boundary correlator in momentum space, which is not invariant under field redefinition \cite{Maldacena:2011nz,Pajer:2020wxk}. Specifically, let's examine a massless free scalar theory in $d=3$. Under a field redefinition:
\eqs{
\phi \to \phi + \alpha \phi^3,
}
the four-point correlator changes to
\eqs{
\la \phi(\vec k_1)  \phi(\vec k_2)  \phi(\vec k_3)  \phi(\vec k_4) \ra \to \la \phi(\vec k_1)  \phi(\vec k_2)  \phi(\vec k_3)  \phi(\vec k_4) \ra -\frac{1}{3} \alpha \sum\limits_{i=1}^{4}(k_i^3).
\label{correlator not inv}
}
These are the local terms or boundary contact terms in momentum space.
More generally, computing the same correlator using different coordinates, gauges, or employing free equations of motion during the derivation, can yield different results, up to local terms. In simpler, lower-point examples, identifying such local terms is feasible. However, in higher-point or more complex scenarios, this process becomes considerably more challenging. In this paper, we aim to construct higher-point correlators recursively using lower-point data. This approach raises the question of which local terms should be retained and which should be disregarded when constructing the higher-point correlators.\\

Another crucial observation is that in a theory with shift symmetry, the correlator should exhibit enhanced soft limits, meaning it should vanish when one of the legs becomes soft \cite{Armstrong:2022vgl}. However, in cases such as the simplest Nonlinear Sigma Model (NLSM), there exist local terms that do not vanish in the soft limit. It's important to note that the presence of non-zero terms in the soft limit does not imply a lack of protection by shift symmetry. Instead, this discrepancy arises because the Noether current associated with the shift symmetry conserves only up to local terms for correlators, precisely due to the Ward identity mentioned above Eq(\ref{WTIDENTITY1}).

None of these problems arise in flat space when working with the S-matrix, thanks to the LSZ reduction formula. In a similar spirit, to overcome all the challenges mentioned above, we introduce the on-shell amplitude Mellin-momentum amplitude below.

Significant progress has been made in this topic on defining such invariant observables. See defining the S-matrix in AdS boundary \cite{Giddings:1999qu}, on-shell correlators in maximally symmetric space \cite{Cheung:2022pdk}, and refining the S-matrix in de-Sitter space to enhance crossing and positivity\cite{Melville:2023kgd,Melville:2024ove}. While much of this research focuses on scalar theory, here in this work we focus on spinning particles, and how to compute spinning correlators recursively and efficiently. Our approach shares the same spirit with these prior works, so it would be very interesting to combine these works to better improve other properties of the observables.

The definition of the Mellin-momentum amplitude is largely inspired by on-shell correlators \cite{Cheung:2022pdk}. The key aspect in proving that the on-shell correlator remains invariant under field redefinitions is that the external legs are on-shell and satisfy Ward identities. The first condition is ensured by using a differential representation, allowing for the amputation of any Witten diagrams, while the second is achieved through the Mellin transform \cite{Sleight:2021plv, Sleight:2019hfp, Sleight:2019mgd}.\\
It has been argued that, for AdS correlators, a natural counterpart to momentum conservation in the S-matrix is provided by the conformal Ward identities first suggested by the study of scattering equation in AdS\cite{Eberhardt:2020ewh,Gomez:2021qfd,Gomez:2021ujt,Armstrong:2022csc,Diwakar:2021juk}. However, as an operator process, implementing this in momentum space is non-trivial, as these identities do not yield zero on the right-hand side. Nonetheless, the conformal Ward identity can be manifested through boundary momentum conservation and the Mellin delta function, as demonstrated in the examples in Appendix \ref{appendixb1} and \ref{appendixb2}. Hence, the boundary momentum conservation and Mellin delta function serve as a natural counterpart to momentum conservation in the S-matrix.
 \begin{table}[htbp]
  \centering
  \caption{Comparison between Minkowski and AdS}
  \label{tab:example}
  \begin{tabular}{|c|c|}
    \hline
    Amplitude in Minkowski space & Mellin-momentum amplitude in AdS  \\
    \hline
    Lorentz-Invariance   & Conformal Invariance  \\     \hline
    Translation symmetry   & Boundary translation + Dilatation   \\
    $e^{ik_{\mu}  x^{\mu}}$   &  $e^{ik_{\mu} x^{\mu}} z^{-2s+d/2}$  \\
    $(2\pi)^{d+1}\delta^{d+1}(\sum \limits_{i=1}^n k_i^{\mu})$   &  
    $(2\pi)^{d}\delta^{d}(\sum \limits_{i=1}^n k_i^{\mu})$\\
     &  $\times 2 \pi i \delta(d+b+\sum \limits_{i=1}^n (2 s_i-d/2))$ \\
     \hline
     On-shell condition & On-shell condition\\
     $k^2+m^2=0$ & $z^2k^2+(d/2-\Delta)^2-4s^2=0$\\
     \hline
     Pole structure & Factorization+OPE\\
     $\frac{1}{k_I^2+m^2}$ & $\frac{1}{\mathcal{D}^{\Delta}_{k}} \quad \frac{1}{(k_I)^2}$\\
     \hline

  \end{tabular}
\end{table}

\subsection{Definition of Mellin-momentum amplitude}
We will be working on the Poincar\'e patch of $AdS_{d+1}$ space with metric
\begin{equation}
\tilde{g}_{mn}dx^{m}dx^{n}=\tfrac{\mathcal{R}^{2}}{z^{2}}(dz^{2}+\eta_{\mu\nu}dx^{\mu}dx^{\nu}),
\end{equation}with $0 <z<\infty$ and $\mathcal{R}=1$. The equation of motion (EoM) operator in momentum space is defined as:
\begin{align}\label{eq:eom}
    &\mathcal{D}_{k}^{\Delta} \phi_{\Delta} (k,z) =0 ,\nonumber \\
    &\mathcal{D}_{k}^{\Delta}  \equiv   z^{2}k^{2}-z^{2}\partial_{z}^{2}-(1-d)z\partial_{z}+\Delta(\Delta-d),
\end{align}
with  $\Delta$ the scaling dimension and $k=|\vec{k}|$ the norm of the boundary momentum. The solution for the bulk-to-boundary propagator is given by 
\eqs{
\phi_{\Delta}(k, z)=\sqrt{\frac{2}{\pi }} z^{d/2} k^{\Delta -\frac{d}{2}} \mathcal{K}_{\Delta -\frac{d}{2}}(z	k)\\
}
with $\mathcal{K}_{\Delta -\frac{d}{2}}(zk)$ being the Bessel K function of the second kind. In the next step, we will consider the Mellin-Fourier transform $\phi_{\Delta}(x,z)\sim e^{ik\cdot x}z^{-2s+d/2}\phi_{\Delta}(s,k)$:
\begin{align}
    &\phi_{\Delta}(k,z)=\int_{-i\infty}^{+i\infty} \frac{ds}{2\pi i}z^{-2s+d/2} \phi_{\Delta}(s,k),\\
    &\phi_{\Delta}(s, k)=\frac{\Gamma\left(s+\frac{1}{2}\left(\frac{d}{2}-\Delta\right)\right) \Gamma\left(s-\frac{1}{2}\left(\frac{d}{2}-\Delta\right)\right)}{2 \Gamma\left(\Delta-\frac{d}{2}+1\right)}\left(\frac{k}{2}\right)^{-2 s+\Delta-\frac{d}{2}}.\label{btBMellin} 
\end{align}
We can now start to define our observables.
The observables in AdS - boundary correlators or wavefunction coefficient in dS (after Wick rotation) can be treated as the boundary correlators of CFT in momentum space,
\eqs{
\Psi_n=\delta^d(\Vec{k}_T) \langle \mathcal{O}(\Vec{k}_1) \dots \mathcal{O}(\Vec{k}_n) \rangle,
}
where $\Vec{k}_T=\vec{k}_1 +\dots \vec{k}_n$. The definition of Mellin-momentum amplitude $\mathcal{A}(zk,s)$ is given as
\begin{align}\label{eq:MMamplitude}
  \Psi_n & =  \int [ds_i]\int \frac{dz}{z^{d+1}} \mathcal{A}_n(zk,s)\prod_{i=1}^n\phi(s_i,k_i)z^{-2 s_i+d/2},
\end{align}
with $\int [ds_i]= \prod_{i=1}^n \int_{-i \infty}^{+i \infty}\frac{d s_i}{(2 \pi i)} $.

The amplitude obeys the following on-shell condition (The Mellin transform of the EoM Eq(\ref{eq:eom}) above):
\begin{align}\label{eq:mellineom}
     (z^{2}k^{2}+(d/2-\Delta)^2-4s^2)\phi_{\Delta}(s,k)=0.
\end{align}
It's worth emphasizing that momentum $k$ will always be associated with a factor $z$ to capture scale invariance, and the amplitude will also depend on the differential operator of $z$ as we will see later.\\
We can also integrate out the $z$ variable at every vertex \cite{Sleight:2021plv}:
\begin{align} \label{eq:mmdelta}
  \int_0^\infty \frac{dz}{z^{d+1}} z^{\sum\limits_{i=1}^{n} (-2s_i+d/2)}z^a=\delta(d-a+\sum\limits_{i=1}^{n}(2s_i-d/2)),
\end{align}
where $a$ is counting the extra factor of $z$ due to scale invariance and we can easily read it off from the Mellin-momentum amplitude. The RHS is referred to Mellin delta function, similar to the momentum conservation that captures translation invariant.
\subsection{Local terms and LSZ reduction}
As shown in \eqref{correlator not inv} for correlators, under the field redefinition the wavefunction coefficients/ boundary correlators are not invariant. However, we can consider the same field redefinition for the Mellin-momentum amplitude,
\eqs{
\phi \to \phi + \alpha \phi^3.
}
The amplitude will transform as,
\eqs{
\mathcal{A}\to \mathcal{A}+\alpha \sum\limits_{i=2}^{4}(z^2  k_1 \cdot k_i +(d/2-2s_1)(d/2-2s_i)).
}
Using boundary momentum conservation, Mellin delta function and on-shell condition Eq (\ref{eq:mellineom}), we can easily show that the second term is zero and hence the Mellin-momentum amplitude is invariant under field redefinition!  We leave the details of computation in Appendix \ref{field_redef}.

These observations can be explained by the generalised LSZ reduction formula in AdS \cite{Giddings:1999qu,Cheung:2022pdk,Melville:2023kgd} with similar argument as QFT in Minkowski space: \emph{local term is not singular on the on-shell poles}. Local term of boundary correlator in position space takes the following general form:
\begin{align}
    \langle \mathcal{O}(x_1) \dots \mathcal{O}(x_n)  \rangle \delta^{d}(x_i-x_j),
    \label{EXlocal}
\end{align}
which vanishes unless the two operators collide. We now consider the Mellin-Fourier transform $\phi(x,z)\sim e^{ik\cdot x}z^{-2s+d/2}\phi(s,k)$. Due to the delta function in \eqref{EXlocal}, it becomes a $n-1$ point correlator. Hence, the contact term can not be written as Eq(\ref{eq:MMamplitude}) where the definition needs $\prod_{i=1}^n \phi(s_i,k_i)$.
\begin{quote}
    Therefore, \emph{contact terms in a boundary correlator do not contribute to the Mellin-momentum amplitude $\mathcal{A}_n$.}
\end{quote}

More rigorously, one can follow the proof in \cite{Cheung:2022pdk} which closely mimics the proof of S-matrix to show that on-shell amplitude is indeed invariant under field redefinition. As emphasized before, the key point is to enforce external leg on-shell and apply conformal ward identities (which is achieved by boundary momentum conservation and Mellin delta function, otherwise conformal ward identities for correlator are usually not zero).
\subsection{Scalar examples}
We will start with a few simple examples to see how our definition of amplitude leads to similar analytic structure as S-matrix in Minkowski space.\\
\textbf{Four-point contact diagram for $\lambda \phi^4$ theory}:

\begin{center}
    \begin{tikzpicture}[scale=0.8]
	\begin{feynman}
		\vertex (a) at (-2, 2);
		\vertex (b) at (-2, -1);
		
		\vertex (c) at (2,2);
		\vertex (d) at (2, -1);
		
		\vertex [blob] (v3) at (0, 0.5);
		
		\diagram* {
			(a) -- [plain, edge label'=$p_2$] (v3),
			(b) -- [plain, edge label'=$p_1$] (v3),
			(v3) -- [plain, edge label=$p_4$] (d),
			(v3) -- [plain, edge label=$p_3$] (c),
		};
	\end{feynman}
	\end{tikzpicture}
\end{center}
 
\eqs{
\mathcal{A}^{\phi^4}_{4,c}=\lambda \times (2\pi)^{d}\delta^{d}(\sum \limits_{i=1}^4 k_i^{\mu})\times 2\pi i\delta(d+\sum \limits_{i=1}^4 (2s_i-d/2)).
\label{scalarC4pt}
}
The amplitude is simply coupling constant with momentum and Mellin conservation.\\
\textbf{Four-point exchange diagram for $\lambda \phi^3$ theory}:

\begin{center}
    \begin{tikzpicture}[scale=0.8]
		\begin{feynman}
			\vertex (a) at (-2, 2);
			\vertex (b) at (-2, -1);
			
			\vertex (c) at (2,2);
			\vertex (d) at (2, -1);
			
			\vertex [blob] (v1) at (-1, 0.5);
			\vertex [blob] (v2) at (1, 0.5);
			
			\diagram* {
				(a) -- [plain] (v1),
				(b) -- [plain] (v1),
				(v2) -- [plain] (c),
				(v2) -- [plain] (d),
				(v2) -- [plain, edge label'=$k_s$] (v1),
			};
		\end{feynman}
	\end{tikzpicture}
 \qquad 
     \begin{tikzpicture}[scale=0.8]
		\begin{feynman}
			\vertex (a) at (-2, 2);
			\vertex (b) at (-2, -1);
			
			\vertex (c) at (2,2);
			\vertex (d) at (2, -1);
			
			\vertex [blob] (v1) at ( 0,1.5);
			\vertex [blob] (v2) at ( 0,-0.5);
			
			\diagram* {
				(a) -- [plain] (v1),
				(c) -- [plain] (v1),
				(v2) -- [plain] (b),
				(v2) -- [plain] (d),
				(v2) -- [plain, edge label'=$k_t$] (v1),
			};
		\end{feynman}
	\end{tikzpicture}\qquad
  \begin{tikzpicture}[scale=0.8]
		\begin{feynman}
			\vertex (a) at (-2, 2);
			\vertex (b) at (-2, -1);
			
			\vertex (c) at (2,2);
			\vertex (d) at (2, -1);
			
			\vertex [blob] (v1) at ( 0,1.5);
			\vertex [blob] (v2) at ( 0,-0.5);
			
			\diagram* {
				(a) -- [plain] (v1),
				(d) -- [plain] (v1),
				(v2) -- [plain] (c),
				(v2) -- [plain] (b),
				(v2) -- [plain, edge label'=$k_u$] (v1),
			};
		\end{feynman}
	\end{tikzpicture}
\end{center}
\begin{eqnarray}
  \mathcal{A}_{4}^{\phi^3}=\lambda^2(2 \pi)^d\delta^d\left(\sum_{i=1}^4 k_i^\mu\right)\left(\frac{1}{\cD_{k_{s}}^\Delta}+\frac{1}{\cD_{k_{t}}^\Delta}+\frac{1}{\cD_{k_{u}}^\Delta}\right),
\end{eqnarray}
where $\cD_{k}^{\Delta}$ is the EoM operator defined in Eq(\ref{eq:eom}), whose inverse is defined as:
\begin{eqnarray}
    (\mathcal{D}(z))^{-1} \mathcal{O}(z)=\int \frac{d y}{y^{d+1}} G(z, y) \mathcal{O}(y)\label{defOneOverD}
\end{eqnarray}
\begin{eqnarray}
    \mathcal{D}_k^{\Delta} G(z, y)=z^{d+1} \delta(z-y) .\label{GreenfnDef}
\end{eqnarray}
This representation closely mimics the flat space structure.\footnote{However, one could argue that for scalar exchange, this is a useless setup since that is nothing but the insertion of the Bulk-to-Bulk propagator. But the power of this representation is coming from particles with spin, whereas we will see later that we can easily determine the rest of the propagator by simply taking residue.} Another subtlety is that we can not replace the $z$ derivative in terms of Mellin variables in the EoM operator \footnote{This can be seen that after Mellin transform this is still not the eigenvalue of the operator and hence we can not invert the Mellin variables.}. For the same reason there is no overall Mellin delta function. One could only employ Mellin delta function in contact interaction or only in the sub-diagram at the 3-point vertex. But as we will see later in the OPE limits $k^2 \to 0$, we can then replace the Mellin variables with the inverse operator since in the limits they are the eigenvalues.

\section{Spinning particles}
\label{section3}
\subsection{Set up}
Next, for the spinning particle, we rescale the field accordingly, such that gluon behaves like a $\Delta=d-1$ scalar, while the graviton behaves like a massless scalar $\Delta=d$. Here we follow the derivation and notation in \cite{Armstrong:2022mfr}, for Yang-Mills theory, with the usual field strength,
\eqs{
\mathbf{F}_{m n}=\partial_m \mathbf{A}_{n}-\partial_n \mathbf{A}_{m}-i[\mathbf{A}_{m},\mathbf{A}_{n}].\label{scalarC4pt}
}
We then rescale the field as $\mathbf{A}_{m}=(\mathcal{R}/z) A_{m}$ and with the gauge condition $k_{\mu}  A^{\mu}=0$. The graviton will be parametrized as $g_{mn}=\tilde{g}_{mn}+\frac{\mathcal{R}^2}{z^2}h_{mn}$. We can then expand this in field equation and obtain the free EoM,
\begin{align} 
    \mathcal{D}_{k}^{d-1} A_{\mu}(k,z)&=0 ,\\
    \mathcal{D}_{k}^{d} h_{\mu \nu}(k,z)&=0. 
\end{align}
Note that for free EoM, our gauge condition also implies all the $z$ components vanish, details can be found in Appendix \ref{appendixA}. Clearly, the solutions are just scalar propagators dressed up with boundary polarization.
\begin{align}
    A_{\mu}(k,z)&=\varepsilon_{\mu}\phi_{d-1}(k,z),\\
    h_{\mu \nu}(k,z)&=\varepsilon_{\mu \nu}\phi_{d}(k,z).
\end{align}
With spinning particles being simply scalar dressed up with polarization, we have disentangled scalar integral and tensor structure in the amplitude and so we can now easily use the same definition Eq(\ref{eq:MMamplitude}) to define our spinning Mellin-momentum amplitude.
\subsection{Feynman rules for Mellin-momentum amplitude}
\label{Ymfeyn}
In this section, we write down a new set of Feynman rules for gluons that makes the flat space structure manifest. We will employ the boundary transverse gauge, $k_{\mu} \cdot A^{\mu}=0$, which allows us to have only scalar-like propagators\cite{Armstrong:2022mfr}. By directly solving the equation of motion, we found the Feynman rules are identical to flat space in the Coulomb gauge with the simple replacement of the EoM operator $\frac{1}{s}$ with $\frac{1}{\mathcal{D}^{d-1}_{k}}$, details can be found in Appendix \ref{appendixA}:
\eqs{\begin{tikzpicture}
  \begin{feynman}
    \vertex (a) ;
    \vertex[right=1.5cm of a] (b);
    \node[above] at (a) {\(\mu\)};
    \node[above] at (b) {\(\nu\)};
    \diagram* {
      (a) -- [gluon] (b),
    };
  \end{feynman}
\end{tikzpicture}: G_{\mu \nu}&=\frac{\Pi_{\mu \nu}}{\mathcal{D}^{d-1}_{k}},  \\
\begin{tikzpicture}
  \begin{feynman}
    \vertex (a) ;
    \vertex[right=1.5cm of a] (b);
    \node[above] at (a) {\(z\)};
    \node[above] at (b) {\(z\)};
    \diagram* {
      (a) -- [gluon] (b),
    };
  \end{feynman}
\end{tikzpicture}: G_{zz} &=\frac{1}{z^2k^2},
\label{YMfeynProp}}
where $\Pi_{\mu \nu}=\eta_{\mu \nu}-\frac{k_{\mu}k_{\nu}}{k^2}$ is the spin-1 projection tensor. Similarly, after employing the Mellin-Fourier transform $\phi(x,z)\sim e^{ik\cdot x}z^{-2s+d/2}\phi(k,s)$, it's easy to derive the color ordered vertex \cite{Raju:2012zs} and see that they are the same as flat space with:
 \eqs{
 \begin{tikzpicture}[baseline=(current bounding box.center)]
  \begin{feynman}
    \vertex (a);
    \vertex [right=1cm of a] (b);
    \vertex [above right=1cm of b] (c);
    \vertex [below right=1cm of b] (d);
    \node[above left] at (a) {\(p_1\)};
    \node[below left] at (a) {\(m\)};
    \node[above right] at (c) {\(p_2\)};
    \node[below right] at (c) {\(n\)};
    \node[above right] at (d) {\(p_3\)};
    \node[below right] at (d) {\(q\)};
    \diagram* {
      (a) -- [gluon] (b),
      (b) -- [gluon] (c),
      (b) -- [gluon] (d),
    };
  \end{feynman}
\end{tikzpicture}:&  \;  V_{mnq}(p_1,p_2,p_3)
   =\frac{1}{2}\left(\eta _{mn}(p_1-p_2)_q +\eta _{nq}(p_2-p_3)_m+\eta _{qm}(p_3-p_1)_n\right),\\
   \begin{tikzpicture}[baseline=(current bounding box.center)]
  \begin{feynman}
    \vertex (a);
    \vertex [above left=1cm of a](b);
    \vertex [above right=1cm of a] (c);
    \vertex [below right=1cm of a] (d);
    \vertex [below left=1cm of a] (e);
    \node[above left] at (b) {\(m\)};
    \node[above right] at (c) {\(n\)};
    \node[below right] at (d) {\(q\)};
    \node[below left] at (e) {\(o\)};
    \diagram* {
      (a) -- [gluon] (b),
      (a) -- [gluon] (c),
      (a) -- [gluon] (d),
      (a) -- [gluon] (e),
    };
  \end{feynman}
\end{tikzpicture}: & \; V_{mnqo}  =\left(\frac{1}{2} \eta _{mq}\eta _{no} -\frac{1}{4}(\eta _{mn}\eta _{qo}+\eta _{mo}\eta _{nq}) ,\right)\label{YMfeynvetex}
}
where $p^m=(z k^{\mu},i(2s-d/2))$. So we can easily uplift any Feynman diagrams from flat space to AdS amplitude. The crucial distinction between flat space and AdS is the $k_I$ pole, which is non-physical in flat space. However, this pole in AdS is precisely the signal of a CFT, which is necessary for the CFT to have an infinite expansion (corresponding to infinite number of descendent operators) in the OPE limit\cite{Arkani-Hamed:2015bza}. 
\subsection{Examples on Spinning Witten diagrams}
\textbf{Three-point contact diagram}: we directly contract the 3 point vertex in \eqref{YMfeynvetex} with polarizations\footnote{Note that we have stripped off the momentum and Mellin conservation, where unlike scalar contact \eqref{scalarC4pt} and four-point gluon contact below \eqref{ampYM4pointContact}, the Mellin delta gets shifted in this case because of extract factors of $z$ and should be $2 \pi i \delta\left(d-1+\sum_{i=1}^3\left(2 s_i-d / 2\right)\right)$.}
\begin{center}
    \begin{tikzpicture}[baseline=(current bounding box.center)]
  \begin{feynman}
    \vertex (a);
    \vertex [right=1.7cm of a] (b);
    \vertex [above right=1.7cm of b] (c);
    \vertex [below right=1.7cm of b] (d);
    \node[above left] at (a) {\(p_1\)};
    \node[below left] at (a) {\(\varepsilon_1\)};
    \node[above right] at (c) {\(p_2\)};
    \node[below right] at (c) {\(\varepsilon_2\)};
    \node[above right] at (d) {\(p_3\)};
    \node[below right] at (d) {\(\varepsilon_3\)};
    \diagram* {
      (a) -- [gluon] (b),
      (b) -- [gluon] (c),
      (b) -- [gluon] (d),
    };
  \end{feynman}
\end{tikzpicture}
\end{center}
 \begin{eqnarray}
    \mathcal{A}_3=z\left(\varepsilon_1 \cdot \varepsilon_2 \varepsilon_3 \cdot k_1+\varepsilon_2 \cdot \varepsilon_3 \varepsilon_1 \cdot k_2+\varepsilon_3 \cdot \varepsilon_1 \varepsilon_2 \cdot k_3\right).
\end{eqnarray}

\textbf{Four-point contact diagram}
\begin{center}
  \begin{tikzpicture}[scale=0.8]
	\begin{feynman}
		\vertex (a) at (-2, 2);
		\vertex (b) at (-2, -1);
		
		\vertex (c) at (2,2);
		\vertex (d) at (2, -1);
		
		\vertex [blob] (v3) at (0, 0.5);
		
    \node[left] at (b) {\(\varepsilon_1\)};
    \node[left] at (a) {\(\varepsilon_2\)};
    \node[ right] at (c) {\(\varepsilon_3\)};
    \node[ right] at (d) {\(\varepsilon_4\)};
		\diagram* {
			(a) -- [gluon, edge label'=$p_2$] (v3),
			(b) -- [gluon, edge label'=$p_1$] (v3),
			(v3) -- [gluon, edge label=$p_4$] (d),
			(v3) -- [gluon, edge label=$p_3$] (c),
		};
	\end{feynman}\
	\end{tikzpicture}
\end{center}This diagram contributes to amplitude via direct contraction of polarizations with the Feynman rule 
\begin{eqnarray}
   \mathcal{A}_{4}^c \eqn
  \left( \frac{1}{2}\varepsilon_1\cdot \varepsilon_3\ \varepsilon_2\cdot \varepsilon_4-\frac{1}{4}\varepsilon_1\cdot \varepsilon_2\ \varepsilon_3\cdot \varepsilon_4-\frac{1}{2}\varepsilon_1\cdot \varepsilon_4\ \varepsilon_2\cdot \varepsilon_3\right).
  \label{ampYM4pointContact}\non 
\end{eqnarray}

\textbf{Four-point Gluon s-channel diagram}\begin{center}
    \begin{tikzpicture}[scale=0.8]
		\begin{feynman}
			\vertex (a) at (-2, 2);
			\vertex (b) at (-2, -1);
			
			\vertex (c) at (2,2);
			\vertex (d) at (2, -1);
			
			\vertex [blob] (v1) at (-1, 0.5);
			\vertex [blob] (v2) at (1, 0.5);
			
    \node[above left] at (b) {\(p_1\)};
    \node[below left] at (b) {\(\varepsilon_1\)};
    \node[above left] at (a) {\(p_2\)};
    \node[below left] at (a) {\(\varepsilon_2\)};
    \node[above right] at (c) {\(p_3\)};
    \node[below right] at (c) {\(\varepsilon_3\)};
    \node[above right] at (d) {\(p_4\)};
    \node[below right] at (d) {\(\varepsilon_4\)};
			\diagram* {
				(a) -- [gluon] (v1),
				(b) -- [gluon] (v1),
				(v2) -- [gluon] (c),
				(v2) -- [gluon] (d),
				(v2) -- [gluon, edge label'=$k_s$] (v1),
			};
		\end{feynman}
	\end{tikzpicture}
\end{center}

We first rewrite the 3-point vertices in components,

\begin{eqnarray}
 V_{\mu \nu \rho}\left(k_1, k_2, k_3\right) 
 \eqn 
 \frac{z}{2}\left(\eta_{\mu \nu}\left(k_1 - k_2\right)_\rho + \eta_{\nu \rho}\left(k_2 - k_3\right)_\mu + \eta_{\rho \mu}\left(k_3 - k_1\right)_\nu\right)
 \non 
 V _{\mu \nu z}(s_1,s_2,s_3)
 \eqn 
i \eta_{\mu \nu }(s_1-s_2)\non  V_{\mu zz}\left(k_1, k_2, k_3\right)
\eqn 
\frac{z}{2}\eta_{zz}(k_2-k_3)_\mu.
\label{vertex31}
\end{eqnarray}
Then from the Feynman rule, we should connect the two 3-point vertices with propagator \eqref{YMfeynProp}, and this gives a transverse part and a longitudinal part:
\begin{eqnarray}
 \mathcal{A}_{4}^s=  \left( \frac{V^{12 \mu} \Pi_{\mu \nu} V^{\nu 34}}{\mathcal{D}_{k_s}^{d-1}}+ \frac{V^{12 z} V^{z 34}}{z^2 k_s^2}\right),
\end{eqnarray}
where we use a short hand notation: $V_{12m}=\varepsilon_1^\mu \varepsilon_2^\nu  V_{\mu \nu m}$ with $m$ could be boundary coordinate $\mu$ or radial coordinate $z$.

Starting from Five-point, we will have two different topologies, the first one is the following diagram:

\begin{center}
   \begin{equation}
        \begin{tikzpicture}
	\begin{feynman}
		\vertex (a) at (-2, -1) {$k_5$};
		\vertex (b) at (-2, 1) {$k_4$};
		\vertex (c) at (0, 1.5) {$k_3$};
		\vertex (d) at (2, 1) {$k_2$};
		\vertex (e) at (2, -1) {$k_1$};
		
		\vertex (x1) at (-1, 0);
		\vertex (x2) at (0, 0);
		\vertex (x3) at (1, 0);
		
		\diagram*{
			(a) -- [gluon] (x1) -- [gluon] (x2),
			(b) -- [gluon] (x1),
			(x2) -- [gluon] (c),
			(x3) -- [gluon] (d),(x2) -- [gluon] (x3),
			(x3) -- [gluon] (e),
		};
	\end{feynman}
\end{tikzpicture} \label{5pttopology}
   \end{equation}
\end{center}
and we label this as $\mathcal{A}_5^{a} $. There are 4 possible combinations of indices for the propagator in the middle:
\eqs{
\mathcal{A}_5^a=&V^{12z}\frac{1}{z^2k^2_{12}} V^{z3z} \frac{1}{z^2k_{45}^2}V^{z45} 
+V^{12z} \frac{1}{z^2k_{12}^2} \frac{ V^{z 3\rho} \Pi_{\rho \sigma} V^{\rho 45} }{\mathcal{D}^{d-1}_{k_{45}}} +\frac{V^{12\mu} \Pi_{\mu \nu} V^{\nu 3z}} {\mathcal{D}^{d-1}_{k_{12}}}\frac{1}{z^2k_{45}^2} V^{z45}+\\&\frac{V_{12}^{\mu}\Pi_{\mu \nu}V_{3}^{\nu \sigma}\Pi_{\sigma \rho}V_{45}^{\rho} }{\mathcal{D}^{d-1}_{k_{12}}\mathcal{D}^{d-1}_{k_{45}}}  ,
\label{3chen3chen3}
}
where $V^{z 3 z}=\frac{1}{2}z \varepsilon_3 \cdot (k_{45}-k_{12})$, and  $V^{z3 \rho}=i \varepsilon_3^\rho (s_3-u)$ contains internal Mellin variables $u$ \cite{Sleight:2019mgd} serving as derivative acting on the bulk-to-bulk propagator. We can either keep it or remove it by the Mellin delta function on the vertex. Both give the same result in momentum space as detailed in Appendix \ref{ymFeynmanrule}.

For another topology, we label this diagram as $\mathcal{A}_5^b$
\begin{center}
   \begin{equation}
        \begin{tikzpicture}
	\begin{feynman}
		\vertex (a) at (-2, -1) {$k_5$};
		\vertex (b) at (-2, 1) {$k_4$};
		\vertex (c) at (0.7, 1.3) {$k_3$};
		\vertex (d) at (1.5, 0) {$k_2$};
		\vertex (e) at (0.7, -1.3) {$k_1$};
		
		\vertex (x1) at (-1, 0);
		\vertex (x2) at (0, 0);
		
		\diagram*{
			(a) -- [gluon] (x1) -- [gluon] (x2),
			(b) -- [gluon] (x1),
			(x2) -- [gluon] (c),
			(x2) -- [gluon] (d),
			(x2) -- [gluon] (e),
		};
	\end{feynman}
\end{tikzpicture}.
\label{5pttopology2}
   \end{equation}
\end{center}
We have
\begin{eqnarray}
   \mathcal{A}_5^b= {V^{12 \mu} \frac{\Pi_{\mu \nu}}{\mathcal{D}^{d-1}_{k_{12}}} V^{\nu 345}}.
\end{eqnarray}
As the polarization only has boundary indices, there will only be transverse propagations, i.e. $\mu,\nu $ rules in \eqref{YMfeynProp}.

After summing over channels, the Feynman rule results agree with the bootstrap result, see file \texttt{ymFeynmanRUle.nb}.
\subsection{Pole structure}
\label{pole structuer}
It will also be useful to understand the pole structure of the amplitude using the propagator above. Especially later on we will use this structure to bootstrap the amplitude. We will use current conservation to show that in the OPE limit, the residue of the OPE pole should vanish. Consider the following general exchange diagram for gluon:\\
\begin{center}
    \begin{tikzpicture}[scale=1.3]
\begin{feynman}
	\node[circle, draw, fill=gray!30, minimum size=20pt] (a) at (0,0) {$J_L$};
	\node[circle, draw, fill=gray!30, minimum size=20pt] (b) at (2,0) {$J_R$};
	
	\diagram* {
		(a) -- [gluon] (b) 
	};
	
	\vertex (gluon1) at (-1.5,1) {};
	\vertex (gluon2) at (-1.57,0.5) {};
	\vertex (gluon3) at (-1.57,-0.5) {};
	\vertex (gluon4) at (-1.5,-1) {};
	
	\diagram* {
		(a) -- [gluon] (gluon1),
		(a) -- [gluon] (gluon2),
		(a) -- [gluon] (gluon3),
		(a) -- [gluon] (gluon4),
	};
	
	\node[rotate=90, transform shape] at ($(gluon2)!0.5!(gluon3)$) {$\cdots$};
	
	\vertex (gluon5) at (3.5,1) {};
	\vertex (gluon6) at (3.57,0.5) {};
	\vertex (gluon7) at (3.57,-0.5) {};
	\vertex (gluon8) at (3.5,-1) {};
	
	\diagram* {
		(b) -- [gluon] (gluon5),
		(b) -- [gluon] (gluon6),
		(b) -- [gluon] (gluon7),
		(b) -- [gluon] (gluon8),
	};
	
	\node[rotate=90, transform shape] at ($(gluon6)!0.5!(gluon7)$) {$\cdots$};
	
	\node[yshift=6pt] at (a.north) {$\vec{k}_L, s_L$};
	\node[yshift=6pt] at (b.north) {$\vec{k}_R, s_R$};
	
	\path (a) -- node[above] {$k$} (b);
\end{feynman}
\end{tikzpicture}
\end{center}
where $k_L^2=k_R^2=k^2$ due to boundary momentum conservation.
This diagram will contribute to the amplitudes as follows:
\eqs{
\mathcal{A}_n&=J_L^{\mu} G_{\mu \nu} J_R^{\nu} + J_L^zG_{zz}J_R^z \\
&=J_L^{\mu}\frac{ \Pi_{\mu \nu} }{\mathcal{D}_k^{d-1}}J_R^{\nu}+J_L^z \frac{1}{z^2 k^2}J_R^z .
}
Considering the following OPE limits,  $k^2 \to 0$, we see the leading contributions correspond to terms with pole $\frac{1}{k^2}$ and taking the residue gives:
\eqs{
\underset{k^2 \to 0}{\mathrm{Res}}\mathcal{A}_n&=k^L_{\mu}J_L^{\mu}\frac{ 1 }{\mathcal{D}_k^{d-1}|_{k^2 \to 0}}k^R_{\nu}J_R^{\nu}+J_L^z \frac{1}{z^2}J_R^z \\
&=k^L_{\mu}J_L^{\mu}\frac{1}{\frac{1}{4}\left(d-4 s_R-4\right)\left(d+4 s_R\right)}k^R_{\nu}J_R^{\nu}+J_L^z \frac{1}{z^2}J_R^z.
}
To show above equation is zero, we first recall that the current conservation is
\eqs{
\nabla_{m}J^m=0,
}
which after applying Mellin-Fourier transform $J_m(x,z) \sim e^{ik\cdot x}z^{-2s+d/2}J_m(s,k)$ yields 
 \begin{eqnarray}
z k^\mu J_\mu+i\left(2 s+d/2\right) J_z=0.
	\end{eqnarray}
 Hence by applying current conservation and on the support of Mellin delta function $\delta(s_R+s_L+1)$ we have:
 \eqs{
\underset{k^2 \to 0}{\mathrm{Res}}\mathcal{A}_n&=-\frac{1}{z^2 k^2} \left(4 s_L+d\right) {J}_L^z \frac{1}{\left(d-4 s_R-4\right)\left(d+4 s_R\right)} \left(4 s_R+d\right) {J}_R^z +J_L^z \frac{1}{z^2k^2}J_R^z, \\
&=0,
 }
So we have  demonstrated here that considering a general n-point diagram with gluon exchange, in OPE limit $k^2 \to 0$, the vanishing of residue is from current conservation. Similar argument should holds also for gravity with the use of conservation of energy momnetum tensor, but we won't go into that in this paper and leave it to future work.

\section{Bootstrap from unitarity and soft limits}
\label{section4}
\subsection{Amplitude Bootstrap}\label{sec: AmpBoot}
In this section, we will use the pole structure of the Mellin-momentum amplitude to bootstrap gluon and graviton amplitudes. To start, our only input for the AdS amplitudes is the following 3-pt on-shell Yang-Mills amplitude:
\begin{equation}
   \mathcal{A}_3=z(\varepsilon_1 \cdot \varepsilon_2 \varepsilon_3 \cdot k_1+\varepsilon_2 \cdot \varepsilon_3 \varepsilon_1 \cdot k_2 + \varepsilon_3 \cdot \varepsilon_1 \varepsilon_2 \cdot k_3).\label{ymamp3}
\end{equation}
Based on the Feynman rules in section \ref{Ymfeyn}, we see that the amplitude can only have two types of poles which we will refer to as factorization pole $\frac{1}{\cD_{k}}$ and OPE pole $\frac{1}{k^2}$, and we remind the readers that both poles are coming from the kinetic terms in the action. This leads us to the following ansatz for the $n$-point amplitude based on their pole structure:
\begin{eqnarray}
	\mathcal{A}_n=\sum_{\mathrm{Channels}}\frac{a_1(12 \ldots n)}{\mathcal{D}_{k_I}^{\Delta} \mathcal{D}_{k_J}^{\Delta} \ldots \mathcal{D}_{k_M}^{\Delta}}+\frac{a_2(12 \ldots n)}{\mathcal{D}_{k_{J}}^{\Delta} \ldots \mathcal{D}_{k_M}^{\Delta}}+\cdots+\frac{b_{\ell_I,...,\ell_M}(12 \ldots n)}{k_I^{2 \ell_I} k_J^{2 \ell_J} \ldots k_M^{2 \ell_M}}+\cdots+c(12 \ldots n).\nonmy\label{generalf}
\end{eqnarray}
Here the $I$ refers to a general subset of external momenta, $\mathcal{D}_{k_{I}}^{\Delta}$ is a bulk-to-bulk propagator with the sum of momenta in the set $I$ flowing through it, and $1/k_I^2\equiv 1/|\sum_{x\in I}\vec{k}_x|^2$. As we will see later, poles in $k_I$ are required by the OPE. The order of these poles can go up to $\ell_{\mathrm{max}}=1$ for gluons and $\ell_{\mathrm{max}}=2$ for gravitons. The coefficients appearing in the numerators of this ansatz are then fixed by imposing various constraints coming from factorization, the OPE, and generalized dimensional reduction. We briefly describe each of them below.\\

\textbf{Factorization}: Unitarity implies that the amplitude will factorize into lower point amplitudes when the exchanged particles are on-shell \cite{Goodhew:2020hob,Goodhew:2021oqg,Meltzer:2020qbr}
\eq{
\mathcal{A}_n \to \sum_h \mathcal{A}_{a}^{h} \frac{1}{\mathcal{D}^{\Delta}_{k_I}} \mathcal{A}_{n-a+2}^{-h}.
}
In particular in our examples:
\eqs{
\text{YM: }	\mathcal{A}_n \to& \mathcal{A}_{a}^\mu \frac{\sum_h  \varepsilon_{\mu }^h \varepsilon^{*h}_\nu}{\mathcal{D}^{d-1}_{k_I}}\mathcal{A}_{n-a+2}^\nu\\ 
\text{GR: }	\mathcal{M}_n \to& \mathcal{M}_{a}^{\mu_1\mu_2}\frac{\sum_h  \varepsilon_{\mu_1\mu_2}^h \varepsilon^{*h}_{\nu_1\nu_2}}{\mathcal{D}^{d}_{k_I}}\mathcal{M}_{n-a+2}^{\nu_1\nu_2},\label{unitary}
	}
 where the polarization sums are
\begin{eqnarray}
	\begin{aligned}
		\sum_{h= \pm} \varepsilon_\mu (k, h) \varepsilon_\nu(k, h)^* 
  & =\eta_{\mu\nu}-\frac{k_\mu k_\nu}{k^2} \equiv \Pi_{\mu\nu} \\
		\sum_{h= \pm} \varepsilon_{\mu \nu}(k, h) \varepsilon_{\rho \sigma}(k, h)^*
  & =\frac{1}{2} \Pi_{\mu \rho} \Pi_{\nu \sigma}+\frac{1}{2} \Pi_{\mu \sigma} \Pi_{\rho \nu}-\frac{1}{d-1} \Pi_{\mu \nu} \Pi_{\rho \sigma}.\label{projections}
	\end{aligned}
\end{eqnarray}
This step will fix all the $a_i$ terms in our ansatz. \\

\textbf{OPE limit}: Consider two operators get close to each other, and they are far from other operators in position space. This implies that taking internal momentum $k_I^2 \to 0$, whose behavior is then controlled by lower-point amplitudes from the usual OPE limit\cite{Arkani-Hamed:2015bza,Sleight:2019hfp,Bzowski:2022rlz,Assassi:2012zq}
Equivalently, as we have shown in section \ref{pole structuer} in OPE limit the residue of the OPE poles for amplitudes must be zero:
\eqs{
 \underset{k_I^2 \to 0}{\mathrm{Res}}\mathcal{A}_n = 0,
}
where $I$ labels internal momentum.
Remarkably, with this condition, all the $b_i$ terms in our ansatz are simply determined by taking the residue of $a_i$:
\eqs{
{b(12\dots n)}=-\underset{k_I^2 \to 0}{\mathrm{Res}}\left( \frac{a_1(12\dots n)}{\mathcal{D}^{\Delta}_{k_I}\mathcal{D}^{\Delta}_{k_J}\dots \mathcal{D}^{\Delta}_{k_M}}+\dots \right).
}

\textbf{Flat space limit}: In the flat space limit, the particles are ignoring the curvature correction, and the Lorentz symmetry will emerge and give us the S-matrix in flat space\cite{Arkani-Hamed:2018bjr,Raju:2012zr,Penedones:2010ue,Li:2021snj}:
\eqs{
\lim_{z \to \infty} \mathcal{A}_n \to z^{\#} A_n,
}
with $z^{\#}$ being the leading power in flat space limit, and it is easy to see that as $z\to \infty$:
\begin{eqnarray}
    s_i\to \frac{ z k_i}{2}, \quad \quad 
    \cD_{k_{ij}}\to z^2S_{ij},\label{Flatspacelimit}
\end{eqnarray}
where the EoM operator reduced to the Mandalstam variables $S_{ij}$ in flat space and hence the flat space amplitude $A_n$.\\
This step fixes all the contact terms $c_i$ in our ansatz for gluons while partly for graviton.\\

In particular, for $n\geq 5$ gluon amplitudes, the $n$-point functions are fully determined by the $(n-1)$ point amplitudes through factorization and the residue of the OPE poles. There are no contact terms beyond four-points, as this would result in an incorrect flat space limit.\\

However, gravity is slightly more subtle, since as a two-derivative theory, the contact terms can potentially receive curvature correction simply from dimensional analysis, but the same procedure applies, and contact terms are fully determined by the flat space limit and external soft limit as followed.\\

\textbf{Adler zero}:
To elaborate, we reduce external legs $i,j$ to scalars by setting: $\varepsilon_i \cdot \varepsilon_j=1, \varepsilon_i \cdot k_a=0, \varepsilon_j \cdot k_a=0$, where $a$ is any leg. After doing so we obtain $\langle \phi \phi hhh\dots \rangle $ where the scalar will couple with graviton as $\nabla^m \phi \nabla^n \phi h_{mn} $. It follows that the amplitude must vanish when taking a soft limit of a scalar momentum, which is an Adler zero in curved space \cite{Armstrong:2022vgl}, so for n-point graviton amplitude:
\eqs{
\mathcal{M}_n |_{\varepsilon_i \cdot \varepsilon_j=1, \varepsilon_i \cdot k_a=0, \varepsilon_j \cdot k_a=0} \xlongrightarrow{\hspace{0.2cm} \vec{k}_i \to 0 \hspace{0.2cm}} 0.
}
This will be our last constraint on the gravity amplitude to fix the contact terms $c(12\dots n)$.

So we have outlined all the conditions used in the Mellin-momentum amplitude bootstrap in AdS for both Yang-Mills (YM) and General Relativity (GR). To conclude this section, the table below compares the bootstrap conditions for gauge fields between the flat-space S-matrix and Mellin-momentum amplitude.
\begin{table}[htbp]
  \centering
  \caption{Comparison between on-shell bootstraps of Minkowski and AdS}
  \label{tab:example}
	\begin{tabular}{ m{5cm}  m{4cm}  m{4cm} }

	 & \textbf{S-matrix} & \textbf{AdS} \\
		\hline
		\textbf{Space-time symmetry} & Manifest  & Not Manifest \\
	
		
		\textbf{Unitarity} & Manifest & Manifest \\
		
		\textbf{Gauge Condition} & $\begin{array}{l}
		     \varepsilon_{\mu} \cdot k^{\mu}=0
  \\
  \text{To preserve}
  \\
  \text{Lorentz symmetry}
		\end{array}$ & 
$\begin{array}{l}
\text{ Boundary Observable}\\ 
\text{ forces } \varepsilon^z = 0 \; \;  \footnotemark
\end{array}$
 \\
		
	
	\end{tabular}
 \end{table}
 \footnotetext{One could also consider covariant gauge condition in AdS \cite{DHoker:1999kzh,Costa:2014kfa}. However, for boundary observable we still need to take the external leg $\varepsilon_z \to 0$ which leaves us with no more gauge redundancy like Coulomb gauge.}
\subsection{Gluon amplitudes}
\subsubsection{Four-point}
Following our amplitude bootstrap procedure, we now start with 3-point Yang-Mills amplitude in Eq \eqref{ymamp3} to bootstrap  the color ordered four-point amplitude:
\begin{equation}\label{eq:4YM}
\mathcal{A}_4=\frac{a(1,2,3,4)}{\mathcal{D}_{k_{s}}^{d-1}} + \frac{b(1,2,3,4)}{k_{s}^2}+c(1,2,3,4)+[2 \leftrightarrow 4 ],
\end{equation}
where the $[2 \leftrightarrow 4 ]$ is to obtain t-channel and $k_s^2=k^2_{12}=(\Vec{k}_1+\Vec{k}_2)^2$. Then by factorization:
\begin{align}
    a(1,2,3,4)
    =
    \sum_{h=\pm} \mathcal{A}_3(1,2,-k_{s}^h) \mathcal{A}_3(k_{s}^{-h},3,4).
\end{align}
It can then be written as:
\begin{eqnarray}
    a(1,2,3,4)=z^2\left(\Pi_{1,1}\varepsilon_{12,34}+W_s\right)
\end{eqnarray}
where $\Pi_{1,1}$ is the spin-1 polarization sum, and $W_s$ are defined as

\begin{align}
\Pi_{1,1}\equiv &\frac{1}{4} (k_{1}^{\mu}-k_2^{\mu})\Pi_{\mu \nu}(k_{3}^{\nu}-k_4^{\nu})=\frac{1}{4}(k_1-k_2)\cdot (k_3-k_4)+\frac{(k_1^2-k_2^2)(k_3^2-k_4^2)}{4k_s^2}, \\
    4W_s=&\varepsilon_1 \cdot \varepsilon_2 (k_1 \cdot \varepsilon_3 k_2 \cdot \varepsilon_4 -k_2 \cdot \varepsilon_3 k_1 \cdot \varepsilon_4 )   \nonumber \\
    +&\varepsilon_3 \cdot \varepsilon_4 (k_3 \cdot \varepsilon_1 k_4 \cdot \varepsilon_2 -k_4 \cdot \varepsilon_1 k_3 \cdot \varepsilon_2 )   \nonumber \\
    +&(k_2 \cdot \varepsilon_1 \varepsilon_2 -k_1 \cdot \varepsilon_2  \varepsilon_1 ) \cdot (k_4 \cdot \varepsilon_3 \varepsilon_4 -k_3 \cdot \varepsilon_4  \varepsilon_3 ) .
\end{align}
It is worth to note that there is a $1/{k_s^2}$ pole in $\Pi_{1,1}$. This structure is what we need in the second step of our bootstrap procedure, which we demand for the amplitude, the residue of this $k_s^2$ pole must be zero, so
\begin{eqnarray}
    b(1,2,3,4)=-\underset{k_s^2 \rightarrow 0}{\operatorname{Res}}\frac{a(1,2,3,4)}{\mathcal{D}_{k_s}^{d-1}}.
\end{eqnarray} 
To be more precise, the leading term with $k_s^2$ above is given by
\begin{eqnarray}
    \frac{a(1,2,3,4)}{\mathcal{D}_{k_s}^{d-1}}\xlongrightarrow{\hspace{0.2cm} k_s^2 \to 0 \hspace{0.2cm}}\frac{ z^2\left(k_1^2-k_2^2\right)\left(k_3^2-k_4^2\right)\varepsilon_{12,34}}{16k_s^2(s_1+s_2-1)(s_3+s_4-1)}
    = 
    \frac{(s_1-s_2)(s_3-s_4)\varepsilon_{12,34}}{z^2k_s^2}.\label{OPE1}
\end{eqnarray}
Now we explain the derivation for the calculation above. First, we remind the reader that the inverse EoM operator is acting on the lower point amplitude, and without taking any limits the Mellin transform is not an eigenfunction of the EoM operator (unlike Fourier transform in flat space), we have to keep the inverse operator as the insertion of bulk-to-bulk propagator. However, in the OPE limits, the EoM operator $\mathcal{D}_k^{\Delta}|_{k^2\to0} \equiv-z\partial_zz\partial_z+d z \partial_z+\Delta(\Delta-d)$ acting on lower three point structure as:
\begin{eqnarray}
			&&\frac{1}{\mathcal{D}_{k_{s}}^{d-1}|_{k_s^2\to 0}} (z z^{-2s_1+d/2}z^{-2s_2+d/2}	)
   \non\eqn
   \frac{1}{2\left(d-2 s_1-2 s_2\right)\left(s_1+s_2-1\right)} z z^{-2s_1+d/2}z^{-2s_2+d/2}	.\label{dtos}
\end{eqnarray}
In the OPE limits, the Mellin transform $z^x$ is then the eigenfunction of $z\partial_z$ and hence it is now indeed the eigenfunction of EoM opeartor in OPE limits. So we can write $s$ in the denominator as the eigenvalue. All the $s_i$ pole here are spurious, and indeed using the on-shell condition \eqref{Onshell}, which takes $z^4\left(k_1^2-k_2^2\right)\left(k_3^2-k_4^2\right) \to \left(s_1^2-s_2^2\right)\left(s_3^2-s_4^2\right)$ and shift the Mellin variables $s$ in the denominator by $+1$ when applying on-shell condition\footnote{The subtlety of shifting $s$ here is that when we use onshell condition for Mellin-momentum amplitudes,
\begin{eqnarray}
    f(s) k^2\to f(s+1)\left(\frac{ (d / 2-\Delta)^2-4 s^2}{z^2}\right) \label{Onshell}
\end{eqnarray}
where $f(s)$ is any rational function involving the Mellin variable $s$, we have to shift $s\to s+1$. And one can derive this by  using the definition of \eqref{btBMellin}.}.
Then we arrive at the result in Eq\eqref{OPE1}. Using Mellin delta function one can eliminate $s_1$ dependence which will be useful in constructing five-point in next section.\\
In summary, our second bootstrap step gives:
\begin{align}
    b(1,2,3,4)
    =
    -\underset{k_{s}^2 \to 0}{\mathrm{Res}}\frac{a(1,2,3,4)}{\mathcal{D}_{k_{s}}^{d-1}}
    =
    -4\left(s_1-s_2\right)\left(s_3-s_4\right)\varepsilon_{12,34}.
\end{align}
Finally, we are only left with the $c(1,2,3,4)$ term in our ansatz, which by dimensional analysis, must be $\varepsilon_{ij,kl}$ with unfixed coefficients. Hence we can easily compare with the flat space limit result and obtain

\begin{eqnarray}
c(1,2,3,4)=\frac{1}{4}(\varepsilon_{13,24} - \varepsilon_{14,23}).
\end{eqnarray}
So adding up all the bootstrap results together and summing over other channels, we have the 4-point color-ordered Yang-Mills amplitude, 
\begin{eqnarray}
    \mathcal{A}_4
    \eqn 
    \frac{z^2\left(\Pi_{1,1} \varepsilon_{12,34}+W_s\right)}{\mathcal{D}_{k_s}^{d-1}}
    -
    \frac{\left(s_1-s_2\right)\left(s_3-s_4\right) \varepsilon_{12,34}}{z^2k_s^2}
    +
    \frac{1}{4}(\varepsilon_{13,24} - \varepsilon_{14,23})+[2 \leftrightarrow 4 ]   \label{A4YM}
\end{eqnarray}
where $[2 \leftrightarrow 4 ]$ is from summing over t-channel.
To sum up, at four-point, we use factorization, OPE (internal soft) limit, and flatspace limit to fully fix gluon tree-level amplitudes. 

This matches our Feynman rules result in section \ref{Ymfeyn}. We will translate the result back to momentum space in Section \ref{backtomomentum}, and it agrees with literature \cite{Albayrak:2018tam}. All processes are given in \texttt{4ym.nb}.

\subsubsection{Five-point}
We now turn to the five-point ansatz:
\begin{equation}
    \mathcal{A}_5
    =
     \frac{a_1(1,2,3,4,5)}{\mathcal{D}_{k_{12}}^{d-1} \mathcal{D}_{k_{45}}^{d-1}}
     + 
     \frac{a_2(1,2,3,4,5)}{\mathcal{D}_{k_{12}}^{d-1}}
     + 
     \frac{b(1,2,3,4,5)}{k_{12}^2k_{45}^2}+Cyc,\label{eq5ym}
\end{equation}
where $Cyc$ means to get a color ordered gluon amplitude we have to sum over cyclic permutations for diagrams in the usual trace basis:
\begin{eqnarray}
\left\{(1,2,3,4,5),
(2, 3, 4, 5, 1),
(3, 4, 5, 1, 2),
(4, 5, 1, 2, 3),
(5, 1, 2, 3, 4)\right\},\label{whatiscyc}
\end{eqnarray}
and factorization fixes all the $a$ terms by recycling the four-point Eq (\ref{eq:4YM}): 
\eqs{
    &
    a_1(1,2,3,4,5)
    =
    \sum_{h} a(1,2,3,k_{45})\cdot \mathcal{A}_3(-k_{45},4,5), \\ 
    &
    a_2(1,2,3,4,5)
    \\ 
    =&
    \sum_{h} \mathcal{A}_3(1,2,-k_{12}) \cdot\left(
    \frac{b(k_{12},3,4,5)}{k_{45}^2}
    +
    c(k_{12},3,4,5)
    + [3 \leftrightarrow 5]
    \right).
}

The detailed expressions are recorded in Appendix \ref{Derivation of double OPE pole}. Having fixed $a_1$ and $a_2$, the next step is to fix the $b's$ which is determined by the vanishing residue of the amplitudes in OPE limits. Following the same step as the four-point, we can easily determine 
it by taking the residue of the OPE poles for $a_1$ and $a_2$, and details are given in Appendix \ref{YMdoubleOPE}. The final result is simply:
    \begin{align}
& -b(1,2,3,4,5)\notag
\\ \nonumber
= &
\underset{
    \begin{array}{l}
    \substack{\scriptscriptstyle 
 k^2_{12} \to 0}
 \\
      \substack{ \scriptscriptstyle  k^2_{45} \to 0}
    \end{array} 
    }
    {\mathrm{Res}}
    \left (
    \frac{a_1(1,2,3,4,5)}{\mathcal{D}_{k_{12}}^{d-1} \mathcal{D}_{k_{45}}^{d-1}}
    +
    \frac{a_2(1,2,3,4,5)}{\mathcal{D}_{k_{12}}^{d-1}}
    +
    \frac{a_2(4,5,1,2,3)}{\mathcal{D}_{{k_{45}}}^{d-1}}
    \right) \notag
    \\
=&\frac{
\left(s_1-s_2\right)\left(s_4-s_5\right)
}{
z^3
}
\varepsilon_1\cdot \varepsilon_2 \varepsilon_4\cdot \varepsilon_5\varepsilon_3\cdot (k_4+k_5) . \label{5pt-bterm}
\end{align}
After getting the double OPE poles, we find that the amplitude already has the correct flat space limit and has no terms with a single OPE pole. After summing over permutations, we obtained the full color ordered five-point amplitude by using only the known four-point data and demanding the vanishing of OPE poles. It's also worth pointing out that simply by dimensional analysis there are no contact terms $c$ in $n\ge 5$ point.

All processes are given in \texttt{5ym.nb}. Additionally, we have confirmed that our five-point result matches the results from the new Feynman rules calculation and the literature \cite{Albayrak:2019asr,Albayrak:2018tam} after applying the mapping in Section \ref{backtomomentum}. Furthermore, the result has been checked to ensure the correct soft limit \cite{Chowdhury:2024wwe}.\\

\subsection{Graviton amplitude}

\subsubsection{Four-point}

Our starting point is the three graviton amplitude, which takes a very simple form in Mellin-momentum amplitude as manifestly the square of the Yang-Mills one \cite{Farrow:2018yni,Lee:2022fgr,Lipstein:2019mpu},

\begin{align}
    \mathcal{M}_3=&\mathcal{A}_3^2 =z^2(\varepsilon_1 \cdot \varepsilon_2 \varepsilon_3 \cdot k_1+\varepsilon_2 \cdot \varepsilon_3 \varepsilon_1 \cdot k_2 + \varepsilon_3 \cdot \varepsilon_1 \varepsilon_2 \cdot k_3)^2.
\end{align}
Our four-point ansatz is given as:

\begin{equation}
\mathcal{M}_4=\frac{a^{\text{GR}}(1,2,3,4)}{\mathcal{D}_{k_{s}}^{d}} + \frac{b^{(2,\text{GR})}(1,2,3,4)}{k_{s}^{2}}+ \frac{b^{(4,\text{GR})}(1,2,3,4)}{k_{s}^{4}}+c^{\text{GR}}(1,2,3,4)+\mathcal{P}(2,3,4),\label{ampgr41}
\end{equation} 

where the $\mathcal{P}(2,3,4)$ denotes permutation to obtain other channels. Similar to the spin-1 case, we first determine $a$ from factorization
\eqs{
    a^\text{GR}(1,2,3,4)&=\sum_{h=\pm} \mathcal{M}_3(1,2,-k_s^h) \cdot \mathcal{M}_3(k_s^{-h},3,4) ,
}
where the spin-2 polarization sum is given by,
\eqs{
   & \sum_{h=\pm} \varepsilon_{\mu \nu}(k,h) \varepsilon_{\rho \sigma}(k,h)^* \\ 
   =&\frac{1}{2} \Pi_{\mu \rho}\Pi_{\nu \sigma}+ \frac{1}{2}\Pi_{\mu \sigma} \Pi_{\rho \nu}-\frac{1}{d-1}  \Pi_{\mu \nu}\Pi_{\rho \sigma}.\label{plsum}
}
We get
\begin{eqnarray}
    \frac{a^\text{GR}(1,2,3,4)}{\mathcal{D}_{k_s}^d}=\frac{\left(\varepsilon_{12,34}z^2 \Pi_{1,1}+4 z^2 W_s\right)^2-\varepsilon_{12,34}^2 z^4\Pi_{2,2}^{\operatorname{Tr}}}{ \mathcal{D}_{k_s}^d},\label{gr4pta1}
\end{eqnarray}
where the $\Pi_{2,2}^{\operatorname{Tr}}$ is the trace part of the spin-2 polarization sum and given in Appendix \ref{polarization sums}. Next, we can easily see that for gravity we have two OPE poles $k_s^{-2}$ and $k_s^{-4}$ coming from the spin-2 polarization sum. Hence in the OPE limit, we should eliminate $k_s^{-2}$ and $k_s^{-4}$ poles, so $b$ should be:
\begin{eqnarray}
     b^{(2,\text{GR})}(1,2,3,4)=-\underset{k_{s}^2 \to 0}{\mathrm{Res}} \frac{a(1,2,3,4)}{\mathcal{D}_{k_{s}}^d},\qquad   b^{(4,\text{GR})}(1,2,3,4)=-\underset{k_{s}^4 \to 0}{\mathrm{Res}} \frac{a(1,2,3,4)}{\mathcal{D}_{k_{s}}^d}.
\end{eqnarray}

Determining the $b$ terms is very similar to the spin-1 case, except that this time we should expand the inverse operator to higher order for the $k_s^2$ pole to capture full contributions\footnote{
We use operator expansion $\frac{1}{{\mathcal{D}_{k_{s}}^d}}\bigg|_{k_s^2 \to 0}= 
\frac{1}{-z \partial_z z \partial_z+d z \partial_z}
-\frac{1}{-z \partial_z z \partial_z+d z \partial_z}
z^2 k_s^2
\frac{1}{-z \partial_z z \partial_z+d z \partial_z}
+O(k_s^4)$. And it acts on $z^2 z^{-2 s_1+d / 2} z^{-2 s_2+d / 2}$ gives \eqref{cDopeexpand}.
}:
\begin{eqnarray}
   \frac{1}{{\mathcal{D}_{k_{s}}^d}}\bigg|_{k_s^2 \to 0}= && \frac{1}{2\left(s_{12}-1\right)\left(d-2 s_{12}+2\right)} -\frac{z^2 k_s^2}{4\left(s_{12}-2\right)\left(s_{12}-1\right)\left(d-2 s_{12}+2\right)\left(d-2 s_{12}+4\right)} \\ \nonmy
   &&+O(k_s^4).\label{cDopeexpand}
\end{eqnarray}
Plugging \eqref{cDopeexpand} into \eqref{gr4pta1}, collecting the coefficients of $k_s^{-4}$ and $k_s^{-2}$, using onshell condition \eqref{Onshell}, and massaging the expression using Mellin delta \eqref{eq:mmdelta}, we get $k_s^{-4}$ contribution
\begin{eqnarray}
  b^{(4,\text{GR})}(1,2,3,4)=-\frac{(d-2)\left(s_1-s_2\right)\left(s_3-s_4\right)\left(k_1^2-k_2^2\right)\left(k_3^2-k_4^2\right)\varepsilon_{12,34}^2}{4(d-1)}.
\end{eqnarray}
Similarly for $k_s^{-2}$ contribution
\begin{eqnarray}  
&&
b^{(2,\text{GR})}(1,2,3,4)
\nonmy\eqnmy
\left[ W_s+\left((k_1-k_2)\cdot (k_3-k_4)+k_s^2\right)\varepsilon_{12,34}\right]\Pi_{1,0}\varepsilon_{12,34}
 \nonmy
   &&+
    \frac{\left(\frac{4(d-2) z^2\left(k_1^2-k_2^2\right)\left(k_3^2-k_4^2\right)\left(s_1-s_2\right)\left(s_3-s_4\right)}{\left(d-2\left(s_1+s_2\right)\right)\left(d-2\left(s_3+s_4\right)\right)}+\frac{2\left(k_3^2-k_4^2\right)\left(d^2-8\left(s_1^2+s_2^2\right)\right)\left(s_3-s_4\right)}{
    d-2\left(s_3+s_4\right)
    }+ 1,2\leftrightarrow 3,4
   \right) }{16(d-1)}
    \varepsilon_{12,34}^2.
    \nonmy  
\end{eqnarray} 

Poles like $d-2\left(s_3+s_4\right)$ in the expression are spurious, which can be easily shown to go away by onshell condition \eqref{Onshell} and Mellin delta  \eqref{eq:mmdelta}, and the spurious pole free expression can be found in \texttt{4gr.nb}.

Finally, we need to determine the terms with no poles that is the $c$ terms in our ansatz and can be split into two terms $c^{\mathrm{GR}}(1,2,3,4)=c_0^{\mathrm{GR}}(1,2,3,4)+c_1^{\mathrm{GR}}(1,2,3,4)$. We would first like to make our expression Lorentz invariant in the flat space limit so we add $c_0^{\mathrm{GR}}(1,2,3,4)$ to cancel out the non-Lorentz-invariant part. Furthermore, the expression $c_0^{\mathrm{GR}}$ should also contain curvature correction. This can be probed by Alder's zero condition in curved space when we do dimension reduction:
$
\varepsilon_1 \cdot \varepsilon_2=1, \varepsilon_1 \cdot k=0, \varepsilon_2 \cdot k=0$.
The details are given in Appendix \ref{four point details}, where we bootstrap the following term according to the flat space limit Lorentz invariance and Adler's zero:
\eqs{
    &c_0^{\mathrm{GR}}(1,2,3,4) \\
    =&\varepsilon^2_{12,34}\frac{8d(s_1-s_2)(s_3-s_4)-4(s_1+s_3-s_2-s_4)^2+d^2}{16(d-1)}.\label{c0explain}
}
Finally, for the rest contact term $c_1$, simply by dimensional analysis, we write down the ansatz with two derivatives and Lorentz invariant in the flat space limit:
\eqs{
    c^\text{GR}_1(1,2,3,4)=&\varepsilon_{ab,cd,ef}(C_1 z^2\varepsilon_m \cdot k_i \varepsilon_n \cdot k_j +C_2\varepsilon_m \cdot \varepsilon_n \mathcal{D}_{k_s}^d)
}
where the EoM operator directly acts on the bulk-to-boundary propagator gives $\mathcal{D}_{k_s}^d=z^2k_s^2+4 (s_1+s_2)(s_3+s_4)$ \footnote{This is also simply two covariant derivatives acting on scalar field.} and comparing with the flat space amplitude determines the unknown coefficient $C_i$ and gives:
\eqs{
    c^\text{GR}_1(1,2,3,4)=& \left[\left( {V_{s}^c}-\frac{\varepsilon_{12,34}}{4}\right)^2-\frac{d-2}{d-1}\frac{\varepsilon_{12,34}^2}{16}\right]\cD_{k_s}^d +V_{s}^c\varepsilon_{12,34}\mathcal{D}_{k_t}^d
     -2z^2V_{s}^cW_s.
}
The full process is given in \texttt{4gr.nb}. \footnote{One might be concerned that the two-derivative contact interaction could have $1/R^2$ correction, which vanishes in the flat space limit. However, such term will violate the Alder zero when taking external soft limit before, hence such terms should not be written down.} After adding different channels the four-graviton amplitude is symmetric under particle exchange. This completely determines the four-graviton amplitude and matches with \cite{Mei:2023jkb}, and \cite{Bonifacio:2022vwa,Armstrong:2023phb} after using the mapping in section \ref{backtomomentum}.

To sum up, we have used factorization, internal (OPE) and external soft limit, and flat space limit to fix the four-point graviton Mellin-momentum amplitudes in $AdS_{d+1}$. We should emphasize that our bootstrap algorithm is quite automatic with no guesswork.

\subsubsection{Five-point}\label{fivegraviton}

To demonstrate the power of our bootstrap algorithm, we derive the first five-graviton amplitude in $\mathrm{AdS}$. We first write down our ansatz according to the pole structure:
\eqs{
   & \mathcal{M}_5=
     \frac{a^\text{GR}_1(1,2;3;4,5)}{\mathcal{D}_{k_{12}}^{d} \mathcal{D}_{k_{45}}^{d}}+ \frac{a^\text{GR}_2(1,2;3,4,5)}{\mathcal{D}_{k_{12}}^{d}}\\ 
     +& \sum_{m_1, m_2=1}^2 \frac{b_1^{\left(m_1, m_2,\text{GR}\right)}(1,2 ; 3 ; 4,5)}{k_{12}^{2 m_1} k_{45}^{2 m_2}}+\sum_{m=1}^2\frac{b_2^{(m,\text{GR})}(1,2;3,4,5)}{k_{12}^{2m}}+c^\text{GR}(1,2,3,4,5)+Perm.\label{ampgr5}
}

To proceed, we should first choose the channel that we need to sum over. Following the permutations in \cite{Bern:2019prr}, the topology of the Feynman diagram with two propagators \eqref{5pttopology} has 15 channels. We pick our 15 channels for $a_1 ,b_1$ as
\begin{align}
a_1^{} \text{ and } b_1: \quad &\{1, 2, 3, 4, 5\} , \{1, 2, 4, 5, 3\} , \{1, 2, 5, 3, 4\} 
,\{2, 3, 1, 4, 5\} , \{1, 5, 4, 3, 2\} ,\non  & \{2, 3, 5, 1, 4\} 
,\{4, 3, 2, 1, 5\} , \{2, 5, 1, 3, 4\} , \{3, 1, 2, 4, 5\} 
,\{1, 5, 3, 2, 4\} ,\non  &\{2, 5, 4, 1, 3\} , \{3, 1, 5, 2, 4\}
,\{4, 1, 3, 2, 5\} , \{3, 5, 2, 1, 4\} , \{4, 2, 1, 3, 5\}.\label{list15}
\end{align}

While $a_2$ has only one factorization pole and hence 10 permutations with topology like \eqref{5pttopology2}
\begin{align}
a_2^{} \text{ and } c_0 :\quad&{\{1, 2, 3, 4, 5\}} , \{2, 3, 1, 4, 5\} , \{4, 3, 2, 1, 5\}
,\{5, 4, 3, 2, 1\} , \{5, 1, 2, 3, 4\} , \non  &\{3, 1, 2, 4, 5\} 
,\{4, 1, 3, 2, 5\} , \{4, 2, 3, 5, 1\} , \{2, 5, 1, 3, 4\} ,\{5, 3, 1, 2, 4\}.\label{list10}
\end{align}
with $c_0$ shares the same list.

The $b_2$ is the single OPE contribution, again with 10 channels but with a slightly different list due to the derivation and will explain later,
\begin{align}
    b_2^{\text{Type I}}: \    &\{1, 2, 3, 4, 5\},
\{2, 3, 1, 4, 5\},
\{3, 1, 2, 4, 5\},
\{4, 1, 3, 2, 5\},
\{4, 2, 1, 3, 5\},
\{4, 3, 2, 1, 5\},
\{5, 4, 3, 2, 1\} \non b_2^{\text{Type II}}:    \ & \{1, 5, 4, 3, 2\},
\{2, 5, 4, 1, 3\},
\{3, 5, 4, 2, 1\}\label{b2channels}.
\end{align}

Now we start to use factorization to determine the $a$'s. We can first directly determine $a_1$ and $a_2$ by recycling our four-point results to obtain all the terms with factorization $1/\cD$ poles:
\eqs{
    &a^\text{GR}_1(1,2;3;4,5)=\sum_h a^\text{GR}(1,2,3,-k_I)\cdot \mathcal{M}_3(k_{I},4,5), \\
    &a^\text{GR}_2(1,2;3,4,5)\\ 
    =&\sum_{h} \mathcal{M}_3(1,2,-k_I) \cdot\left(\sum_{m=1}^2\frac{b^{(m,\text{GR})}(k_I,3,4,5)}{k_{45}^{2m}}+c^\text{GR}(k_I,3,4,5)+\text{Cyc} (3,4,5)\right),\label{a1a2gr}
}
where $\text{Cyc} (3,4,5)$ means $\{3,4,5\},\ \{5,3,4\},\ \{4,5,3\}$ and the $\cdot$ contains polarization sums \eqref{plsum}. 
Inheriting from 3 and four-point symmetric property, $a_1^\text{GR}$ is symmetric under particle exchange of $1 \leftrightarrow 2$ and $4 \leftrightarrow 5$, and invariant under reversal:
\begin{eqnarray}
    a_1^\text{GR}(A,B;C;D,E)=a_1^\text{GR}(\{B,A\};C;\{E,D\})=  a_1^\text{GR}(E,D;C;B,A).
\end{eqnarray}
By construction, $a_2$ has the following symmetry,
\begin{eqnarray}
    a_2^\text{GR}(A,B;C,D,E)=a_2^\text{GR}(\{A,B\},\{C,D,E\}).
\end{eqnarray}

This will be important in fixing the OPE poles and simplifying computation. Next collecting the double OPE pole structure, we have
    \begin{align}
&-b_1^{\left(m_1, m_2,\text{GR}\right)}(1,2;3;4,5)\notag\\ 
=&\underset{
    \begin{array}{l}
    \substack{\scriptscriptstyle k^{2m_1}_{{12}} \to 0}\\
      \substack{ \scriptscriptstyle 
 {k^{2m_2}_{{45}} \to 0}}
    \end{array} }{\mathrm{Res}}\left(\frac{a^\text{GR}_1(1,2;3;4,5)}{\mathcal{D}_{k_{{12}}}^{d} \mathcal{D}_{k_{{45}}}^{d}}+\frac{a^\text{GR}_2(1,2;3,4,5)}{\mathcal{D}_{k_{{12}}}^{d}}+\frac{a^\text{GR}_2(4,5;1,2,3)}{\mathcal{D}_{{k_{{45}}}}^{d}} \right ).
\end{align}
The $1/\cD_{k_{ij}}$ should expand to quadratic order of $k_{ij}$ as in \eqref{cDopeexpand}. 
We find that because of symmetry in $a_1$ and $a_2$, $b_1$ has reversal symmetry $b_1^{\left(m_1, m_2, \mathrm{GR}\right)}(1,2 ; 3 ; 4,5)=b_1^{\left(m_1, m_2, \mathrm{GR}\right)}(5,4;3;2,1)$
The detailed expression for $b$ is in \texttt{5gr.nb}. Unlike YM, gravity has also terms with single OPE pole. The terms seem complicated, but we can still simply take residue to determine it:
    \eqs{
&-b_2^{m,\text{GR}}(1,2;3,4,5)^{\text{Type I}}\\ 
=&\underset{\substack{\scriptscriptstyle 
 k^{2m}_{{12}} \to 0}}{\mathrm{Res}}\left\{ \left(\frac{a^{\text{GR}}_1(1,2;3;4,5)}{\mathcal{D}_{k_{{12}}}^{d} \mathcal{D}_{k_{{45}}}^{d}}+\sum_{m_1, m_2=1}^2 \frac{b_1^{\left(m_1, m_2, \mathrm{GR}\right)}(1,2 ; 3 ; 4,5)}{k_{12}^{2 m_1} k_{45}^{2 m_2}}+\text{Cyc} (3,4,5)\right)\right.\\
&\left.+\frac{a^\text{GR}_2(1,2;3,4,5)}{\mathcal{D}_{k_{{12}}}^{d}}+\left(\frac{a^\text{GR}_2(3,4;5,1,2)}{\mathcal{D}_{k_{{34}}}^{d}}+\text{Cyc} (3,4,5)\right)\right \},
\label{b2def}}where $\text{Cyc} (3,4,5)$ means $\{3,4,5\},\ \{5,3,4\},\ \{4,5,3\}$ which is defined as canonical permutation, i.e. generally for $C,D,E$, we have
\begin{eqnarray}
\sum_{m_1, m_2=1}^2    \frac{b_1^{\left(m_1, m_2, \mathrm{GR}\right)}(A,B;C,D,E)}{k_{AB}^{2 m_1} k_{DE}^{2 m_2}}+\frac{b_1^{\left(m_1, m_2, \mathrm{GR}\right)}(A,B;D,E,C)}{k_{AB}^{2 m_1} k_{CE}^{2 m_2}}+\frac{b_1^{\left(m_1, m_2, \mathrm{GR}\right)}(A,B;E,C,D)}{k_{AB}^{2 m_1} k_{CD}^{2 m_2}}.\label{conf}
\end{eqnarray} 
This combination of $b_1$  gives the $b_2$ of Type I. Computational details are given in appendix \ref{GRsingleope}. At this point, one is tempted to add up all cyclic permutations to get full $b_2$. But it turns out that we can at best do 7 permutations out of 10 as the first line in \eqref{b2channels}. 

The reason is that fixing the $a_1$ $a_2$ and $b_1$ list as in \eqref{list15} \eqref{list10}, the first line of $b_2$ in \eqref{b2channels} will not require new $b_1$ apart from the 15 in \eqref{list15} to determine it
\footnote{
The symmetry of $a_1$ and $a_2$ usually save one from requiring new permutations of indices in $a_1$ $a_2$ to determine $b_2$ in given channel.
}. However, if one keeps going with $b_2$ determined from canonical permutation in \eqref{conf}, one will require new $b_1$ that the ordering of 1,2,3,4,5 is different from anyone in \eqref{list15}.

The solution is to have a second way of determining $b_2$ from a new $b_1$ combination as shown below. With this, we can determine all $b_2$ using only $b_1$ within the list of \eqref{list15}.
  \eqs{
&-b_2^{m}(1,5;4,3,2)^{\text{Type II}}\\ 
=&\underset{\substack{\scriptscriptstyle 
 k^{2m}_{{12}} \to 0}}{\mathrm{Res}}\left\{ \frac{a_1(5,1;2;4,3)}{\mathcal{D}_{k_{{15}}}^{d} \mathcal{D}_{k_{{34}}}^{d}}+\frac{a_1(1,5;4;3,2)}{\mathcal{D}_{k_{{15}}}^{d} \mathcal{D}_{k_{{23}}}^{d}}+\frac{a_1(1,5;3;2,4)}{\mathcal{D}_{k_{{15}}}^{d} \mathcal{D}_{k_{{24}}}^{d}}+\right.\\
&\left.\left(\frac{b_1(5,1,2,3,4)}{k_{15}^{2m_1} k_{34}^{2m_2}} +\frac{{b_1(1,5,4,3,2)}}{k_{15}^{2m_1} k_{23}^{2m_2}}+\frac{{b_1(1,5,3,2,4)}}{k_{15}^{2m_1} k_{24}^{2m_2}}\right)+\right.\\
&\left.\frac{a_2(5,1;2,3,4)}{\mathcal{D}_{k_{{15}}}^{d}}+\left(\frac{a_2(2,3;4,5,1)}{\mathcal{D}_{k_{{23}}}^{d}}+\frac{a_2(4,2;3,5,1)}{\mathcal{D}_{k_{{24}}}^{d}}+\frac{a_2(3,4;2,1,5)}{\mathcal{D}_{k_{{34}}}^{d}}\right)\right \}.
\label{b2def2}}

Some computational details are given in Appendix \ref{GRsingleope2}. With the list \eqref{b2channels}, one finally has full $b_2$ contributions.

Having $a_1$ $a_2$ $b_1$ $b_2$'s and sum over all channels, we now have the total contribution that has no $k_I^2$ poles in OPE limit. As a consistency check there should not be any $k_I^2$ pole in the flat space limit.

As before, we now need to determine the contact terms $c$ with no poles. The procedure is basically the same as four-point by using flat space limit and Alder zero condition and gives:

\begin{align}
c_0(1,2 ; 3,4,5)= & \frac{d^2{-{4\left(s_1-s_2\right)^2}}-2 d\left(s_1+s_2\right)}{32(d-1)} \varepsilon_{12,12,34,35,45} .\label{ctter}
\end{align}
As a consistency check summing over all the contributions we have now, the expression is Lorentz invariant in flat space limit and vanishing soft limits after dimension reduction.

Finally, there is only one contribution left which corresponds to five graviton contact interactions $c_1$ and it takes the following form. We are only left with unfixed coefficients that can be readily determined by comparison with the flat space amplitude\footnote{
The full result for $c_1$ is in \texttt{GR5flatminusGR5ADSflat.txt}.},
\eqs{
    c_1=&\varepsilon_{ab,cd,ef,rs}(C_1 z^2\varepsilon_m \cdot k_i \varepsilon_n \cdot k_j +C_2\varepsilon_m \cdot \varepsilon_n \mathcal{D}_{k_{ij}}^d).\label{fiveptc1}
}
Summing over permutation for all the terms bootstrapped above, we obtained the first five-graviton amplitude in $AdS_{d+1}$ and it shares the similar analytic structure of S-matrix in flat space. We have checked the symmetry properties by swapping particles. The final result is given in \texttt{GR5fullADS.txt} and the computation process is coded in Mathematica file \texttt{5gr.nb}. 

So in the end, we bootstrap 5-point graviton amplitudes in $AdS_{d+1}$ from factorization soft limit (internal OPE limit, external soft limit) and high energy (flat space) limit.

\begin{eqnarray}
    &&\mathcal{M}_5^{GR}\non \eqn c_1+ \sum_{\vec{X}\in \text{list}_{a_1,b_1}}a_1(\vec{X})+b_1(\vec{X})+\sum_{\vec{X}\in \text{list}_{a_2,c_0}}a_2(\vec{X})+c_0(\vec{X})+\non &&\sum_{\vec{X}\in \text{list}_{b_2^{\text {Type I}}}}b_2^{\text {Type I}}(\vec{X})+\sum_{\vec{X}\in \text{list}_{b_2^{\text {Type II}}}}b_2^{\text {Type II}}(\vec{X}) .
    \label{gr5amp}
\end{eqnarray}

\section{Scalar Integrals and Cosmological correlators in momentum space}\label{backtomomentum}
Ultimately, our interest still lies in correlators composed of pure kinematic momentum. Mellin-momentum amplitude not only serves as a convenient framework for understanding amplitude structure but also serves as a useful computational tool for cosmological correlators. In this section, we will translate the Mellin-momentum amplitude results above back into momentum space by computing scalar integrals and provide a straightforward algebraic algorithm to demonstrate that this transition is transparent and simple for $n$-point. Finally, we will show that this directly gives us the cosmological correlators without any further computation.\\
The wavefunction coefficients in dS space can be computed from Witten diagrams in AdS after wick rotation $z \to i \eta$ \cite{Maldacena:2002vr,McFadden:2009fg,McFadden:2010vh,Maldacena:2011nz,Ghosh:2014kba} which is precisely the AdS correlators we studied here. This is particularly straightforward for gluons and gravitons, whose correlators are simply rational functions without IR divergences. In this case, the Wick rotation is trivial, with $z \to i \eta$ equivalent to $k_i \to i k_i$ in the bulk-to-boundary propagator. Thus, after performing the bulk integral over $z$, our result matches the wavefunction coefficients, up to an overall normalization factor. For the more careful readers, we refer the detailed discussion and derivation in \cite{Bzowski:2023nef} which we also record the formula here:
\eqs{
\psi^{\mathrm{dS}}_n\left(\boldsymbol{q}_1, \ldots, \boldsymbol{q}_n\right)=\left.(-i)^{n(d-\Delta)}\left\langle\left\langle\mathcal{O}\left(\boldsymbol{q}_1\right) \ldots \mathcal{O}\left(\boldsymbol{q}_n\right)\right\rangle\right\rangle^{\mathrm{AdS}}\right|_{L_{A d S} \rightarrow i L_{d S}}.
}

\begin{figure}
    \centering
    \includegraphics[width=0.75\linewidth]{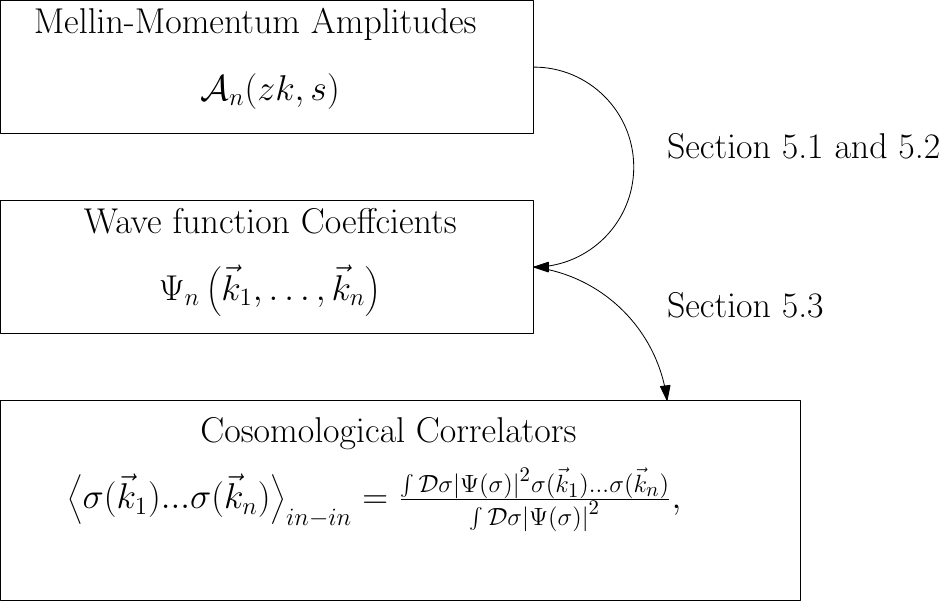}
    \caption{From On-shell Mellin-momentum amplitude in AdS to Cosmological correlators}l
    \label{fig:fromAmpToCC}
\end{figure}
\subsection{Yang-Mills}
Using the definition of Mellin-momentum amplitude \eqref{eq:MMamplitude}, we can easily obtain the following map for the scalar integral/ form factor in $d=3$:
\eqs{
\text{Amplitudes}& \to \text{Wavefunction coefficients} \\  
A(k,\varepsilon)& \to A(k,\varepsilon) \mathcal{C}_{YM} \\
\frac{A(k,\varepsilon)}{\mathcal{D}^{d-1}_{k_I}} 
& \to 
A(k,\varepsilon)\mathcal{I}^{YM}_I
\\
    \frac{A(k,\varepsilon)}{\mathcal{D}^{d-1}_{k_J}\dots \mathcal{D}^{d-1}_{k_M}} 
    & \to
    A(k,\varepsilon)\mathcal{I}^{YM}_{J\dots M}\\
s_i &\rightarrow \frac{z k_i}{2}+\frac{1}{4}.
\label{ymbacktomomentum}
}
The mapping of the Mellin variable is simply coming from the definition that the $z$ derivative acts on the bulk-to-boundary propagator. \footnote{This mapping for YM is slightly different from our previous work, as it turns out that at six points and beyond there will be internal Mellin variables, and the previous mapping is not enough.}
\subsubsection{Contact integrals}
Since scalar part of YM bulk-to-boundary propagator is the same as conformally coupled scalar, which is just plane wave in $d=3$, the scalar integral could be rather trivial as just pulling down the total energy in the exponent. Consider the following $n$-points contact integrals, we have
\eqs{
\mathcal{C}_{YM}^{n;1}=&\int_0^\infty \frac{dz}{z^4}z^{4-n}\prod_{i=1}^n \phi_{\Delta=2}(z,k_i) =\frac{1}{E_t},\\
\mathcal{C}_{YM}^{n;2}=&\int_0^\infty \frac{dz}{z^4}z^{3-n}(2s_i-1/2)\prod_{i=1}^n \phi_{\Delta=2}(z,k_i) = \frac{k_i}{E_t},\\
}
where we denote total energy as $E_t=\sum_{i=1}^{n}k_i$, and the second integral is a simple example that involves Mellin variables. The Mellin variable $s_i$ should be understood as $\frac{1}{4}(d-2z\partial_z)$ acting on the bulk-to-boundary propagator $\phi_i$.\footnote{
This is because a single Mellin variable $s$ is pull down as 
 $ \frac{1}{4}(d-2z\partial_z)z^{-2s+d/2}$.} Then by definition when $s$ acts on the bulk-to-boundary propagator we have 
 $s \phi_{\Delta=2}(z,k)\equiv \frac{1}{4}\left(3-2 z \partial_z\right) z e^{-z k}=(\frac{z k}{2}+\frac{1}{4}) ze^{-zk} $. 
  So to sum up, we have  $s_i \to \frac{z k_i}{2}+\frac{1}{4}$.
  
 \subsubsection{Exchange integrals}
 
We will be using the following representation of the bulk-to-bulk propagator,\footnote{
The usual Bulk-to-Bulk propagator is\cite{Liu:1998ty}
\begin{eqnarray}
    G_{\Delta}(\vec{k},z_1,z_2)=\left(z_1z_2\right)^{d / 2}  \int_0^{\infty} d p  \frac{p}{p^2+k^2} J_{\Delta-d/2}\left(p z_1\right) J_{\Delta-d/2}\left(p z_2\right).\label{oldrep}
\end{eqnarray} 
This can also be shown to be equivalent by using the Mellin representation of Bessel function $J$ and $K$ and $\Gamma(x)\Gamma(1-x)=\csc(\pi x)$)
}
 \eqs{
G_\Delta(\vec{k},z_1,z_2)=-i\int_{0}^{\infty}\frac{dp}{2\pi i}\frac{p^{d+1-2\Delta}}{k^2+p^2} \left( \phi_{\Delta}(z_1,i p)-\phi_{\Delta}(z_1,-i p)\right) \left( \phi_{\Delta}(z_2,i p)-\phi_{\Delta}(z_2,-i p)\right),\label{newrep}
}
where $\phi_{\Delta}$ is the bulk-to-boundary propagator  $\phi_{\Delta}(z,k)=\sqrt{\frac{2}{\pi}}z^{d/2}k^{\Delta-d/2} K_{\Delta-d/2}(kz)$ with $k=|\vec{k}|$.
For $n$-point amplitude with only one inverse operator $\frac{1}{\mathcal{D}^{\Delta}_{k_I}}$, the scalar exchange integral involves for exchange diagram with $n_I$ legs at left is,
\eqs{
\mathcal{I}^{YM}_I=&\int \frac{d z_1}{z_1^4} \frac{d z_2}{z_2^4} \left( z_1^{4-n_I}\prod_{i=1}^{n_I}\phi_{\Delta=2}\left(z_1, k_i\right) \right)  G_{\Delta=2}(\vec{k}_I,z_1,z_2)\left( z_2^{4-(n-n_I)}\prod_{i=n_I+1}^{n}\phi_{\Delta=2}\left(z_2, k_i\right) \right) \\
=&\int _{0} ^{\infty} \frac{dp}{2 \pi i} \frac{-i}{k_I^2+p^2}  \bar{\mathcal{C}}_{YM}(k_1, \dots,ip) \bar{\mathcal{C}}_{YM}(ip, \dots,k_n) 
\label{eq133}
}
where $\vec{k}_I=\vec{k}_1+...+\vec{k}_{n_I}$ and we define shifted functions: $\Bar{\mathcal{C}}=\mathcal{C}(k_1,\dots,ip)-\mathcal{C}(k_1,\dots,-ip)$ and $\Bar{\mathcal{I}}=\mathcal{I}(ip\dots,k_n)-\mathcal{I}(-ip\dots,k_n)$. The integral for $p$ is from minus infinity to plus infinity and can be regarded as sum of all residues above the real line\cite{Raju:2012zs}. Summing up those residues, one will end up with the final line which coinsides with usual result. We will give example later.
\\
Similarly, when we have more than one inverse operator/bulk-to-bulk propagator $\frac{1}{\mathcal{D}_{k_{I}} \dots \mathcal{D}_{k_{M}}}$, we have the following recursive integrals
\eqs{
\mathcal{I}^{YM}_{IJ \dots M}&\equiv \int_{0}^{\infty} \frac{dp}{2\pi i} \frac{-i}{k_I^2+p^2}\Bar{\mathcal{C}}_{YM}(k_1,\dots,ip)\Bar{\mathcal{I}}^{YM}_{J\dots M}(ip,\dots,k_n). \label{exchangeYM}
}
\\
Finally we should note that, in $n<6$ point, we can just use the recursion in \cite{Mei:2024abu}. However, we find that at six-point we will have internal Mellin variables, so we need this new version to make things work.
\subsubsection{Examples on recursive integral}
We start with a four point YM exchange  scalar integral which corresponds to map single inverse operator to momentum space
\eqs{
\frac{1}{\mathcal{D}_{k_{12}}^{d-1} }\to \mathcal{I}_{\{12\}}^{YM}(k_1,k_2,k_3,k_4,k_{12}).
}
With this, by definition
\eqs{
\mathcal{I}_{\{12\}}^{YM}(k_1,k_2,k_3,k_4,k_{12}) 
=
\int \frac{d z_1}{z_1^4} \frac{d z_2}{z_2^4} 
\prod_{i=1}^{2}\phi_{\Delta=2}\left(z_1, k_i\right)
z_1
G_{\Delta=2}(k_{12},z_1,z_2) 
z_2
\prod_{j=3}^{4}\phi_{\Delta=2}\left(z_2, k_j\right)
}
where the additional $z_1$ and $z_2$ are from scaling of 3 point YM vertex. The alternative way of evaluating it that we advocate here is as follows:
starting with 3 point contact
\begin{eqnarray}
    \mathcal{C}_{Y M}^{3 ; 1}=\int_0^{\infty} \frac{d z}{z^4} z^{4-n} \prod_{i=1}^3 \phi_{\Delta=2}\left(z, k_i\right)=\frac{1}{k_1+k_2+k_3},
\end{eqnarray}
then using \eqref{newrep}, the integral is just taking residues of all the poles on the upper half-plane $p \to ik_{12}, i(k_1+k_2),i(k_3+k_4)$ of the shifted 3 point contact:
\begin{eqnarray}
 &  \mathcal{I}_{\{12\}}^{Y M}(k_1,k_2,k_3,k_4,k_{12})
\nonmy
 =&
  \underset{p}{\mathrm{Res}} \frac{1}{(k_{12}^2+p^2)}\left(\frac{1}{k_1+k_2+ip}-\frac{1}{k_1+k_2-ip}\right)
\left(\frac{1}{k_3+k_4+ip}-\frac{1}{k_3+k_4-ip}\right)
\nonmy
=&
\frac{1}{(k_1+k_2+k_3+k_4)(k_{12}+k_1+k_2)(k_{12}+k_3+k_4)}.
\end{eqnarray}
Now we can move to more complicated example:
\\
\begin{eqnarray}
  \frac{1}{\mathcal{D}_{k_{12}}^{d-1} \mathcal{D}_{k_{45}}^{d-1} }\to \mathcal{I}_{\{12\}\{45\}}^{YM}(k_1,k_2,k_3,k_4,k_5,k_{12},k_{45}).
\end{eqnarray}
\\
Recycling four-point and three-point data, we can compute five-point exchange diagram with two propagator as
\\
\begin{eqnarray}
    \mathcal{I}_{\{12\}\{45\}}^{Y M} = \int_{0}^{\infty} \frac{d p}{2 \pi i} \frac{-i}{k_{12}^2+p^2} \overline{\mathcal{C}}^{3,1}_{Y M}\left(k_1,k_2, i p\right) \overline{\mathcal{I}}_{\{45\}}^{Y M}(ip,k_3,k_4,k_5).
\end{eqnarray}
\\
Summing up the residues at 
$p\to i k_{12} , i (k_1+k_2) , i (k_3+k_4+k_5), i (k_3+k_{45}) $, one gets the result matched with literature 
\begin{eqnarray}
 &&\mathcal{I}_{\{12\}\{45\}}^{Y M}
 \nonmy
 \eqnmy
 \frac{k_{12}+k_{45}+k_1+k_2+2 k_3+k_4+k_5}{E_t \left(k_{12}+k_1+k_2\right)
   \left(k_{12}+k_3+k_4+k_5\right) \left(k_{45}+k_1+k_2+k_3\right) \left(k_{45}+k_4+k_5\right)
   \left(k_{12}+k_{45}+k_3\right)},
   \nonmy 
\end{eqnarray}
with $E_t=k_1+k_2+k_3+k_4+k_5$.
\\
\subsubsection{Examples with Mellin variable}
Apart from purely exchange diagrams, one may have additional $s$ variables enter the game. In practive, this occurs when we convert $a_2$ terms back to momentum space. And one will encounter terms that have Mellin variable of quadratic power 
\\
\begin{eqnarray}
    \frac{\left(\varepsilon_1 \cdot k_2\right)\left(\varepsilon_2 \cdot \varepsilon_3\right)\left(\varepsilon_4 \cdot \varepsilon_5\right)\left(s_4-s_5\right)\left(2\left(1+2 s_3+s_4+s_5\right)-d\right)}{2 z  \cD_{k_{12}} k_{45}^2}.
\end{eqnarray}
\\
In this case, one should first use on-shell condition 
$s_a^2=\frac{1}{16}\left((d-2 \Delta)^2+4 z^2 k_a^2\right)$
to lower explicit Mellin variable power to 1 and then use the map 
$s_i \rightarrow \frac{z k_i}{2}+\frac{1}{4} \text {.}$
Eventually for this example, one gets back to momentum space as
\begin{eqnarray}
    \frac{\left(\varepsilon_1 \cdot k_2\right)\left(\varepsilon_2 \cdot \varepsilon_3\right)\left(\varepsilon_4 \cdot \varepsilon_5\right)\left(k_4-k_5\right)\left(2 k_3+k_4+k_5\right)}{4\left(k_1+k_2+k_3+k_4+k_5\right)\left(k_1+k_2+k_{12}\right)\left(k_3+k_4+k_5+k_{12}\right) k_{45}^2}.
\end{eqnarray}

\subsection{Gravity}
For Gravity in $d=3$, we found a similar recursion for $n$-point scalar integral as well. 
\eqs{
\text{Amplitudes}
& \to
\text{Wavefunction coefficients} \\ 
z^2 M(k,\varepsilon)
& \to
M(k,\varepsilon) \mathcal{C}_{G R}^{n ; 1}\\
\prod\limits_{m=1}^{l}\left(-2s_{m}+d/2\right) M(k,\varepsilon) & \to
M(k,\varepsilon)\mathcal{C}_{G R}^{n ; 2} \label{maptomomentumGR}\\ 
\frac{M(k,\varepsilon)}{\mathcal{D}^d_{k_I}} 
& \to 
M(k,\varepsilon)\mathcal{I}_I
\\
    \frac{M(k,\varepsilon)}{\mathcal{D}^d_{k_J}\dots \mathcal{D}^d_{k_M}} 
    & \to
    M(k,\varepsilon)\mathcal{I}_{J\dots M}.
}
\subsubsection{Contact integrals}
The scalar part of GR bulk-to-boundary propagator is the same as massless scalar. The possible contact integrals one would encounter are:
\eqs{
\mathcal{C}_{GR}^{n;1}=&\int \frac{dz}{z^4}z^{2}\prod_i^n \phi_{\Delta=3}(z,k_i) =\left(\sum\limits_{m=0}^{n-2}m!\sum\limits_{1\le i_1<\ldots<i_{m+2}}^{n} \frac{k_{i_1} \ldots k_{i_{m+2}}}{E_t^{m+1}}\right)-E_t,
\label{contactSIGR}
}
\eqs{
\mathcal{C}_{G R}^{n ; 2}=\int \frac{dz}{z^4} \prod_{m=1}^l\left(-2 s_m+d / 2\right)\phi_{\Delta=3}(z,k_i) =\Bigg(\sum_{m=0}^{n-l} {(2l-4) !} \sum_{
\begin{array}{c}
  \substack{\scriptscriptstyle l+1\le i_{l+1}<} \\  \substack{\scriptscriptstyle  \ldots<i_{m+l}}
\end{array}
}^n 
\frac{k_{i_{l+1}} \ldots k_{i_{m+l}}}{E_t^{(2l-4)+m+1}}\Bigg)(-1)^lk_1^2...k_l^2,\label{contact2}
}
and we refer the reader to Appendix \ref{Appendix:GRcontact} for the derivation.

\subsubsection{Gravity exchange integral}
We again use the following representation of bulk-to-bulk propagator with $\Delta=d=3$,
\eqs{
G_\Delta(\vec{k},z_1,z_2)=-i\int_{0}^{\infty}\frac{dp}{2\pi i}\frac{p^{d+1-2\Delta}}{k^2+p^2} \left( \phi_{\Delta}(z_1,i p)-\phi_{\Delta}(z_1,-i p)\right) \left( \phi_{\Delta}(z_2,i p)-\phi_{\Delta}(z_2,-i p)\right),\label{newrep1}
}
where $\phi_{\Delta}(z,k)=\sqrt{\frac{2}{\pi}}z^{d/2}k^{\Delta-d/2} K_{\Delta-d/2}(kz)$.
Consider an exchange diagram with $n_I$ legs at left. Replacing the propagator with \eqref{newrep1}, one will end up with two contact scalar integral we determined earlier:
\eqs{
\mathcal{I}^{GR}_I=&\int \frac{d z_1}{z_1^4} \frac{d z_2}{z_2^4} \left( z_1^{2}\prod_{i=1}^{n_I}\phi_{\Delta=3}\left(z_1, k_i\right) \right)  G_{\Delta=3}(\vec{k}_I,z_1,z_2)\left( z_2^{2}\prod_{i=n_I+1}^{n}\phi_{\Delta=3}\left(z_2, k_i\right) \right) \\
=&\int _{{0}} ^{\infty} \frac{dp}{2 \pi i} \frac{-i p^{-2}}{k_I^2+p^2}  \bar{ \mathcal{C}}_{G R}^{n ; 1}(k_1, \dots,ip) \bar{\mathcal{C}}_{G R}^{n ; 1}(ip, \dots,k_n) ,
}
where $\overline{\mathcal{C}}_{G R}^{n ; 1}\left(k_1, \ldots, i p\right)={\mathcal{C}}_{G R}^{n ; 1}\left(k_1, \ldots, i p\right)-{\mathcal{C}}_{G R}^{n ; 1}\left(k_1, \ldots,- i p\right)$.
Furthermore, we can recursively build up exchange diagrams with more propagators as
\eqs{
\mathcal{I}^{GR}_{IJ \dots M}&\equiv\int_{0}^{\infty} \frac{dp}{2\pi i} \frac{-i p^{-2}}{k_I^2+p^2}\Bar{\mathcal{C}}_{G R}(k_1,\dots,ip)\Bar{\mathcal{I}}^{GR}_{J\dots M}(ip,\dots,k_n).\label{exchange}
}

These cover all $n$-point scalar integrals for gravity scalar integral in $d=3$. Therefore, if one is provided with an $n$-point Mellin-momentum amplitude, one can simply follow the map to obtain the wavefunction coefficients, requiring only the computation of a finite number of residues without the need for any new time integrals. 
\subsubsection{Examples on recursive integral}
We start with a four point GR s-channel exchange scalar integral
\eqs{
\frac{1}{\mathcal{D}_{k_{12}}^{d} }\to \mathcal{I}_{\{12\}}^{GR}(k_1,k_2,k_3,k_4,k_{12}),
}
which has the form
\begin{eqnarray}
{\mathcal{I}}_{\{12\}}^{GR}\left(k_1, k_2, k_3, k_4,k_{12}\right)
=
\int \frac{d z_1}{z_1^4} \frac{d z_2}{z_2^4}   z_1^2 z_2^2 
\prod_{i=1}^{2}\phi_{\Delta=3}\left(z_1, k_i\right) G_{\Delta=3}(\vec{k}_{12},z_1,z_2)
\prod_{j=3}^{4}\phi_{\Delta=3}\left(z_2, k_j\right).
\end{eqnarray}
 With representation \eqref{newrep1}, any scalar integral in $d=3$ can be derived recursively starting from 3 point contact integral: 
\begin{eqnarray}
  \begin{aligned}
      \mathcal{C}_3(k_1,k_2,k_3)&=\int \frac{d z_1}{z_1^4}z_1^2\phi_{\Delta=3}\left(z_1, k_1\right) \phi_{\Delta=3}\left(z_1, k_2\right)\phi_{\Delta=3}\left(z_1, k_3\right)\\&=\frac{k_2 k_3 k_1}{\left(k_1+k_2+k_3\right)^2}-k_1-k_2-k_3+\frac{k_1 k_2+k_3 k_2+k_1 k_3}{k_1+k_2+k_3}.
  \end{aligned}
\end{eqnarray}
In this case, we have 
\begin{eqnarray}
 &{\mathcal{I}}_{\{12\}}^{GR}\left(k_1, k_2, k_3, k_4,k_{12}\right)\notag \\
 =& 
 \frac{1}{2i}\underset{p}{\mathrm{Res}}
 \frac{p^{-2}}{k_{12}^2+p^2}\left(\mathcal{C}_3(k_1,k_2,ip)-\mathcal{C}_3(k_1,k_2,-ip)\right)\left(\mathcal{C}_3(k_3,k_4,ip)-\mathcal{C}_3(k_3,k_4,-ip)\right)
 \notag\\
 =&
 \frac{2 k_1 k_2 k_3 k_4\left(E_L E_R+E k_{12}\right)}{E_L^2 E^3 E_R^2}+\frac{k_1 k_2\left(E_L (k_3+k_4)+E k_{12}\right)}{E_L^2 E^2 E_R}+\frac{k_3 k_4\left(E k_{12}+E_R (k_1+k_2)\right)}{E_L E^2 E_R^2}+\frac{E_L E_R-k_{12}^2}{E_L E E_R},
 \non
\end{eqnarray} 
where the residues are the same as YM and $E_L=k_1+k_2+k_{s},$ $E_R=k_3+k_4+k_{s}$, and $E=k_1+k_2+k_3+k_4$.
We should emphasize that there is no integral to do just taking residues of integral variable $p$. This agrees with conventional scalar integral computation using \eqref{oldrep} in $d=3$. Once we have it, the five-point exchange with two propagators 
\begin{eqnarray}
  \frac{1}{\mathcal{D}_{k_{12}}^{d} \mathcal{D}_{k_{45}}^{d} }\to \mathcal{I}_{\{12\}\{45\}}^{GR}(k_1,k_2,k_3,k_4,k_5,k_{12},k_{45}),
\end{eqnarray}
can be expressed by 
\begin{eqnarray}
&\mathcal{I}_{\{12\},\{45\}}^{GR}\left(k_1, k_2, k_3, k_4,k_5,k_{12},k_{45}\right)
\non
    =&\int_{-\infty}^{\infty} \frac{d p_1}{2 \pi i} \frac{-ip_1^{-2}}{k_{12}^2+p_1^2} \overline{\mathcal{C}}_1\left(k_1, k_2, i p_1\right) \overline{\mathcal{I}}_{\{45\}}^{GR}\left(i p_1, k_3, k_4, k_5\right),
    \non
   =&
    \underset{p_1}{\mathrm{Res}}\; \underset{p_2}{\mathrm{Res}}
    \; \frac{64 k_3^3 p_1^4 p_2^4 \left(k_1^2+4 k_2 k_1+k_2^2+p_1^2\right) \left(k_4^2+4 k_5 k_4+k_5^2+p_2^2\right)}{  \left(k_3^2+\left(p_1-p_2\right){}^2\right){}^2\left(k_3^2+\left(p_1+p_2\right){}^2\right){}^2 \left(\left(k_1+k_2\right){}^2+p_1^2\right){}^2
   \left(k_{12}^2+p_1^2\right)\left(\left(k_4+k_5\right){}^2+p_2^2\right){}^2 \left(k_{45}^2+p_2^2\right)
   },\non 
\end{eqnarray}
where we define a shifted function: $\Bar{\mathcal{C}}=\mathcal{C}(k_1,\dots,ip)-\mathcal{C}(k_1,\dots,-ip)$ and $\Bar{\mathcal{I}}=\mathcal{I}(ip\dots,k_n)-\mathcal{I}(-ip\dots,k_n)$.

\subsection{Cosmological correlators}
After finishing discussing the scalar integral we can get the wave function coefficients for YM and GR explicitly. Then in the following, we will review how to get cosmological correlators starting from wave function coefficients of arbitrary spins.
We use subscript $in-in$ to distinguish the CFT correlator with the cosmological correlator ($in-in$ correlator)
\begin{equation}
\left\langle \sigma(\vec{k}_1)...\sigma(\vec{k}_n)\right\rangle_{in-in} =\frac{\int\mathcal{D}\sigma \left|\Psi(\sigma)\right|^{2} \sigma(\vec{k}_1)...\sigma(\vec{k}_n)}{\int\mathcal{D}\sigma\left|\Psi(\sigma)\right|^{2}},
\label{psi2}
\end{equation}
where the wavefunction expanding in field profiles on the future space-like boundary takes the form
\begin{equation}
\Psi(\sigma)\propto\exp\left[-\frac{1}{2}\int\sigma_{1}\sigma_{2}\Psi_{2}-\frac{1}{3!}\int\sigma_{1}\cdots\sigma_{3}\Psi_{3}-\frac{1}{4!}\int\sigma_{1}\cdots\sigma_{4}\Psi_{4}+\cdots\right],
\end{equation}
and we use the shorthand notation
\begin{equation}
\int\sigma_{1}\cdots\sigma_{n}=\int\prod_{i=1}^{n}\frac{d^{3}k_{i}}{(2\pi)^{3}}\sigma\left(\vec{k}_{i}\right)\delta^{3}\left(\sum_{i=1}^{n}\vec{k}_{i}\right),
\end{equation}
and the wavefunction coefficients are,
\begin{equation}
\Psi_{n}=\Psi_{n}\left(\vec{k}_{1},\dots,\vec{k}_{n}\right),
\end{equation}
which are given in \eqref{eq:MMamplitude}. Hence once we determine the Mellin-momentum amplitude we can use the map in section \ref{backtomomentum}, which computes all the scalar integrals to obtain the wavefunction coefficients.\\

In order to get an intuitive relation between wavefunction coefficients and cosmological correlators,
we can rephrase the path integral in~\eqref{psi2} in terms of an effective action:
\begin{equation}
\left\langle \sigma(\vec{k}_1)...\sigma(\vec{k}_n)\right\rangle_{in-in}=\frac{\int\mathcal{D}\sigma e^{-S(\sigma)}\sigma(\vec{k}_1)...\sigma(\vec{k}_n)}{\int\mathcal{D}\sigma e^{-S(\sigma)}},
\end{equation}
where
\begin{equation}
S(\sigma)=\frac{1}{2}\int\sigma_{1}\sigma_{2}\operatorname{Re} \Psi_2+\frac{1}{3!} \int \sigma_1 \sigma_2\sigma_3\operatorname{Re} \Psi_3+\frac{1}{4!}\int\sigma_{1}\cdots\sigma_{4}\operatorname{Re} \Psi_4+\frac{1}{6!}\int\sigma_{1}\cdots\sigma_{6}\operatorname{Re} \Psi_6+\cdots
\end{equation}

The Feynman rules of this action are easy to read off:
\begin{equation}
\begin{gathered}
\begin{tikzpicture}[baseline]

\draw (0, 0) -- (2, 0);
\node at (1, -0.2) {$\vec{k}$};

\end{tikzpicture}\end{gathered}
=\frac{1}{\mathrm{Re}\Psi_2}~, \quad 
\begin{gathered}
\begin{tikzpicture}[baseline]

\draw (0, 0) -- (-0.7, -0.7);
\draw (0, 0) -- (-0.7, 0.7);
\draw (0, 0) -- (1, 0);

\end{tikzpicture}\end{gathered}
=\mathrm{Re}\Psi_3~, \quad 
\begin{gathered}
\begin{tikzpicture}[baseline]

\draw (-0.7, -0.7) -- (0.7, 0.7);
\draw (-0.7, 0.7) -- (0.7, -0.7);

\end{tikzpicture}\end{gathered}
=\mathrm{Re}\Psi_4, \quad \dots
\label{cosFeyn}
\end{equation}

The effective action used to compute the \textit{in-in} correlators that has an infinite number of vertices, each of which has a perturbative expansion in the bulk coupling. We
can then compute the \textit{in-in} correlators using the effective action, whose vertices
will contain different powers of the bulk coupling and then keep all contributions
that contain a given order in the coupling. 
In practice, we can form Feynman diagrams out of the rules written in \eqref{cosFeyn}. And the additional rule is to also dress the diagrams with
external propagators.
As an example, at tree-level, the three-point \textit{in-in} correlator comes from a three-point
contact diagram where we only keep the tree-level contribution to
the coefficient $\operatorname{Re} \Psi_3$. 
\begin{equation}
    \begin{gathered}
\begin{tikzpicture}[baseline]

\draw (0, 0) -- (-0.7, -0.7);
\draw (0, 0) -- (-0.7, 0.7);
\draw (0, 0) -- (1, 0);

\end{tikzpicture}\end{gathered} \nonumber
\end{equation}

\begin{equation}
\left\langle \sigma(\vec{k}_1)\sigma(\vec{k}_2)\sigma(\vec{k}_3)\right\rangle_{in-in}=\frac{\mathrm{Re}\Psi_3}{\prod\limits_{a=1}\limits^3 \mathrm{Re}\Psi_2(\vec{k}_a)}
\end{equation}

As a slightly interesting example, a tree-level four point \textit{in-in} correlator comes from  four-point wave function coefficient $\operatorname{Re} \Psi_4$, together  with exchange diagram in three channels with propagator as
 $1/\operatorname{Re} \Psi_2\left(k\right)$

\begin{equation}
    \begin{gathered}
\begin{tikzpicture}[baseline]

\draw (-0.3, 0) -- (-0.7, -0.7);
\draw (-0.3, 0) -- (-0.7, 0.7);
\draw ( 0.3,0) -- (0.7, 0.7);
\draw ( 0.3,0) -- (0.7,- 0.7);
\draw ( 0.3,0) -- ( -0.3,0);
\end{tikzpicture}
\qquad 
\begin{tikzpicture}[baseline]

\draw ( 0,-0.3) -- (-0.7, -0.7);
\draw ( 0,-0.3) -- (0.7,- 0.7);
\draw (0,0.3) -- (0.7, 0.7);
\draw (0, 0.3) -- (-0.7, 0.7);
\draw ( 0,0.3) -- ( 0,-0.3);
\end{tikzpicture}
\qquad 
\begin{tikzpicture}[baseline]

\draw ( 0,-0.3) -- (-0.7, -0.7);
\draw ( 0,-0.3) -- (0.7,0.7);
\draw (0,0.3) -- (0.7,- 0.7);
\draw (0, 0.3) -- (-0.7, 0.7);
\draw ( 0,0.3) -- ( 0,-0.3);
\end{tikzpicture}
\qquad 
\begin{tikzpicture}[baseline]

\draw (0, 0) -- (-0.7, -0.7);
\draw (0, 0) -- (-0.7, 0.7);
\draw (0, 0) -- (0.7, 0.7);
\draw (0, 0) -- (0.7,- 0.7);
\end{tikzpicture}
\end{gathered} \nonumber
\end{equation}

\begin{eqnarray}
    \begin{aligned}
\left\langle \sigma({\vec{k}_1})\sigma({\vec{k}_2})\sigma({\vec{k}_3} ) \sigma({\vec{k}_4})\right\rangle_{i n-i n}= & {\Bigg[4 \sum_{s,t,u} \sum_{h} \frac{1}{\operatorname{Re}\Psi_2(\vec  k_s)} \operatorname{Re} \Psi_3\left(\vec{k}_1, \vec{k}_2,-\vec k_s\right) \operatorname{Re} \Psi_{3}\left(\vec{k}_3, \vec{k}_4, \vec k_s\right)} \\
& -2 \operatorname{Re} \Psi_{ 4}
\Bigg] \prod_{a=1}^4 \frac{1}{\operatorname{Re} \Psi_2\left(  \vec  k_a\right)},
\end{aligned}
\end{eqnarray}
%
where $\underset{h}{\sum}$ means we also need to sum over the internal helicity for spinning particles.\\

Finally for the purpose of this paper, the tree-level five-point cosmological correlator comes from a five-point wave function coefficient $\operatorname{Re} \Psi_5$ together with exchange diagrams of two topologies:

\begin{tikzpicture}[baseline]
    \coordinate (A) at (18:1.2);
    \coordinate (B) at (90:1.2);
    \coordinate (C) at (162:1.2);
    \coordinate (D) at (234:1.2);
    \coordinate (E) at (306:1.2);
    \coordinate (O) at (0,0); 

    \draw (O) -- (A);
    \draw (O) -- (B);
    \draw (O) -- (C);
    \draw (O) -- (D);
    \draw (O) -- (E);
\end{tikzpicture}
\qquad 
\begin{tikzpicture}[baseline]
    \coordinate (A) at (18:1.2);
    \coordinate (B) at (85:1.2);
    \coordinate (C) at (162:1.2);
    \coordinate (D) at (234:1.2);
    \coordinate (E) at (316:1.2);
    \coordinate (O1) at (18:0.3);
    \coordinate (O2) at (18:-0.3); 

    \draw (O1) -- (A);
    \draw (O1) -- (B);
    \draw (O2) -- (C);
    \draw (O2) -- (D);
    \draw (O1) -- (E);
    \draw (O1) -- (O2);
\end{tikzpicture}
+ \textit{ cyclic }
\qquad 
\begin{tikzpicture}[baseline]
    \coordinate (A) at (18:1.2);
    \coordinate (B) at (90:1.2);
    \coordinate (C) at (162:1.2);
    \coordinate (D) at (234:1.2);
    \coordinate (E) at (306:1.2);
    \coordinate (O1) at (162:-0.6);
    \coordinate (O2) at (18:-0.6);
    \coordinate (Ox) at (0,0); 

    \draw (O1) -- (A);
    \draw (Ox) -- (B);
    \draw (O2) -- (C);
    \draw (O2) -- (D);
    \draw (O1) -- (E);
    \draw (O1) -- (Ox);
    \draw (O2) -- (Ox);
\end{tikzpicture}
+ \textit{ cyclic },\\
with the expression given as
\begin{eqnarray}
    \begin{aligned}
&
\left\langle \sigma(\vec{k}_1)\sigma(\vec{k}_2)\sigma(\vec{k}_3)\sigma(\vec{k}_4)\sigma(\vec{k}_5)\right\rangle_{in-in}= \Bigg[  \operatorname{Re} \Psi_5+ 
\\&
\sum_h \frac{1}{\operatorname{Re} \Psi_2\left(\vec k_{12}\right)} \operatorname{Re} \Psi_{3}\left(\vec{k}_1, \vec{k}_2,-\vec k_{12}\right) \operatorname{Re} \Psi_{4}\left( \vec k_{12},\vec{k}_3, \vec{k}_4, \vec{k}_5\right)+cyclic
\\ &
{ \sum_{h_1,h_2} \frac{1}{\operatorname{Re} \Psi_2\left(\vec k_{12}\right)}\frac{1}{\operatorname{Re} \Psi_2\left(\vec k_{45}\right)} \operatorname{Re} \Psi_{3}\left(\vec{k}_1, \vec{k}_2,-\vec  k_{12}\right) \operatorname{Re} \Psi_{3 }\left( \vec k_{12},\vec{k}_3, \vec k_{45}\right)\operatorname{Re} \Psi_{3}\left(-\vec {k}_{45}, \vec{k}_4,\vec k_5\right)+cyclic}
\\ 
& 
\Bigg] \prod_{a=1}^5 \frac{1}{\operatorname{Re} \Psi_2\left(\vec  k_{a}\right)} .
\end{aligned}
\end{eqnarray}

\section{Conclusion}
\label{section6}

In this paper, we build upon the concept of on-shell Mellin-momentum amplitudes for cosmological correlators, an idea first introduced in \cite{Mei:2023jkb}. To demonstrate the power of this approach, we have conducted a detailed bootstrap derivation of the full Mellin-momentum amplitudes for both Yang-Mills (YM) and General Relativity (GR) at the five-point in Anti-de Sitter (AdS) space \cite{Mei:2024abu}. Our findings reveal that the analytic structure of Mellin-momentum amplitudes for 
$n$-points is exceptionally straightforward and recursively computable, similar to the process for amplitudes in flat space. This finding supports the notion that these amplitudes in (A)dS spaces serve a role analogous to that of the S-matrix in flat space.\\
Practically, the accuracy of our bootstrapped result for the five-graviton amplitude is inherently verifiable, as any discrepancies would appear as non-physical poles. Moreover, the structure of the five-graviton amplitude supports the extension of the double copy construction, as proposed in \cite{Mei:2023jkb}, to higher-point amplitudes—a prospect that warrants further investigation.
\\
In flat space, the bootstrap method effectively makes Lorentz invariance, locality, and unitarity manifest while leaving a gauge redundancy. However, the situation is slightly different in cosmology, where our observables are boundary correlators. This necessitates setting the bulk coordinate $\varepsilon_{z} \to 0$, so our bootstrap strategy would be making locality, and unitarity manifest with no gauge redundancy while leaving the spacetime symmetry not manifest.\\

Although the amplitude serves as the integrand for AdS boundary correlators or wave function coefficients for cosmological correlators, as demonstrated in the main text, it is still convenient to go back to momentum space. In three dimensions ($d=3$), we show that higher-point scalar integrals can be iteratively constructed from their lower-point counterparts. This capability simplifies the translation of our amplitude results into the Yang-Mills and, most crucially, the Gravity Quadrispectrum for cosmological correlators. We also hope that this paper will act as an introductory guide for the amplitude community interested in cosmological correlators. In this formalism, newcomers can easily adopt Feynman rules, closely resembling those in flat space, to compute their favorite amplitudes and then map these to the corresponding correlators. Alternatively, they can follow the amplitude bootstrap approach detailed in this paper to build their own favorite amplitude.
\\

There are several directions we can explore. Our first dream for the future is to discover an all multiplicity formula like Park-Taylor formula\cite{Parke:1986gb} for Yang-Mills in (A)dS. Especially, in this paper we discover that the above five-point factorization is indeed enough to fix the Yang-Mills amplitude. So for such a highly constrained theory, it's natural to expect that some form of an all multiplicity formula exists. It would be very interesting to combine the recent development of twistors in (A)dS space\cite{Baumann:2024ttn}, which makes conformal symmetry manifest.  We are also keen to extend our gravity calculations to loop-level. Once employing our bootstrap approach to determine the Mellin-momentum amplitude, we are then left with scalar loop integrals. Particularly, it was shown in \cite{Chowdhury:2023arc} that the scalar loop integral for the in-in correlator is closer to the S-matrix in flat space. More broadly, our bootstrap strategy does not rely on spacetime symmetry \cite{Pajer:2020wxk}, positioning our AdS study as a foundational example applicable to various curved backgrounds, including FLRW spacetime and black hole scenarios. Lastly, utilizing the techniques developed herein, we have successfully derived the soft limit of YM and GR \cite{Chowdhury:2024wwe}\footnote{There is a similar result for fermion \cite{Chowdhury:2024snc} in correlator version}. In the upcoming work, we willdemonstrate how this limit can facilitate the construction of higher-point helicity amplitudes within the Mellin-momentum formalism \cite{Mei:2025abx}.

\begin{center}
\textbf{Acknowledgements}
\end{center}

We thank Tim Adamo, Chandramouli Chowdhury, David Stefanyszyn, Einan Gardi, Arthur Lipstein, Paul McFadden, Silvia Nagy, Enrico Pajer, Santiago Agui Salcedo, Charlotte Sleight, Massimo Taronna, Mao Zeng and Bin Zhu, for useful discussions. JM is supported by the European Union (ERC, UNIVERSE PLUS, 101118787), Durham-CSC Scholarship. Views and opinions expressed are however those of the authors only and do not necessarily reflect those of the European Union or the European Research Council Executive Agency. Neither the European Union nor the granting authority can be held responsible for them.  YM is supported by a Edinburgh Global Research Scholarship.

\newpage
\appendix

\section{Example for field redefinition}
\label{field_redef}
Consider field redefinition for a massless free field in the bulk with conformal dimension $\Delta=3$ and $d=3$. The original Lagrangian is given by
\begin{eqnarray}
	\mathcal{L}= \partial_m \phi \partial^m \phi 
\end{eqnarray}
where $m$ runs over $(z,x)$ coordinates.
And then under field redefinition 
$\phi \rightarrow \phi+\alpha \phi^3$ we have the Lagrangian changing to
\begin{eqnarray}
	\mathcal{L}\to \mathcal{L} +6 \alpha \phi^2 \partial_m \phi \partial^m \phi+O(\phi^6).
\end{eqnarray}

\subsection{Correlator under field redefinition}

The new interaction $6 \alpha \phi^2 \partial_m \phi \partial^m \phi$ will change the four-point correlator as:
\begin{eqnarray}
\left\langle\phi\left(k_1\right) \phi\left(k_2\right) \phi\left(k_3\right) \phi\left(k_4\right)\right\rangle \rightarrow\left\langle\phi\left(k_1\right) \phi\left(k_2\right) \phi\left(k_3\right) \phi\left(k_4\right)\right\rangle+\delta\left\langle\phi\left(k_1\right) \phi\left(k_2\right) \phi\left(k_3\right) \phi\left(k_4\right)\right\rangle
\end{eqnarray}
where $\delta\left\langle\phi\left(k_1\right) \phi\left(k_2\right) \phi\left(k_3\right) \phi\left(k_4\right)\right\rangle$ 
should be a local term. Considering Fourier transformation, the field in the bulk is 
$\phi(x,z)=\int d^dk e^{-ik\cdot x}\phi(k,z)$
with 
$\phi(k,z)=(1+kz)e^{-kz}$
and under field redefinition, the new Feynman rule directly gives:
\begin{eqnarray}
&&\delta \left\langle\phi\left(k_1\right) \phi\left(k_2\right) \phi\left(k_3\right) \phi\left(k_4\right)\right\rangle\nonmy \eqnmy 6\alpha	\int_0^\infty \frac{d z}{z^4} \left\{-z^2 \prod_{i=1}^4\phi(k_i,z) \left(\vec{k}_1\cdot (\vec{k}_2+\vec{k}_3+\vec{k}_4)+\vec{k}_2\cdot (\vec{k}_3+\vec{k}_4)+\vec{k}_3\cdot \vec{k}_4 \right)+\right.\nonmy && \left. z^2 \sum_{1\le i < j\le 4}\partial_{z_i} \partial_{z_j}\prod_{i=1}^4 \phi\left(k_i, z\right)\right\}\nonmy \eqnmy -3 \alpha\left(k_1^3+k_2^3+k_3^3+k_4^3\right).
\end{eqnarray}
So we have shown that
\begin{eqnarray}
\left\langle\phi\left(k_1\right) \phi\left(k_2\right) \phi\left(k_3\right) \phi\left(k_4\right)\right\rangle \rightarrow\left\langle\phi\left(k_1\right) \phi\left(k_2\right) \phi\left(k_3\right) \phi\left(k_4\right)\right\rangle-3 \alpha\left(k_1^3+k_2^3+k_3^3+k_4^3\right)
\end{eqnarray}
and $k_i^3$ is analytic in 3 other momenta and so is a local term. The correlator is not strictly invariant under field redefinition or say it is only invariant up to boundary local terms.

\subsection{Amplitude under field redefinition}

Let us suppose additional $6 \alpha \phi^2 \partial_m \phi \partial^m \phi$ produces changes in Mellin-momentum amplitude as 
\begin{eqnarray}
\cA
\rightarrow
\cA
+
\delta \cA.
\end{eqnarray}
From directly reading off the Feynman rule, 
\begin{eqnarray}
	&\delta \mathcal{A} &
 =
 6\alpha	\Bigg\{-z^2\left(\frac{1}{2}\left(k_1^2+k_2^2+k_3^2+k_4^2\right)\right)+  \sum_{1\le i < j\le 4}(-2s_i+d/2)(-2s_j+d/2)\Bigg\}\delta(d-2\left(s_{1234}\right))
 \nonmy
 \eqnmy 
 6\alpha	
 \left\{d^2-3 d\left(s_{1234}\right)+2\left(s_{1234}\right)^2\right\}\delta(d-2\left(s_{1234}\right))
 \nonmy\eqnmy 
 0,
 \label{fieldred}
\end{eqnarray}
where in the second equation we have used the on-shell condition for massless field $k_i^2 z^2=\frac{16 s_i^2-d^2}{4}$. And on the support of the Mellin delta function, we see that the change of the Mellin-momentum amplitude is indeed zero.

\section{Examples on conformal ward identity}
For AdS correlators, the conformal Ward identities provide a natural analogue to momentum conservation in the S-matrix, as mentioned in the main text. In the first part of this section, using the example of a massless $\phi^4$ theory, we will illustrate how these identities do not lead to a vanishing right-hand side. We will then demonstrate how this issue is resolved through the enforcement of on-shell condition, boundary momentum conservation and the use of the Mellin delta function.

The second part of the section is to show how the Mellin-momentum amplitude is exactly invariant under special conformal identity. We use the example of $\mathcal{A}_{\left\langle J_1 \varphi_2 \varphi_3\right\rangle} $to demonstrate and on the same time it is compared with $\left\langle J_1 \varphi_2 \varphi_3\right\rangle$ as computed in \cite{Baumann:2020dch}.
\subsection{Correlator: conformal ward identity on contact diagram}
\label{appendixb1}
We start by defining following \cite{Gomez:2021qfd}
\begin{eqnarray}
    \mathcal{C}_n^{\Delta} \equiv \int \frac{d z}{z^{d+1}}
    \prod_{a=1}^n \phi_\Delta\left(k_a, \eta\right),
\end{eqnarray}
where 
$ \mathcal{C}_n^{\Delta}$
is 
 the scalar contact integral with conformal dimension $\Delta $ and $\phi_\Delta \left(k_a, \eta\right)$ is bulk-to-boundary propagator. The generators in the conformal group are: 
 \begin{eqnarray}
     \begin{aligned}
P^\mu & =k^\mu, \\
D & =k^\mu \partial_\mu+(d-\Delta), \\
K_\mu & =k_\mu \partial^\nu  \partial_\nu-2 k^\nu  \partial_\nu \partial_\mu-2(d-\Delta) \partial_\mu,
\end{aligned}
 \end{eqnarray}
which act on the bulk-to-boundary propagator gives:
\begin{eqnarray}
    \begin{aligned}
P^\mu \phi_\Delta & =k^i \phi_\Delta ,\\
D \phi_\Delta & =z \frac{\partial}{\partial z} \phi_\Delta ,\\
K_\mu \phi_\Delta & =-z^2 k_i \phi_\Delta.
\end{aligned}
\label{scwiDiffonBtb}
\end{eqnarray}
 So defining 
 \begin{eqnarray}
     \mathcal{D}_a \cdot \mathcal{D}_b=\frac{1}{2}\left(P_a^i K_{b i}+K_{a i} P_b^i\right)+D_a D_b,
 \end{eqnarray}
 using 
 \begin{eqnarray}
     \left(\mathcal{D}_a \cdot \mathcal{D}_b\right) \phi_\Delta^a \phi_\Delta^b=z^2\left[\partial_z \phi_\Delta^a \partial_z \phi_\Delta^b-\left(\vec{k}_a \cdot \vec{k}_b\right) \phi_\Delta^a \phi_\Delta^b\right].\label{mandelINVads}
 \end{eqnarray}
One can immediately see that the calculation for massless and $d=3$ is basically the same as our previous field redefinition examples:
\begin{eqnarray}
\left(\mathcal{D}_1 \cdot \mathcal{D}_1+\mathcal{D}_1 \cdot \mathcal{D}_2+\mathcal{D}_1 \cdot \mathcal{D}_3+\mathcal{D}_1 \cdot \mathcal{D}_4\right) \mathcal{C}_1 =-k_1^3,
\label{eq144}
\end{eqnarray}
which is a boundary local term. So one can see that the conformal ward identity is not vanish but up to local terms.
\subsection{Amplitude: conformal ward identity on contact diagram}
\label{appendixb2}
Using Mellin transformation of bulk-to-boundary propagator in \eqref{btBMellin}, we have
\begin{eqnarray}
    \begin{aligned}
P^\mu \phi_\Delta & =k^\mu \phi_\Delta \\
D \phi_\Delta & =\left(\frac{d}{2}-2 s\right)\phi_\Delta \\
K_\mu \phi_\Delta & =-z^2 k_\mu \phi_\Delta.
\end{aligned}
\label{scwiDiffonBtbMellin}
\end{eqnarray}
Then \eqref{mandelINVads} becomes,
 \begin{eqnarray}
     \left(\mathcal{D}_a \cdot \mathcal{D}_b\right) \phi_\Delta^a \phi_\Delta^b=\left[\left(\frac{d}{2}-2 s_a\right)\left(\frac{d}{2}-2 s_b\right)-z^2\left(\vec{k}_a \cdot \vec{k}_b\right) \right]\phi_\Delta^a \phi_\Delta^b.
 \end{eqnarray}
That gives Mellin-momentum amplitude  version of \eqref{eq144}:
\eqs{&(\mathcal{D}_1 \cdot \mathcal{D}_1+\mathcal{D}_1 \cdot \mathcal{D}_2+\mathcal{D}_1 \cdot \mathcal{D}_3+\mathcal{D}_1 \cdot \mathcal{D}_4) \mathcal{C}_4 \\
=&\left[\begin{array}{l}
    \Delta_1\left(\Delta_1-d\right) \left(\frac{d}{2} - 2s_1\right)+\left(\frac{d}{2} - 2s_2\right) - z^2 \left( \boldsymbol{k}_1 \cdot \boldsymbol{k}_2 \right)\\
+ \left(\frac{d}{2} - 2s_1\right)\left(\frac{d}{2} - 2s_3\right) - z^2 \left( \boldsymbol{k}_1 \cdot \boldsymbol{k}_3 \right) 
+ \left(\frac{d}{2} - 2s_1\right)\left(\frac{d}{2} - 2s_4\right) - z^2 \left( \boldsymbol{k}_1 \cdot \boldsymbol{k}_4 \right)
\end{array}
\right] \mathcal{A}_{4} .
}
For direct comparison, we can restrict ourselves to massless scalar field $\Delta=d$, and get,
\eqs{&(\mathcal{D}_1 \cdot \mathcal{D}_1+\mathcal{D}_1 \cdot \mathcal{D}_2+\mathcal{D}_1 \cdot \mathcal{D}_3+\mathcal{D}_1 \cdot \mathcal{D}_4) \mathcal{A}_{4} \\
=&[\frac{3 d^2}{4}-d \left(3 s_1+s_2+s_3+s_4\right)+ z^2k_1^2+4 s_1
   \left(s_2+s_3+s_4\right)] \mathcal{A}_{4} \\
   =&
   \frac{1}{2} \left(d-4 s_1\right) \left(d-2
   \left(s_1+s_2+s_3+s_4\right)\right)\mathcal{A}_{4}
   \\
=&0.
}
In the second line, we applied momentum conservation, while in the third line, we utilized the massless on-shell condition \eqref{eq:mellineom} for $z^2k_1^2$. The final line follows from the fact that this contact Mellin-momentum amplitude is evaluated on the support of the Mellin delta function \eqref{eq:mmdelta}, given by $\delta \left(d - 2(s_1 + s_2 + s_3 + s_4)\right)$. Thus, the on-shell condition, boundary momentum conservation, and the Mellin delta function imply that the AdS analogues of the Mandelstam invariants are exactly zero.
\subsection{Correlator: Review on special conformal ward identity with spin}
 The special conformal generator $\tilde{K}$ on the boundary spacetime  acts on conformal scalar $\varphi$, and spin-1 field $J$ as \cite{Baumann:2020dch} 
\begin{eqnarray}
\begin{aligned}
	\tilde{K}_{\varphi}^\mu \varphi & =-K^\mu \varphi \\
	\widetilde{K}_J^\mu J_\nu \varepsilon^\nu & =\left(-\varepsilon^\nu K^\mu+2 \varepsilon^\mu \frac{k^\nu}{k^2}\right) J_\nu, 
\end{aligned}\label{scwiONspinJJJTTT}
\end{eqnarray}
where $K_\mu$ acting on explicit tensor indices is given as,
\begin{eqnarray}
  K^\mu O_{\nu_1 \cdots \nu_{\ell}}=\left[2(\Delta-3) \partial_{k_\mu}+k^\mu \partial_{k^\rho} \partial_{k_\rho}-2 k^\rho \partial_{k^\rho} \partial_{k^\mu}-2 \partial_{k_\rho} \Sigma^\mu{}_{ \rho}\right] O_{\nu_1 \cdots \nu_{\ell}},
\end{eqnarray}
with 
\begin{eqnarray}
    \Sigma_{\mu \rho} O_{\nu_1 \cdots \nu_{\ell}}=\ell\left(O_{\mu\left(\nu_1 \ldots\right.} \eta_{\left.\nu_{\ell}\right) \rho}-O_{\rho\left(\nu_1 \ldots\right.} \eta_{\left.\nu_{\ell}\right) \mu}\right).
    \label{big sigma}
\end{eqnarray}
Now we consider boundary correlator
$\langle J_1\varphi _2\varphi_3\rangle $ which is 
\begin{eqnarray}
    \langle J_1\varphi_2 \varphi_3\rangle
    =
    \frac
    {\varepsilon_1 \cdot k_2}
    {k_1+k_2+k_3}.
\end{eqnarray}
After working out the algebra, the change $\delta  \langle J_1\varphi_2 \varphi_3\rangle$ under special conformal transformation is
\begin{eqnarray}
\left(\tilde{K}_{\varphi_3}^\mu+\tilde{K}_{\varphi_2}^\mu+\tilde{K}_{J_1}^\mu\right)   \langle J_1\varphi_2 \varphi_3\rangle
    =
    - 
    \frac{(k_2-k_3)\varepsilon_1 ^\mu}{k_1^2}
\end{eqnarray}
which is a boundary local term.
\subsection{Amplitude: Special conformal ward identity with spin}
Starting from the definition of amplitude \eqref{eq:MMamplitude},
we first record how the generators cross the bulk-to-boundary propagator, and then the generators $\hat{K}$ directly acting on the amplitudes have the following form
\begin{eqnarray}
\begin{aligned}
	\hat{K}_{\varphi}^\mu \mathcal{A} &
 =
 -\left\lbrace{\color{black} -z^2k^\mu +2\left(2s-d /2\right)\partial_{k_\mu} +k^\mu\partial_{k^\rho} \partial_{k_\rho}-2 k^\rho\partial_{k^\rho} \partial_{k_\mu}}\right\rbrace \mathcal{A}
 \\
	\hat{K}_J^\mu \mathcal{A}_\nu \varepsilon^\nu
 &
 =
 \left[-\varepsilon^\nu \left\lbrace{\color{black} -z^2k^\mu +2\left(2s-d /2\right)\partial_{k_\mu} +k^\mu\partial_{k^\rho} \partial_{k_\rho}-2 k^\rho\partial_{k^\rho} \partial_{k_\mu}-\color{black}2({v}^\rho +\partial_{k_\rho})\Sigma^\mu{} _{\rho }}\right\rbrace +\right.
 \\
 &
 \qquad\qquad\left.+{\color{black}2 \varepsilon^\mu \frac{k^\nu}{k^2}}\right] \mathcal{A}_\nu ,
\end{aligned}\non 
\end{eqnarray}
where $s$ is the Mellin variable associated to the external leg and 
${v}^\mu=k^\mu \frac{\left(\Delta-d / 2-2 s\right)}{k^2}$ from $\partial_{k^\mu} \phi_{\Delta}\left(s, z k\right)$. \\
Starting from the Mellin-momentum amplitude corresponds to $\langle J_1 \varphi_2 \varphi_3\rangle$:
\begin{eqnarray}
    \mathcal{A}_{\langle J_1\varphi_2 \varphi_3\rangle}=z \varepsilon_1 \cdot k_2,
\end{eqnarray}
 each generator acting on the amplitude gives
%
\begin{eqnarray}
z\hat{K}_{\varphi_3}^\mu \varepsilon_1 \cdot k_2\eqnmy z^3 k_3^\mu\varepsilon_1 \cdot k_2 	\nonmy z\hat{K}_{\varphi_2}^\mu \varepsilon_1 \cdot k_2\eqnmy z^3 k_2^\mu\varepsilon_1 \cdot k_2 -2z\left(2 s_2-d / 2\right)\varepsilon_1^\mu\nonmy
 z\hat{K}_{J_1}^\mu \varepsilon_1 \cdot k_2\eqnmy z^3 k_1^\mu\varepsilon_1 \cdot k_2+{\color{black}2z\varepsilon_1^\nu v_1^\rho\Sigma^\mu{}_{  \rho}k_2{}_\nu}+2z\varepsilon_1^\mu\frac{k_1\cdot k_2}{k_1^2}.\label{KJ1}
\end{eqnarray}
Here we know from \eqref{big sigma} that $\Sigma_{ \mu \rho}k^\nu= k_\mu\delta_{\rho}^\nu- k_\rho\delta_{\mu}^\nu.$
Plugging this in third line of \eqref{KJ1} we find:
\begin{eqnarray}
	z\hat{K}_{J_1}^\mu \varepsilon_1 \cdot k_2
 \eqnmy   z^3 k_1^\mu\varepsilon_1 \cdot k_2 -{\color{black}z\varepsilon_1^\nu\left(-k_1^2-k_2^2+k_3^2\right) \frac{\left(d/2-2 s_1-2\right)}{k_1^2}}.
 \label{mid}
\end{eqnarray}
Adding \eqref{mid} and the first two line of \eqref{KJ1} together
\begin{eqnarray}
(\hat{K}_{\varphi_3}^\mu+\hat{K}_{\varphi_2}^\mu+\hat{K}_{J_1}^\mu)	\mathcal{A}_{\left\langle J_1 \varphi_2 \varphi_3\right\rangle}\eqnmy-2 z\left(2 s_2-d / 2\right) \varepsilon_1^\mu+\nonmy&&-{\color{black}z\varepsilon_1^\mu\left(d/2-2 s_1-2\right)+2z\varepsilon_1^\mu\left(k_2^2-k_3^2\right) \frac{\left(d/2-2 s_1-2\right)}{k_1^2}}.
\nonmy
\label{OnshellSCWI}
\end{eqnarray}
Using on-shell condition on numerator (see \eqref{Onshell}) and on denominator i.e.
\begin{eqnarray}
    \frac{X}{k^2}= \frac{z^2 X|_{s\to s-1}}{4(s-1)^2-(\Delta-d/2)^2}\label{InverseOnshell}
\end{eqnarray}
and in
${d=3}$,
we have 
\begin{eqnarray*}
\left(\tilde{K}_{\varphi_3}^\mu+\tilde{K}_{\varphi_2}^\mu+\tilde{K}_{J_1}^\mu\right) \mathcal{A}_{\left\langle J_1 \varphi_2 \varphi_3\right\rangle}
\eqnmy 
z\varepsilon_1^\mu\left({\color{black} -2s_1-1/2}+2(3/2-2s_2)\right) -{\color{black}8z\varepsilon_1^\mu \frac{\left(s_2^2-s_3^2\right)}{4s_1-5}}
\nonmy \eqnmy
z \varepsilon_1^\mu\left(\frac{\left(4 s_1+4 s_2-4 s_3-5\right)\left(5-(4 s_1+4 s_2+4 s_3)\right)}{2\left(4 s_1-5\right)}\right).
\end{eqnarray*}
Noticing there is an additional power of $z$ on the support of Mellin delta \eqref{eq:mmdelta}, and thus we have: 
$$5-4(s_1+s_2+s_3)=0.$$
Finally we find
\begin{eqnarray}
(\tilde{K}_{\varphi_3}^\mu+\tilde{K}_{\varphi_2}^\mu+\tilde{K}_{J_1}^\mu)	\mathcal{A}_{\left\langle J_1 \varphi_2 \varphi_3\right\rangle}=0.
\end{eqnarray}
So using on-shell condition, three-momentum conservation and Mellin delta function, we show that three-point Mellin-momentum amplitude $\mathcal{A}_{\left\langle J_1 \varphi_2 \varphi_3\right\rangle}$ is exactly invariant under special conformal transformation.

\section{Derivation of Propagators and Feynman rules}
\label{appendixA}

\subsection{Yang-Mills}
We follow the derivation and notation in \cite{Armstrong:2022mfr}, for Yang-Mills theory, with the usual field strength,
\eqs{
\mathbf{F}_{m n}=\partial_m \mathbf{A}_{n}-\partial_n \mathbf{A}_{m}-i[\mathbf{A}_{m},\mathbf{A}_{n}].
}
We then rescale the field as $\mathbf{A}_{m}=(\mathcal{R}/z) A_{m}$ and also work in the transverse boundary gauge $k_{\mu}  A^{\mu}=0$. Now both the field $A_{\mu}$ and $J_{\mu}$ are dimensionless, and the full equation of motion reads:
\eqs{ 
\mathcal{D}_{k}^{d-1} A_{\mu}&=i zk_{\mu}[(d-2)-z\partial_z] A_z+J_{\mu}, \\
z^2k^2 A_z&=J_z.
}
Note that by contracting $k_{\mu}$ with the EoM and then applying the gauge condition $k_{\mu}  A^{\mu}=0$, it implies the usual current conservation
\eqs{
zk^{\mu}  J_{\mu}&=i (z\partial_z-d) J_z.\label{current_conservation}
}
Now we can use the current conservation to reorganize the equations into the following:
\eqs{
\mathcal{D}_{k}^{d-1} A_{\mu}&=\Pi_{\mu \nu} J^{\nu}, \\
z^2k^2 A_z&=J_z,
\label{emo2}
}
where the spin 1 projection tensor: $\Pi_{\mu \nu}=(\eta_{\mu \nu}-\frac{k_{\mu} k_{\nu}}{k^2})$. So now we are looking for the solution of bulk-to-bulk propagator
\eqs{
A_m(z)=\int\frac{dy}{y^{d+1}} G_{mn}(z,y) J^n(y),
}
and then comparing with \eqref{emo2}, the bulk-to-bulk propagator is given by
\eqs{
G_{\mu \nu}&=\frac{\Pi_{\mu \nu}}{\mathcal{D}_k^{d-1}},  \\
G_{zz} &=\frac{1}{z^2k^2}.
}
Equivalently, the action that determines the propagator is given by,
\eqs{
S^{YM}&=\int \frac{dz}{z^{d+1}} \left( A_{\mu}J^{\mu}+A_zJ^z \right) \\
&=\int \frac{dz}{z^{d+1}} \left( J^{\mu}\frac{\Pi_{\mu \nu}}{\mathcal{D}_k^{d-1}}J^{\nu}+J^z \frac{1}{z^2k^2} J^z \right).
}

\subsection{Gravity Propagator}\label{grpropagatorF}
The graviton will be parametrized as $g_{mn}=\tilde{g}_{mn}+\frac{\mathcal{R}^2}{z^2}h_{mn}$, and we will use the following gauge condition,
\eqs{i \eta ^{\mu \nu} k_{\mu} h_{z\nu}&=\frac{1}{2} \eta^{\mu \nu} \partial_z h_{\mu \nu}+\frac{d}{2z}h_{zz} \\
\eta^{\nu \rho} k_{\rho} h_{\mu \nu}&=\frac{1}{2} k_{\mu} (\eta ^{\nu \rho} h_{\nu \rho}+\beta h_{zz}).
}
\footnote{Here $\beta$ is just a parameter that will not appear in the final answer but serve as a consistency check.}
We can then expand it in the Einstein field equation,
\eqs{
\begin{aligned}
k^2 h_{z z} &=\frac{2\kappa}{d-1}[(d-2)T_{zz}-\eta^{\mu \nu}T_{\mu \nu}] ,\\
k^2 h_{z \mu} &=2\kappa T_{z \mu}+\frac{i}{2 z}\left(d-2-\beta z \partial_z\right) k_\mu h_{z z} ,\\
\mathcal{D}_k^{d} h_{\mu \nu} &=2\kappa z^2 T_{\mu \nu}-\frac{2\kappa z^2}{(d-1)} \eta_{\mu \nu}(T_{zz}+\eta ^{\rho \sigma}T_{\rho \sigma})\\
&+\left[(\beta-1) z^2 k_\mu k_\nu+\eta_{\mu \nu}\left(z \partial_z-d\right)\right] h_{z z} \\
&+i z \left[(d-1)-z \partial_z \right]\left(k_\mu h_{\nu z}+k_\nu h_{\mu z}\right), \\
k^2 h_{\mu} ^{\mu}&= \frac{2 \kappa}{d-1}((-\beta)T_{\mu}^{\mu}+(2-2d-2\beta +d \beta) T_{zz}).
\end{aligned}
}
As before, applying the gauge condition implies the energy-momentum conservation:
\eqs{
g^{mn}\nabla_m T_{nz}&=0 \Rightarrow (z\partial_z -d+2) T_{zz} +i z k^{\mu} T_{\mu z} +T_{\mu \nu} \eta^{\mu \nu}=0,\\
g^{mn}\nabla_m T_{n\mu}&=0 \Rightarrow (z\partial_z -d+1)T_{z\mu} +i z k^{\nu} T_{\nu \mu} =0.
}
We will do the tensor decomposition on the spin 1 and spin 2 fields to obtain the transverse and traceless part\cite{Liu:1998ty},
\eqs{h_{z \mu}&= h_{z \mu}^{\bot} -\frac{i k_{\mu}}{2k^2}(\partial_z h+\frac{d}{z}h_{zz}),\\
h_{\mu \nu}&=h_{\mu \nu}^{\bot}+\eta_{\mu \nu}\frac{h-\beta h_{zz}}{2(d-1)}+\frac{k_{\mu} k_{\nu}}{k^2} \frac{(d-2)h+d \beta h_{zz}}{2(d-1)},
}
where $k^{\mu} h_{z\mu}^{\bot}=0$ and $k^{\mu} h_{\mu \nu}^{\bot}=0=\eta^{\mu \nu} h_{\mu \nu}^{\bot}$. So this means only the following equations are kinematic, everything else is just constraint equations,
\eqs{k^2 h^{\bot}_{z\mu}&=2\kappa \Pi_{\mu \nu} T_z^{\nu}, \\
\mathcal{D}_k^d h^{\bot}_{\mu \nu} &=2\kappa z^2\Pi_{2,2}^{\mu \nu, \rho \sigma} T_{\rho \sigma}, \\
k^2 h_{z z} &=\frac{2\kappa}{d-1}[(d-2)T_{zz}-\eta^{\mu \nu}T_{\mu \nu}], \\
k^2 h_{\mu} ^{\mu}&= \frac{2 \kappa}{d-1}((-\beta)T_{\mu}^{\mu}+(2-2d-2\beta +d \beta) T_{zz}).
}
So action that determines the bulk-to-bulk propagator is now given by
\eqs{
S^{EG}&=\int \frac{dz}{z^{d+1}} (h_{\mu \nu} T^{\mu \nu} -2 h_{z \mu} T^{z \mu} +h_{zz}T^{ zz}) \\
&=\int \frac{dz}{z^{d+1}}  (T_{\mu \nu}\frac{ \Pi_{2,2} ^{\mu \nu, \rho \sigma} }{\mathcal{D}_k^{d}}T_{\rho \sigma} -2\frac{T_{z\mu}\Pi_{\mu \nu}T_{z\nu}}{z^2k^2} \\
&+ \frac{2((d-2)T_{zz}T_{zz}-T_{zz}T-TT_{zz})}{(d-1)z^2k^2} + \frac{2d(T+(2-d)T_{zz}+z T'_{zz})(T+(2-d)T_{zz}+z T'_{zz})}{(d-1)z^4k^4} ).
\label{qwqwert}
}
When graviton is coupled to the conformally coupled scalar, the stress tensor obey the traceless condition $T_{zz}+T=0$, which simplifies the expression and in $d=3$
\eqs{S^{\mathrm{conformal}}&=\int\frac{dz}{z^4} \left (\frac{T_{\mu \nu} \Pi_{2,2} ^{\mu \nu, \rho \sigma} T_{\rho \sigma}}{\mathcal{D}_k^{\Delta=3}} -2\frac{T_{z\mu}\Pi_{\mu \nu}T_{z\nu}}{z^2k^2}+ 3\frac{TT}{k^2} + \frac{3}{z^2}\frac{(2T-z T')(2T-z T')}{z^4k^4} \right) \\
&=\int\frac{dz}{z^4} \left (\frac{T_{\mu \nu} \Pi_{2,2} ^{\mu \nu, \rho \sigma} T_{\rho \sigma}}{\mathcal{D}_k^{\Delta=3}} -2\frac{T_{z\mu}\Pi_{\mu \nu}T_{z\nu}}{z^2k^2}-3\frac{T\mathcal{D}^{\Delta=2}_k T}{z^4k^4} \right),
}
where $\mathcal{D}_{k}^{\Delta=2}$ is the EoM operator for $\Delta=2$ scalar acting on $T$.\footnote{Note for conformally coupled scalar, the $\frac{TT}{k^4}=\Pi_{2,0}$ in \cite{Baumann:2020dch}}

\section{Polarization sums } \label{polarization sums}
In this appendix, we provide the details of the polarization sums employed in this paper. Following the boundary transverse gauge \cite{Armstrong:2022mfr}: (This is the same as in QFT textbook \cite{Weinberg:1995mt} with Coulomb gauge.)
\begin{align}
    \sum_{h=\pm} \varepsilon_{\mu}(k,h) \varepsilon_{\nu}(k,h)^* &=\eta_{\mu \nu}-\frac{k_{\mu}k_{\nu}}{k^2} \equiv \Pi_{\mu \nu}, \\
    \sum_{h=\pm} \varepsilon_{\mu \nu}(k,h) \varepsilon_{\rho \sigma}(k,h)^* &=\frac{1}{2} \Pi_{\mu \rho}\Pi_{\nu \sigma}+ \frac{1}{2}\Pi_{\mu \sigma} \Pi_{\rho \nu}-\frac{1}{d-1}  \Pi_{\mu \nu}\Pi_{\rho \sigma},
\end{align}
which are transverse and traceless projection tensor. Let's return to QED in Coulomb gauge for a moment. The polarization tensor above which appears in the photon propagator is not Lorentz invariant on its own, but we can restore Lorentz invariance to obtain the covariant photon propagator. This is the same logic that we use to derive all of the polarization sums below by demanding conformal invariance.\\
Let us explicitly write out the polarization sums at 4-point, see also \cite{Arkani-Hamed:2018kmz,Baumann:2020dch,Baumann:2021fxj} for the case of conformally
coupled scalar.
\begin{align}
\Pi_{1,1}\equiv &\frac{1}{4} (k_{1}^{\mu}-k_2^{\mu})\Pi_{\mu \nu}(k_{3}^{\nu}-k_4^{\nu})=\frac{1}{4}(k_1-k_2)\cdot (k_3-k_4)+\frac{(k_1^2-k_2^2)(k_3^2-k_4^2)}{4k_s^2}, \\
\Pi_{1,0}\equiv & -\frac{(s_1-s_2)(s_3-s_4)}{z^2 k_s^2}.
\end{align}
Next, we write the spin-2 polarization sums in a way that makes its double copy structure clear.
\begin{align}
\Pi_{2,2}\equiv & \frac{1}{16}(k_{1}^{\mu}-k_2^{\mu})(k_{1}^{\nu}-k_2^{\nu})(\frac{1}{2} \Pi_{\mu \rho}\Pi_{\nu \sigma}+ \frac{1}{2}\Pi_{\mu \sigma} \Pi_{\rho \nu}-\frac{1}{d-1}  \Pi_{\mu \nu}\Pi_{\rho \sigma})(k_{3}^{\rho}-k_4^{\rho})(k_{3}^{\sigma}-k_4^{\sigma})\\
=&\Pi_{1,1}^2-\Pi_{2,2}^{\mathrm{Tr}}, \\
\Pi_{2,1}\equiv & 2\Pi_{1,1} \Pi_{1,0},
\end{align}
\begin{align}
\Pi_{2,2}^{\mathrm{Tr}}\equiv & \frac{(k_{1}^{\mu}-k_2^{\mu})\Pi_{\mu \nu}(k_{1}^{\nu}-k_2^{\nu})(k_{3}^{\rho}-k_4^{\rho})\Pi_{\rho \sigma}(k_{3}^{\sigma}-k_4^{\sigma})}{16(d-1)}, \\ 
\Pi_{2,0}\equiv & -\frac{d(k_1^2-k_2^2)(k_3^2-k_4^2)(s_1-s_2)(s_3-s_4)}{4(d-1)k_s^4} +\frac{(d-2)z^2(k_1^2-k_2^2)(k_3^2-k_4^2)(s_1-s_2)(s_3-s_4)}{4(d-1)k_s^2(d-2s_{12})(d-2s_{34})} \nonumber \\
&+\frac{(k_1^2-k_2^2)(s_1-s_2)(d^2-8(s_3^2+s_4^2))}{8(d-1)k_s^2(d-2s_{12})}+\frac{(k_3^2-k_4^2)(s_3-s_4)(d^2-8(s_1^2+s_2^2))}{8(d-1)k_s^2(d-2s_{34})} \nonumber \\
&+\frac{(z^2k_s^2+4s_{12}s_{34})}{(d-1)}+\frac{4(s_1-s_2)^2+4(s_3-s_4)^2-d^2}{16(d-1)},
\end{align}
where $s_{ij}=s_i+s_j$.

 Note that there are still terms with Mellin variables in the denominator that naively violate locality. However, they will all cancel after using the Mellin delta function.\\

\section{Example with internal Mellin variable for YM amplitude} \label{ymFeynmanrule}

When computing YM amplitude at higher point, one will encounter an internal Mellin variable which corresponds to the $z$ derivative of bulk-to-bulk propagator. We will now show how to deal with this.\\
In YM five-point Mellin-momentum amplitudes, using the Feynman rule we encounter the following contribution in the main text \eqref{3chen3chen3}
$$\frac{V^{12 \mu} \Pi_{\mu \nu} V^{3 \nu z}}{\mathcal{D}_{k_{12}}^{d-1}} \frac{1}{z^2 k_{45}^2} V^{45 z}.$$ 
The non-trivial part is
\begin{eqnarray}
	z V^{3 \nu z} \frac{1}{z^2 k_{45}^2\mathcal{D}_{k_{12}}^{d-1}} V^{45 z}
 =
 \frac{(s_4-s_5)(u-s_3)}{zk_{45}^2\mathcal{D}_{k_{12}}^{d-1}},
\end{eqnarray}
where the additional $z$ factor comes from the scaling in vertex $V^{12 \mu}$.
Recalling that each Mellin variable is $z \partial_z$\footnote{The $u$ notation is from the $z \partial_z$ acting on the Mellin representation of bulk-to-bulk propagator \cite{Sleight:2020obc}}. Then this represents scalar integral 
\begin{eqnarray}
\frac{1}{4}\int_0^{\infty} d z_2	\int_{0}^{\infty} dz_1 z_1 \phi_1(z_1) \phi_2(z_1) (\partial_4^{z_2}-\partial_5^{z_2})(\partial_3^{z_2}-\partial_G^{z_2})\frac{G_{k_{12}}(z_1,z_2)}{k_{45}^2}  \phi_3 (z_2)\phi_4(z_2)\phi_5(z_2),
\end{eqnarray}
where the remaining $z$ factor will contribute to the 3-point vertex for particles 1 and 2. Using exchange to contact representation of bulk-to-bulk propagator in section \ref{backtomomentum}, we have
\begin{eqnarray}
&&\frac{1}{4k_{45}^2}\int_0^{\infty} d z_2\int_{0}^{\infty} dz_1 z_1 \phi_1(z_1) \phi_2(z_1)\nonmy && (\partial_4^{z_2}-\partial_5^{z_2})(\partial_3^{z_2}-\partial_G^{z_2})\nonmy&&\int_0^{\infty} \frac{d p}{2 \pi i} \frac{-p^{d+1-2 \Delta}}{k^2+p^2}\left(\phi_{\Delta}\left(z_1, i p\right)-\phi_{\Delta}\left(z_1,-i p\right)\right)\left(\phi_{\Delta}\left(z_2, i p\right)-\phi_{\Delta}\left(z_2,-i p\right)\right)  \phi_3 (z_2)\phi_4(z_2)\phi_5(z_2).\nonmy
\end{eqnarray}
Using the map
\begin{eqnarray}
	\partial_z \phi_{\Delta=2}(z,k)=\left( \frac{1}{z}-k \right) \phi_{\Delta=2}(z,k).
\end{eqnarray}
After some algebra, we have

\eqs{
&\frac{1}{4k_{45}^2} \int_{0}^{\infty} \frac{dp}{2\pi i} \frac{-1}{k_{12}^2+p^2} \left( \mathcal{C}_{3}(ip,k_1,k_2)- \mathcal{C}_{3}(-ip,k_1,k_2)\right)\\
& \times \left ( (k_4-k_5)(k_3-ip)\mathcal{C}_4(ip,k_3,k_4,k_5)-(k_4-k_5)(k_3+ip)\mathcal{C}_4(-ip,k_3,k_4,k_5)   \right)\\
=&-\frac{\left(k_4-k_5\right) \left(2 k_3+k_4+k_5\right)}
{4k_{45}^2\left(k_1+k_2+k_3+k_4+k_5\right) \left(k_1+k_2+k_{12}\right) \left(k_3+k_4+k_5+k_{12}\right)}.
}
This exactly matches with explicit computation when using the Mellin sub-delta function to convert $u$ to $s_3$ $s_4$ $s_5$. In particular, noticing sub Mellin delta function \\  $2\left(s_1+s_2+s_3+\Bar{u}+1\right)=d$ on sub-diagram $V^{12 z} \frac{1}{z^2k_{12}^2} {V^{3 z \rho}} $, we have
\begin{eqnarray}
&&\frac{\left(s_4-s_5\right)\left(u-s_3\right)}{z k_{45}^2 \mathcal{D}_{k_{12}}^{d-1}}=	-\frac{\left(s_4-s_5\right)\left(-d+2(1+2s_3+s_4+s_5)\right)}{2z k_{45}^2\mathcal{D}_{k_{12}}^{d-1}}
\nonmy
&\underset{\eqref{ymbacktomomentum}}{\begin{array}{c}
     \to\\
     \text{to momentum space}
\end{array}} &
-\frac{\left(k_4-k_5\right)\left(2 k_3+k_4+k_5\right)}{4k_{45}^2\left(k_1+k_2+k_3+k_4+k_5\right)\left(k_1+k_2+k_{12}\right)\left(k_3+k_4+k_5+k_{12}\right)}.
\non
\end{eqnarray}
One final comment would be that when we go to six points, it's not always possible to turn all the internal Mellin variables into external ones and hence the method here of dealing with internal Mellin variables is more general.

\section{Details on bootstrapping Five point Yang-Mills }
\label{Derivation of double OPE pole}
In this section, we will give detailed expressions of $a_1$ and $a_2$ after working out the polarization sums, and then we will show how we arrive at double OPE pole contributions based on this.
\subsection{Actual expression of $a_1$ and $a_2$} 
We have for $a_1$ and $a_2$ in equation \eqref{eq5ym}
\begin{eqnarray}
    \frac{a_1(1,2,3,4,5)}{{\mathcal{D}_{k_{12}}^{d-1} \mathcal{D}_{k_{45}}^{d-1}}}\eqn
    \frac{
    z^3\left(k_1^2-k_2^2\right) k_3^2
    \left(k_4^2-k_5^2\right)
    \varepsilon_1 \cdot \varepsilon_2
    \left(k_1 \cdot \varepsilon_3+k_2 \cdot \varepsilon_3\right)
    \varepsilon_4 \cdot \varepsilon_5
    }
    {8 k_{12}^2 k_{45}^2 \cD_{k_{12}}^{d-1} \cD_{k_{45}}^{d-1}}+ 
    \non &&
    \frac{
    z^3\left(k_1^2-k_2^2\right)\varepsilon_1 \cdot \varepsilon_2X_1^{(54321)}
    }
    {
    8 k_{12}^2 \cD_{k_{12}}^{d-1} \cD_{k_{45}}^{d-1}
    }
    +
    \frac{
    z^3\left(k_4^2-k_5^2\right)\varepsilon_4 \cdot \varepsilon_5X_1^{(12345)}
    }
    {
    8 k_{45}^2 \cD_{k_{12}}^{d-1} \cD_{k_{45}}^{d-1}
    },
    \label{a1}
\end{eqnarray}
and
\begin{eqnarray}
    \frac{a_2(1,2,3,4,5)}{{ \cD_{k_{12}}^{d-1}}}
    \eqn 
    -\frac{{z}\left({k}_1^2-{k}_2^2\right) \varepsilon_1 \cdot \varepsilon_2\left(\varepsilon_3 \cdot \varepsilon_5\varepsilon_4 \cdot {k}_{12}-\varepsilon_3 \cdot \varepsilon_4\varepsilon_5 \cdot {k}_{12}\right)}{8 {k}_{12}^2 {\cD}^{d-1}_{{k}_{12}}}-
    \non &&
   \frac{\left({s}_4-{s}_5\right)\left(-{d}+2\left(1+2 {~s}_3+{s}_4+{s}_5\right)\right)\left({k}_1^2-{k}_2^2\right)\varepsilon_1 \cdot \varepsilon_2\varepsilon_3 \cdot {k}_{12} \varepsilon_4 \cdot \varepsilon_5}{4 {z} {k}_{12}^2 {k}_{45}^2 {\cD}_{{k}_{12}}^{d-1}}+
    \non && 
    \begin{aligned}
& 
\frac{
\left(s_4-s_5\right)\left(-d+2\left(1+2 s_3+s_4+s_5\right)\right)\varepsilon_4 \cdot \varepsilon_5
}
{
4 z{k}_{45}^2 {\cD}_{{k}_{12}}^{{d}-1}
}
W_{s_1}^{(12345)} 
+ 
\frac{z W_{s_2}^{(12345)}}{8 \cD_{k_{12}}^{d-1}} 
+ \left(3 \leftrightarrow 5 \right)
\end{aligned}\non 
 \label{a2}
\end{eqnarray}
where $\left(3 \leftrightarrow 5 \right)$ is from the t-channel contribution in four-point kinematic and $X_1$ $W_{s_1}$ and $W_{s_2}$ are given as
\begin{eqnarray}
\begin{aligned}
X_1^{(12345)}=&
2 \varepsilon_1 \cdot \varepsilon_3 \varepsilon_2 \cdot k_1 k_3^2-2 \varepsilon_1 \cdot k_2 \varepsilon_2 \cdot \varepsilon_3 k_3^2
+\\&
\varepsilon_1 \cdot \varepsilon_2\left(2 \varepsilon_3 \cdot k_2 k_3^2+\left(\varepsilon_3 \cdot k_4+\varepsilon_3 \cdot k_5\right)\left(k_1^2-k_2^2+k_3^2\right)\right)
\end{aligned}
\end{eqnarray}

\begin{eqnarray}
    \begin{aligned}
    W_{s_1}^{(12345)}=&-2 \varepsilon_1 \cdot \varepsilon_3 \varepsilon_2 \cdot k_1+2 \varepsilon_1 \cdot k_2 \varepsilon_2 \cdot \varepsilon_3+\varepsilon_1 \cdot \varepsilon_2\left(\varepsilon_3 \cdot k_1-\varepsilon_3 \cdot k_2\right)\\
W_{s_2}^{(12345)}=& 2 \varepsilon_1 \cdot \varepsilon_5 \varepsilon_2 \cdot k_1 \varepsilon_3 \cdot \varepsilon_4-2 \varepsilon_1 \cdot \varepsilon_4 \varepsilon_2 \cdot k_1 \varepsilon_3 \cdot \varepsilon_5+ \\
& \varepsilon_1 \cdot k_2\left(-2 \varepsilon_2 \cdot \varepsilon_5 \varepsilon_3 \cdot \varepsilon_4+2 \varepsilon_2 \cdot \varepsilon_4 \varepsilon_3 \cdot \varepsilon_5\right)+\varepsilon_1 \cdot \varepsilon_2 \varepsilon_3 \cdot \varepsilon_5 \varepsilon_4 \cdot k_1 +\\
&- \varepsilon_1 \cdot \varepsilon_2 \varepsilon_3 \cdot \varepsilon_5 \varepsilon_4 \cdot k_2-\varepsilon_1 \cdot \varepsilon_2 \varepsilon_3 \cdot \varepsilon_4 \varepsilon_5 \cdot k_1+\varepsilon_1 \cdot \varepsilon_2 \varepsilon_3 \cdot \varepsilon_4 \varepsilon_5 \cdot k_2.
\end{aligned}
\end{eqnarray}

Next, we will use these results to derive double OPE pole contributions.
\subsection{Derivation of double OPE pole contribution for YM five-point amplitude}
\label{YMdoubleOPE}

From the last section preparation, we can determine double OPE pole from \eqref{a1} and \eqref{a2} by taking the residue as follows:
\begin{eqnarray}
&&-b(1,2,3,4,5)\non
&=& \underset{
    \begin{array}{l}
    \substack{\scriptscriptstyle 
 k^2_{12} \to 0}
 \\
      \substack{
      \scriptscriptstyle  k^2_{45} \to 0
      }
    \end{array} }
    {\mathrm{Res}} 
    \left(
    \frac{
    a_1(1,2,3,4,5)
    }
    {
    \mathcal{D}_{k_{12}}^{d-1} \mathcal{D}_{k_{45}}^{d-1}
    }
    +
    \frac{a_2(1,2,3,4,5)}{\mathcal{D}_{k_{12}}^{d-1}}
    +
    \frac{a_2(4,5,1,2,3)}{\mathcal{D}_{k_{45}}^{d-1}}\right)
    \non \eqn
    -\frac{
    z^3 \varepsilon_1 \cdot \varepsilon_2\left(k_4 \cdot \varepsilon_3+k_5 \cdot \varepsilon_3\right) \varepsilon_4 \cdot \varepsilon_5\left(k_1^2-k_2^2\right) k_3^2\left(k_4^2-k_5^2\right)
    }
    {
    8 \cD_{k_{12}}^{d-1} \cD_{k_{45}}^{d-1}
    }
    \non &&
    -\frac{
    \varepsilon_1 \cdot \varepsilon_2\left(k_1 \cdot \varepsilon_3+k_2 \cdot \varepsilon_3\right) \varepsilon_4 \cdot \varepsilon_5\left(k_1^2-k_2^2\right)\left(s_4-s_5\right)\left(-d+2\left(1+2 s_3+s_4+s_5\right)\right)
    }{
    4 z \cD_{k_{12}}^{d-1}
    }+ 
    \non &&
    (1 \leftrightarrow 5 ,4\leftrightarrow 2).
\end{eqnarray}
To proceed, we first use as in \eqref{dtos} to convert $\cD$ to $s$ for $\mathcal{D}_{k_{12}}^{d-1}$  and $\mathcal{D}_{k_{45}}^{d-1}$
\begin{eqnarray}
   \frac{1}{\mathcal{D}_{k_{ij}}^{d-1}}\to  \frac{1}{2\left(d-2 s_i-2 s_j\right)\left(s_i+s_j-1\right)}.
\end{eqnarray}
After that, we use on-shell condition to eliminate all $k^2$ dependence with $\Delta=d-1$,
\begin{equation*}
    f(s) k^2 \rightarrow f(s+1)\left(\frac{(d / 2-\Delta)^2-4 s^2}{z^2}\right).
\end{equation*}
In the double OPE limit, the whole expression of $b$ is a contact interaction and we can use overall Mellin delta function. Then finally evaluating the support of the Mellin delta function
\begin{eqnarray}
    2+4 s_1+4 s_2+4 s_3+4 s_4+4 s_5=3 d,
\end{eqnarray} 
we get 
\begin{align}
& -b(1,2,3,4,5)\notag
\\ \nonumber
= &
\underset{
    \begin{array}{l}
    \substack{\scriptscriptstyle 
 k^2_{12} \to 0}
 \\
      \substack{ \scriptscriptstyle  k^2_{45} \to 0}
    \end{array} 
    }
    {\mathrm{Res}}
    \left (
    \frac{a_1(1,2,3,4,5)}{\mathcal{D}_{k_{12}}^{d-1} \mathcal{D}_{k_{45}}^{d-1}}
    +
    \frac{a_2(1,2,3,4,5)}{\mathcal{D}_{k_{12}}^{d-1}}
    +
    \frac{a_2(4,5,1,2,3)}{\mathcal{D}_{{k_{45}}}^{d-1}}
    \right) \notag
    \\
=&\frac{
\left(s_1-s_2\right)\left(s_4-s_5\right)
}{
z^3
}
\varepsilon_1\cdot \varepsilon_2 \varepsilon_4\cdot \varepsilon_5\varepsilon_3\cdot (k_4+k_5) .
\end{align}

\section{Details on bootstrapping graviton amplitudes}

\subsection{Fixing additional term in four-point GR amplitude from Adler's zero condition}
\label{four point details}
On fixing $a$ and $b$: for gravity, i.e.  $\frac{a(1,2,3,4)}{\mathcal{D}_{k_s}^d}+\frac{b(1,2,3,4)}{k_s^{2 m}}$, it is almost Lorentz invariant when we consider flatspace limit using equation \eqref{Flatspacelimit}. 
This corresponds to the fact that we miss the contribution of the graviton propagator in \eqref{qwqwert} with no OPE poles. We can bootstrap out the missing term as follows.
Adding up the following term the 
$\frac{a(1,2,3,4)}{\mathcal{D}_{k_s}^d}+\frac{b(1,2,3,4)}{k_s^{2 m}}+c_0^{\mathrm{GR}}(1,2,3,4)$ will be Lorentz invariant in flat space limit:
\begin{eqnarray}
  c^\text{GR}_0(1,2,3,4)|_{\text{flat space limit}} 
  =
  \frac{
  z^2\left(-\left((d-2) k_1^2\right)-2(d-2) k_3 k_1+d k_2^2+2 d k_2 k_3+2 k_3^2\right)
  }{
  8(d-1)
}
\varepsilon_{12,34}^2.\notag
\end{eqnarray}

To fill up with the AdS corrections, we first go to AdS by $k_i \to \frac{2s_i+x_i}{z}$,
\begin{eqnarray*}
   c^\text{GR}_0(1,2,3,4)\eqn -\frac{\left((d-2) s_1^2+2(d-2) s_3 s_1-d s_2\left(s_2+2 s_3\right)-2 s_3^2\right)}{2(d-1)}\varepsilon_{12,34}^2+f(s,x)\varepsilon_{12,34}^2,
\end{eqnarray*}
where the $f(s,x)$ captures subleading terms in the flatspace limit which corresponds to AdS curvature corrections.

There should not be any corrections involving cubic power of $s$ (dual to $z\partial z$ in coordinate) because we only consider 2 derivative gravity. The quadratic power of $s$ is already fixed by the flatspace limit.

Next, to determine curvature corrections that are linear and zeroth order in 
$s$'s, we want part of our amplitude to have a longer wavelength so it can feel the curvature.
This boiled down to the Adler zero condition for the bootstrap algorithm: we do a generalized dimension reduction for 1 and 2: $\varepsilon_1 \cdot \varepsilon_2=1, \varepsilon_1 \cdot k=0, \varepsilon_2 \cdot k=0$ and  after dimension reduction the amplitude is zero in the soft limit of $k_1$,
\begin{eqnarray*}
  \lim\limits_{k_1\to 0} \left[ \frac{a^\text{GR}(1,2,3,4)}{\mathcal{D}_{k_s}^d}+\sum_{m=1}^2\frac{b^{(m,\text{GR})}(1,2,3,4)}{k_s^{2 m}}+c^\text{GR}_0(1,2,3,4)\right]\bigg|_{\varepsilon_1 \cdot \varepsilon_2=1, \varepsilon_1 \cdot k=0, \varepsilon_2 \cdot k=0}=0.
\end{eqnarray*}
After some linear algebra 
\footnote{
Technically we take terms quadratic in $x$ as uniformly a constant: $x_i x_k=f$. Then we get a set of linear equations in $x$'s after extracting the coefficients of $s$'s
}, we find
\begin{eqnarray}
  c_0(1,2,3,4)^{\mathrm{GR}}=  \frac{\left(2(d-2) s_1^2-(d-2) s_1\left(d-4 s_3\right)-2 d s_2^2+d s_2\left(d-4 s_3\right)+2 s_3\left(d-2 s_3\right)\right)}{4(d-1)}\varepsilon_{12,34}^2.\label{kwef}
\end{eqnarray}
Putting it in a more compact form by Mellin delta, we have
\eqs{
    &c_0^{\mathrm{GR}}(1,2,3,4) \\
    =&\varepsilon^2_{12,34}\frac{8d(s_1-s_2)(s_3-s_4)-4(s_1+s_3-s_2-s_4)^2+d^2}{16(d-1)}.\label{c0explain1}
}
\subsection{Five point GR single OPE pole}

\subsubsection{$b_2^{\text{Type I}}$}

\label{GRsingleope}

 In practice, instead of \eqref{b2def} which determines the single OPE pole contribution of type I, one can just pick a single channel 
 \begin{eqnarray}
 &&-   b_{21}^{(m),I,\text{GR}}(1,2;3,4,5)\non \eqn \underset{\substack{\scriptscriptstyle 
 k^{2m}_{{12}} \to 0}}{\mathrm{Res}}\left\{ \frac{a_1^\text{GR}(1,2 ; 3 ; 4,5)}{\mathcal{D}_{k_{12}}^d \mathcal{D}_{k_{45}}^d}
 +
 \frac{a_{2s}^\text{GR}(1,2 ; 3,4,5)}{\mathcal{D}_{k_{12}}^d}+\frac{a_{2s}^\text{GR}(5,4 ; 3,2,1)}{\mathcal{D}_{k_{45}}^d}
 +
 \sum_{m_1, m_2=1}^2\frac{b_1^{(m_1,m_2,\text{GR})}(1,2 , 3 , 4,5)}{k_{12}^{2 m_1} k_{45}^{2 m_2}}\right \},\non \label{b21def}
\end{eqnarray}
where
$a_{2 s,\text{GR}}(1,2 ; 3,4,5)=\sum_h \mathcal{M}_3\left(1,2,-k_I\right) \cdot\left(\frac{b^\gr\left(k_I, 3,4,5\right)}{k_{45}^{2 m}}+c^\gr\left(k_I, 3,4,5\right)\right)$.

Then one can get the single channel single OPE contribution of type I by doing a sum of the canonical order of the last three labels
\begin{eqnarray}
    b_2^{(m,\gr)}(1,2 ; 3,4,5)^{\text {Type I }}=b_{21}^{(m,I,\gr)}(1,2 , 3,4,5)+b_{21}^{(m,I,\gr)}(1,2 , 4,5,3)+b_{21}^{(m,I,\gr)}(1,2 , 5,3,4),
    \non 
\end{eqnarray}
and one should use the Mellin delta function and on-shell condition to eliminate spurious poles. 

\subsubsection{$b_2^{\text{Type II}}$}

\label{GRsingleope2}

In practice, instead of \eqref{b2def2} which determines the single OPE pole contribution of type II, we compute a new $b_{21}$ 
\begin{eqnarray*}
   && -b_{21}^{(m), II,\text{GR}}(1,5 ; 4,3,2)
  \non
  \eqn 
     \underset{\substack{\scriptscriptstyle 
 k^{2m}_{{12}} \to 0}}{\mathrm{Res}} \left\{\frac{a_1(1,5 ; 4 ; 3,2)}{\mathcal{D}_{k_{15}}^d \mathcal{D}_{k_{23}}^d}+\frac{a_{2 s}(5,1 ; 4,2,3)}{\mathcal{D}_{k_{15}}^d}+\frac{a_{2 s}(2,3 ; 4,5,1)}{\mathcal{D}_{k_{23}}^d}+\sum_{m_1, m_2=1}^2\frac{b_1^{\left(m_1, m_2, \mathrm{GR}\right)}(1,5,4,3,2)}{k_{15}^{2 m_1} k_{23}^{2 m_2}}\right\}.
\end{eqnarray*}
To get the single OPE contribution of type II, we need the following combination 
\begin{eqnarray*}
    b_2^{(m,\text{GR})}(1,5 ; 4,3,2)^{\text {Type II}}
    =
    b_{21}^{(m, I,\text{GR})}(5,1,2,3,4)+b_{21}^{(m, II,\text{GR})}(1,5,4,3,2)+b_{21}^{(m, II,\text{GR})}(1,5,3,2,4),
\end{eqnarray*}
with $ b_{21}^{(m, I,\text{GR})}(5,1,2,3,4)$ given in former section.

\section{Identity operator}
\begin{center}
    \begin{tikzpicture}[scale=1.3]
\begin{feynman}
	\node[circle, draw, fill=gray!30, minimum size=20pt] (a) at (0,0) {$J_L$};
	\node[circle, draw, fill=gray!30, minimum size=20pt] (b) at (2,0) {$J_R$};
	
	\diagram* {
		(a) -- [gluon] (b) 
	};
	
	\vertex (gluon1) at (-1.5,1) {};
	\vertex (gluon2) at (-1.57,0.5) {};
	\vertex (gluon3) at (-1.57,-0.5) {};
	\vertex (gluon4) at (-1.5,-1) {};
	
	\diagram* {
		(a) -- [gluon] (gluon1),
		(a) -- [gluon] (gluon2),
		(a) -- [gluon] (gluon3),
		(a) -- [gluon] (gluon4),
	};
	
	\node[rotate=90, transform shape] at ($(gluon2)!0.5!(gluon3)$) {$\cdots$};
	
	\vertex (gluon5) at (3.5,1) {};
	\vertex (gluon6) at (3.57,0.5) {};
	\vertex (gluon7) at (3.57,-0.5) {};
	\vertex (gluon8) at (3.5,-1) {};
	
	\diagram* {
		(b) -- [gluon] (gluon5),
		(b) -- [gluon] (gluon6),
		(b) -- [gluon] (gluon7),
		(b) -- [gluon] (gluon8),
	};
	
	\node[rotate=90, transform shape] at ($(gluon6)!0.5!(gluon7)$) {$\cdots$};
	
	\node[yshift=6pt] at (a.north) {$\vec{k}_L, s_L$};
	\node[yshift=6pt] at (b.north) {$\vec{k}_R, s_R$};
	
	\path (a) -- node[above] {$\frac{1}{\cD_{k_R}}$} (b);
\end{feynman}
\end{tikzpicture}
\end{center}

In higher point computation in Mellin-momentum amplitude,  due to boundary momentum conservation, we find that when taking limits e.g. OPE or soft limits, different channels of diagrams will combine together. Some exchange terms would turn out to be contact. The usual form that may be suspected to be an identity operator would be
\eqs{
&
z_L^a
\frac{
    z_L^2k_L^2
    -
    \left(a+\frac{n_L d}{2}-\Delta-2 s_{L}\right)
    \left(a+\frac{(n_L-2) d}{2}-\Delta-2 s_{L}\right)}{\mathcal{D}^{\Delta}_{k_R}}
}
where $k_L^2=k_R^2$, $a$ is the power of $z$, $n_L$ denotes the number of legs on the left, and $s_L$ sums all the left Mellin variables. And we should think of the inverse operator is acting on the RHS of the sub-amplitude.
This can also be understood from IBP, as we explain in next section and we provide examples in \ref{ymgridexample}

\subsection{Proof}

Consider having a numerator which is of the form in general
$f(s_L,k_L)$
and denominator
$\frac{1}{\cD_{k_R}^\Delta}$.
We have the Mellin-momentum amplitude as 

\begin{eqnarray}
    \mathcal{A}=
    {f}(s_L,k_L)  z_L^a {A}_L
  \frac{  1}{\cD_{k_R}^\Delta}
    {A}_Rz_R^b
    \label{cqqw}
\end{eqnarray}
where we have taken out explicit $z$ dependence for Mellin-momentum amplitudes and write it as $\cA_L=z_L^a {A}_L$.
Here, the split of $z_L$ and $z_R$ are determined by primitive amplitudes, e.g. $\cA_3$\footnote{
For YM $\cA_3$ carries one power of $z$ while power of two for GR.
}
that $\cA_R$ and $\cA_L$ are built upon.
And the $s_L$ is directly acting on bulk-to-boundary propagator.

We claim that
\begin{eqnarray}
   {f}(s_L,k_L)  z_L^a {A}_L
  \frac{  1}{\cD_{k_R}^\Delta}
    {A}_Rz_R^b
    =
    A_LA_R z^{a+b}\label{qwqd}
\end{eqnarray}
iff 
\begin{eqnarray}
\boxed
{
f\left(s_L,k_L\right)=
    z^2k_L^2
    -
    \left(a+\frac{n_L d}{2}-\Delta-2 s_{L}\right)
    \left(a+\frac{(n_L-2) d}{2}-\Delta-2 s_{L}\right).
    }
    \label{eiq}
\end{eqnarray}
where $s_L=s_{1...n_L}\equiv \sum_{j=1}^{n_L} s_j$ and there is no splitting of $z$ on the RHS in \eqref{qwqd} as it is already a contact term.

To show this, we can first proof 
\begin{eqnarray}
    \boxed{\int \frac{d z_L}{z_L^{d+1}} \Phi\left(z_L, k_i\right)\left[\mathcal{D}^{z_L \Delta}_{k_L}\right]\left[G_{\Delta}\left(k_R, z_L, z_R\right)\right]
    =
   \int \frac{d z_L}{z_L^{d+1}}
    \left[
    G_{\Delta}\left(k_R, z_L, z_R\right)
    \right]  \left[
    {\mathcal{D}}^{z_L \Delta}_{k_L}\right]
    \Phi(z_L,{k_i})},\nonmy
    \label{mideq}
\end{eqnarray}
where 
\begin{eqnarray}
    \mathcal{D}_k^{\Delta} \equiv z^2 k^2-z^2 \partial_z^2-(1-d) z \partial_z+\Delta(\Delta-d).
\end{eqnarray}

\textbf{Proof:}
adding two total derivative terms and after some algebra, we get, 
\begin{eqnarray}
   && \int \frac{d z_L}{z_L^{d+1}}
   \Phi(z_L,{k_i})
    {\mathcal{D}}^{z_L}{ }_{k_L}^{\Delta}
    G_{\Delta}\left(k_R, z_L, z_R\right)
+
    \nonmy
    && 
    +
    \partial_{z_L}\left[z_L^2 \frac{1}{z_L^{ d+1}}
   \Phi(z_L,{k_i})
    \partial_{z_L} G(z_R,z_L,k_R)\right]
    -
    \partial_{z_L}\left[z_L^2 \frac{1}{z_L^{ d+1}}
     \partial_{z_L} \Phi(z_L,{k_i})
   G(z_R,z_L,k_R)\right]
   \nonmy
   \eqnmy
   \int \frac{d z_L}{z_L^{d+1}}
    G_{\Delta}\left(k_R, z_L, z_R\right) 
    {\mathcal{D}}^{z_L}{ }_{k_L}^{\Delta}
    \Phi(z_L,{k_i}).
  \qquad \qquad \qquad\qquad \qquad \qquad \qquad  \qed
\end{eqnarray}

Now we move back to proof \eqref{eiq}.
Starting from \eqref{qwqd}, we have equivalently 
\begin{eqnarray}
    \mathcal{A}=z_L^a A_L\mathcal{D}_{k_L}^{\Delta} \frac{1}{\mathcal{D}_{k_R}^{\Delta}} A_R z_R^b.
    \label{des}
\end{eqnarray}

To determine $f\left(s_L, k_L\right)$
we can introduce back the $z$ integrals when we go back to momentum space from \eqref{des},
\begin{eqnarray}
   && \int \frac{d z_L}{z_L^{d+1}} \frac{d z_R}{z_R^{d+1}} z_L^a A_L\left( \prod_{i=1}^{n_I} \phi_{\Delta}\left(z_L, k_i\right)\right) {\mathcal{D}}_{{z_L},k_L}^{\Delta}G_{\Delta}\left(k_R, z_L, z_R\right)
    \left(\prod_{i=n_I+1}^n \phi_{\Delta}\left(z_R, k_i\right)\right)A_R z_R^b
    \nonmy
    \eqnmy
    \int \frac{d z_L}{z_L^{d+1}} \frac{d z_R}{z_R^{d+1}} G_{\Delta}\left(k_R, z_L, z_R\right){\mathcal{D}}_{{z_L},k_L}^{\Delta}\left( z_L^a\prod_{i=1}^{n_I} \phi_{\Delta}\left(z_L, k_i\right)\right) 
    \left(\prod_{i=n_I+1}^n \phi_{\Delta}\left(z_R, k_i\right)\right) z_R^b A_L A_R
    \non
\end{eqnarray}
where in the last line we used \eqref{mideq}. Writing 
$
z_L^a\prod_{i=1}^{n_I} \phi_{\Delta}\left(z_L, k_i\right)
\propto
z_L^a\prod_{i=1}^{n_I} z^{-2s_i+d/2}$, we found that $\mathcal{D}_{z_L, k_L}^{\Delta}$ creates the following $f$ in \eqref{cqqw}

\begin{eqnarray*}
    f\left(s_L,k_L\right)=
    z^2k_L^2-
    \left(a+\frac{n_L d}{2}-\Delta-2 \sum_{j=1}^{n_L} s_j\right)
    \left(a+\frac{(n_L-2) d}{2}-\Delta-2 \sum_{j=1}^{n_L} s_j\right).\qed 
\end{eqnarray*}

\subsection{Examples}
\label{ymgridexample}
\subsubsection{Yang-Mills}
Identity operator for four-point YM amplitude in s-channel is 
\begin{eqnarray}
    1=\frac{z^2k_{12}^2+2(s_1+s_2)(d-2-2s_1-2s_2)}{\cD_{k_{34}}^{d-1}}.
\end{eqnarray}
This corresponds to $f\left(s_L, k_L\right)|_{a=-1,n_L=2,\Delta=d-1}$. The reason $a=-1$ is that the numerator has $z$ to power 2 while originally the YM vertex on the 12 side has only $z$ to power 1. One has to write a $1/z$ to compensate the $z$ power.\\
In the special case of $d=3$, this identity turns into 
\begin{eqnarray}
    1=\frac{z^2k_{12}^2-z^2(k_1+k_2)^2}{\cD_{k_{34}}^{d-1}}.
\end{eqnarray}
After using the mapping to correlator we can see that the result is not just $\frac{1}{E_t}$ but also contains a local term $\frac{1}{k_{12}+k_3+k_4}$. This is as expected because in momentum space the form of correlators is not unique but up to local terms.

\subsubsection{Gravity}
Identity operator for four-point GR amplitude in s-channel is 
\begin{eqnarray}
    1=\frac{z^2k_{12}^2+2(s_1+s_2)(d-2s_1-2s_2)}{\cD_{k_{34}}^{d}}
\end{eqnarray}
This corresponds to $f\left(s_L, k_L\right)|_{a=0,n_L=2,\Delta=d}$. The reason $a=0$ is that the numerator has $z$ to power 2 while originally the 3 point GR vertex on the 12 side has same power. So no compensation of $z$ is needed.

\section{Derivation for GR scalar integrals} \label{Appendix:GRcontact}
\textbf{Derivation of Eq (\ref{contactSIGR})}:
\eqs{
\mathcal{C}_{GR}^{n;1}=&\int \frac{dz}{z^4}z^{2}\prod_i^n \phi_{\Delta=3}(z,k_i).
}
In $d=3$, we can express the contact integration as
\begin{eqnarray}
    \mathcal{C}_{G R}^{n ; 1}=	\int_0^{\infty} \frac{d z}{z^4} z^{2 } e^{-z \sum_{j=1}^n k_n} \prod_{j=1}^n\left(k_j z+1\right),
    \label{contacintg}
\end{eqnarray}
where the $z^2$ comes from the additional scaling factor for gravity contact interaction vertex.

The first two inverse powers of $z$ diverge separately, but we can regularized them by putting analytic regulator on the $z$ power. 
And we find after adding them together we get finite contribution:
\begin{eqnarray}
	\int_0^{\infty} \frac{d z}{z^{4+\epsilon} }z^2 (1+z\sum_{j=1}^{n}k_j) e^{-z \sum_{j=1}^n k_n}=-E_t+O(\epsilon).
\end{eqnarray}
This accounts for the second term in \eqref{contactSIGR}.
For other finite integration, we have
\begin{eqnarray}
	\int_0^{\infty} z^{m} e^{-E_t z}=\frac{m!}{E_t^{m+1}},
\end{eqnarray}
whose coefficients are the corresponding cyclic sum of $k$'s: $\sum_{1 \leq i_1<\ldots<i_{m+2}}^n=k_{i_1} \ldots k_{i_{m+2}}$.\footnote{For example, for $m=0$ at contact 3 point it is $k_1k_2+k_1k_3+k_2k_3$.}

\textbf{Derivation of Eq (\ref{contact2}):}
\eqs{
\mathcal{C}_{G R}^{n ; 2}=\int \frac{dz}{z^4} \prod_{m=1}^l\left(-2 s_m+d / 2\right)\phi_{\Delta=3}(z,k_i).
}
By definition 
\begin{eqnarray}
\left(-2 s_m + \frac{d}{2}\right)\eqn z \frac{\partial}{\partial z} \Bigg|_{\text{acting on only the m th}\text{ bulk-to-boundary propagator,}}
\non 
\left(-2 s_m + \frac{d}{2}\right)e^{-z k_m}(1+z k_m) 
\eqn - z^2 k_m^2 e^{-z k_m},
\end{eqnarray}
where the second line is the first line acting on scalar part of graviton bulk-to-boundary propagator in $d=3$.

So the Mellin $s$ variable
$\prod_{m=1}^l\left(-2 s_m+d / 2\right) M(k, \varepsilon) \rightarrow M(k, \varepsilon) \mathcal{C}_{G R}^{n ; 2}$
changes the $\left(k z+1\right)$ to $z^2k^2$ in the integrand in \eqref{contacintg}:
\begin{eqnarray}
\mathcal{C}_{G R}^{n ; 2}=
\int_0^{\infty} \frac{d z}{z^4} z^{2l} e^{
-z \sum_{i=1}^n k_n
}
\prod_{j=l+1}^n\left(k_j z+1\right) (-1)^lk_1^2 \ldots k_l^2.
\end{eqnarray}
For number of derivatives greater than two ($l \geq 2$), the power of $z$ in the integrand gives $\int_0^{\infty} z^{(2 l-4) +m} e^{-E_t z}=\frac{(2 l-4+m) !}{E^{(2l-4)+m+1}}$.
The power of $m$ can be at most $n-l$, so we eventually end up with
\eqs{
\mathcal{C}_{G R}^{n ; 2}=\Bigg(\sum_{m=0}^{n-l} {(2l-4+m) !} \sum_{\begin{array}{c}
  \substack{\scriptscriptstyle l+1\le i_{l+1}<} \\  \substack{\scriptscriptstyle  \ldots<i_{m+l}}
\end{array}}^n \frac{k_{i_{l+1}} \ldots k_{i_{m+l}}}{E_t^{(2l-4)+m+1}}\Bigg)(-1)^lk_1^2...k_l^2,
}
where when $m=0$, the numerator above $E_t$ is just 1. This identity holds for all the Mellin variables with power $1$, as in YM, if we have a Mellin variable of $s$ power greater than 2, one has to use on-shell condition to lower the power.
\bibliography{refs.bib} 
\bibliographystyle{JHEP}

\end{document}